\documentclass[10pt,twoside]{report}

\usepackage{graphicx}
\usepackage{amsfonts,amssymb}
\usepackage[dvips,breaklinks=true,bookmarksdepth=section]{hyperref}
\usepackage[dvips]{bookmark}
\usepackage{apalike}
\usepackage{listings}
\def\ddt#1{\frac{\partial #1}{\partial t}}
\def\ddth#1{\frac{\partial #1}{\partial \theta}}
\def\ddp#1{\frac{\partial #1}{\partial p}}
\def\ddx#1{\frac{\partial #1}{\partial x}}
\def\ddy#1{\frac{\partial #1}{\partial y}}
\def\unsurn{\frac{1}{N}}

\def\undemi{\frac{1}{2}}
\def\Dth{\Delta\theta}
\def\Dp{\Delta p}
\def\zb{\bar z}
\def\Imean{\bar I}
\def\bmean{\bar b}
\def\etal{{\it et al}}
\def\H{\mathcal{H}}

\makeatletter
\newcommand\l@abstract[2]{\vskip 6pt plus 2pt minus 2pt\hbox{\hbox to .05\linewidth{} \minipage{.85\textwidth} {\footnotesize \indent #1}\endminipage} \vskip 6pt plus 2pt minus 2pt}
\newcommand\l@defs[2]{\par \hbox{\hbox to .05\linewidth{} \minipage{.85\textwidth} {\footnotesize \indent {\bf Definitions:\ }#1}\endminipage}\vskip 6pt plus 2pt minus 2pt\par}
\def\toclevel@abstract{5} 
\def\toclevel@defs{5} 
\def\@linkbordercolor{1 1 1}
\def\@citebordercolor{1 1 1}
\makeatother

\newcommand\myabstract[2]{%
  \addcontentsline{toc}{abstract}{#1}%
  \addcontentsline{toc}{defs}{#2}%
    \rule{\linewidth}{1pt}\par\vskip 6pt%
  {\small #1}\par \noindent\rule{\linewidth}{1pt}\vskip 12pt%
}

\begin{document}
\title{Vlasov dynamics of 1D models with long-range interactions}
\author{Pierre de Buyl}

%\maketitle

\begin{titlepage}
%  \maketitle
\pdfbookmark[2]{Vlasov dynamics of 1D models with long-range interactions}{Vlasov dynamics of 1D models with long-range interactions}
  
  \begin{center}
    {\large Universit{\'e} Libre de Bruxelles \\ Center for Nonlinear Phenomena and Complex Systems}
  \end{center}
\bigskip

  \begin{center}
    \rule{\linewidth}{1pt}\par\vskip 6pt%
    {\huge Vlasov dynamics of 1D models with long-range interactions}
    \rule{\linewidth}{1pt}\par\vskip 6pt%

    \bigskip
    {\large Pierre de Buyl}
    \includegraphics[width=\linewidth,draft=false]{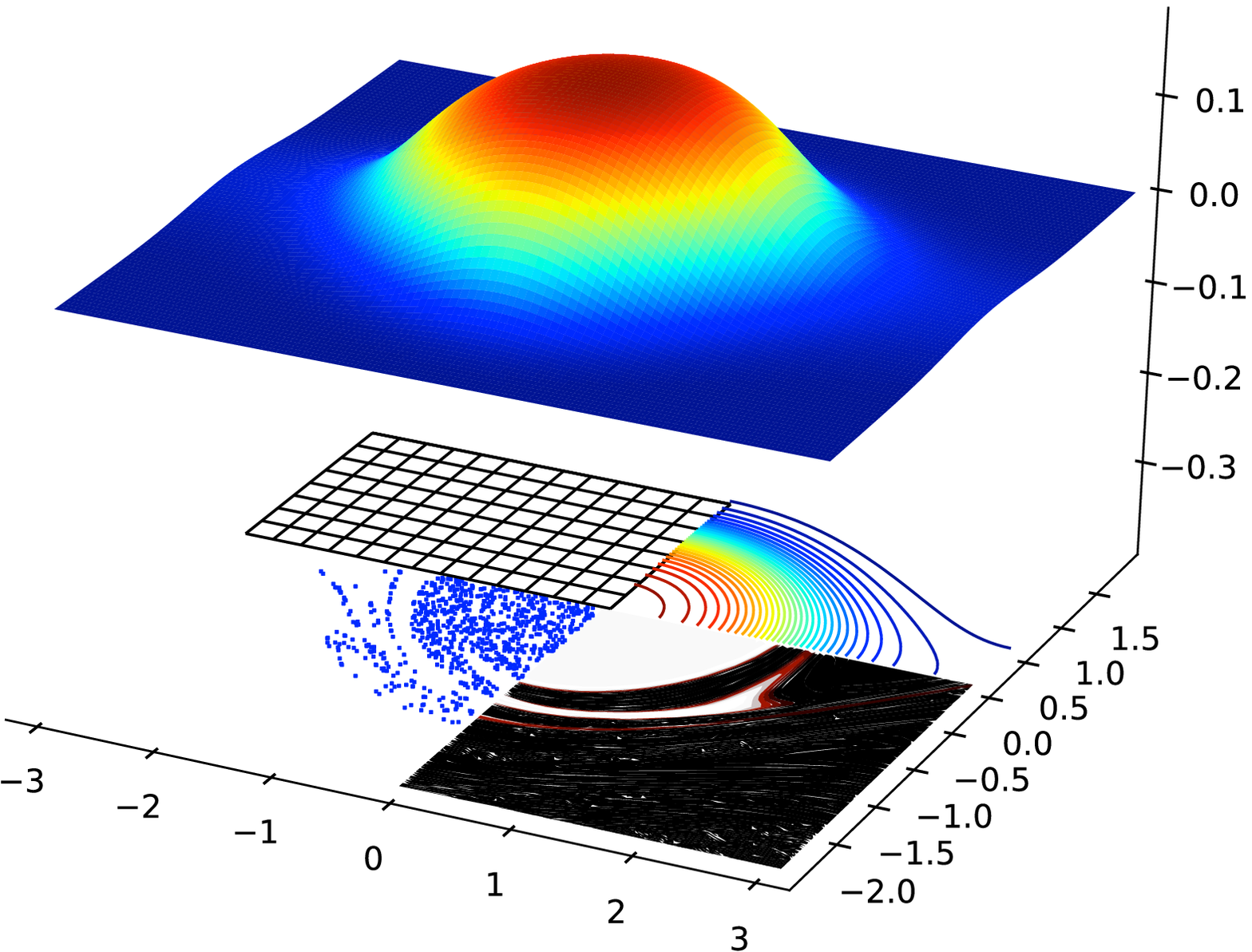}
  \end{center}
  \bigskip
  \bigskip

  \begin{center}
    {\large Academic year 2009-2010}
  \end{center}
  \begin{center}

    {\large Dissertation presented in partial fulfilment of the requirements for the degree of Doctor of Philosophy in Physics\\
      {\bf Supervisor~: }Prof. Pierre Gaspard}
  \end{center}
\end{titlepage}

\cleardoublepage
\vfill
\begin{flushright}
Thoughtcrime does not entail death: thoughtcrime is death.\\
%If all records told the same tale - then the lie passed into history and became truth.\\
Georges Orwell, Nineteen Eighty Four
\end{flushright}
\cleardoublepage
\vfill
\begin{flushright}
  {\`a} Sophie,\\
  pour {\it toujours}.
\end{flushright}

%\listoffigures

\cleardoublepage
\pdfbookmark[2]{Abstract}{Abstract}\section*{Abstract}

Gravitational and electrostatic interactions are fundamental examples of systems with long-range interactions.
Equilibrium properties of simple models with long-range interactions are well understood and exhibit exotic behaviors~: negative specific heat and inequivalence of statistical ensembles for instance.

The understanding of the dynamical evolution in the case of long-range interacting systems still represents a theoretical challenge. Phenomena such as {\it out-of-equilibrium phase transitions} or {\it quasi-stationary states} have been found even in simple models.

The purpose of the present thesis is to investigate the dynamical properties of systems with long-range interactions, specializing on one-dimensional models. The appropriate kinetic description for these systems is the Vlasov equation. A statistical theory devised by D.~Lynden-Bell is adequate to predict in some situations the outcome of the dynamics.
A complementary numerical simulation tool for the Vlasov equation is developed.

A detailed study of the {out-of-equilibrium phase transition} occuring in the Free-Electron Laser is performed and the transition is analyzed with the help of Lynden-Bell's theory.
Then, the presence of stretching and folding in phase space for the Hamiltonian Mean-Field model is studied and quantified from the point of view of fluid dynamics.
Finally, a system of uncoupled pendula for which the asymptotic states are similar to the ones of the Hamiltonian Mean-Field model is introduced. Its asymptotic evolution is predicted via both Lynden-Bell's theory and an exact computation. This system displays a fast initial evolution similar to the violent relaxation found for interacting systems. Moreover, an out-of-equilibrium phase transition is found if one imposes a self-consistent condition on the system.

In summary, the present thesis discusses original results related to the occurence of quasi-stationary states and out-of-equilibrium phase transitions in 1D models with long-range interaction.
The findings regarding the Free-Electron Laser are of importance in the perspective of experimental realizations of the aforementioned phenomena.

\clearpage

\section*{R{\'e}sum{\'e}}

Les interactions gravitationnelles et {\'e}lectrostatiques sont deux exemples fondamentaux de syst{\`e}mes en interaction de longue port{\'e}e. Les propri{\'e}t{\'e}s d'{\'e}quilibre de mod{\`e}les simples en interaction de longue port{\'e}e sont bien comprises et r{\'e}v{\`e}lent des comportemens exotiques~: capacit{\'e} sp{\'e}cifique n{\'e}gative et in{\'e}quivalence des ensembles statistiques par exemple.

La compr{\'e}hension de l'{\'e}volution dynamique dans le cas de syst{\`e}mes en interaction de longue port{\'e}e repr{\'e}sente encore actuellement un d{\'e}fi th{\'e}orique. Des mod{\`e}les simples pr{\'e}sentent des propri{\'e}t{\'e}s telles que des {\it transitions de phase hors d'{\'e}quilibre} ou des {\it {\'e}tats quasi-stationnaires}.

Le but de la pr{\'e}sente th{\`e}se est d'{\'e}tudier les propri{\'e}t{\'e}s dynamiques de syst{\`e}mes en interaction de longue port{\'e}e pour des mod{\`e}les {\`a} une dimension. La description cin{\'e}tique ad{\'e}quate est donn{\'e}e par l'{\'e}quation de Vlasov. Une th{\'e}orie statistique propos{\'e}e par D.~Lynden-Bell est appropri{\'e}e pour pr{\'e}dire dans certaines situations l'aboutissement de la dynamique. Un outil de simulation pour l'{\'e}quation de Vlasov compl{\`e}te cette approche.

Une {\'e}tude d{\'e}taill{\'e}e de la transition de phase dans le Laser {\`a} Electrons Libres est pr{\'e}sent{\'e}e et la transition est analys{\'e}e {\`a} l'aide de la th{\'e}orie de Lynden-Bell.
Ensuite, la pr{\'e}sence d'{\'e}tirement et de repliement est {\'e}tudi{\'e}e dans le mod{\`e}le Hamiltonian Mean-Field en analogie avec la dynamique des fluides.
Enfin, un syst{\`e}me de pendules d{\'e}coupl{\'e}s dont les {\'e}tats asymptotiques sont similaires {\`a} ceux du mod{\`e}le Hamiltonian Mean-Field est introduit. Son {\'e}volution asymptotique est pr{\'e}dite par la th{\'e}orie de Lynden-Bell {\it et} par une approche exacte. Ce syst{\`e}me pr{\'e}sente une {\'e}volution initiale rapide similaire {\`a} la relaxation violente pr{\'e}sente dans des mod{\`e}les plus compliqu{\'e}s. De plus, une transition de phase hors d'{\'e}quilibre est trouv{\'e}e si une condition d'auto-consistence est impos{\'e}e.

En r{\'e}sum{\'e}, la pr{\'e}sente th{\`e}se comporte des r{\'e}sultats originaux li{\'e}s {\`a} la pr{\'e}sence d'{\'e}tats quasi-stationnaires et de transitions de phase hors d'{\'e}quilibre dans des mod{\`e}les unidimensionnels en interaction de longue port{\'e}e.
Les r{\'e}sultats concernant le Laser {\`a} Electrons Libres offrent une perspective de r{\'e}alisation exp{\'e}rimentale des ph{\'e}nom{\`e}nes d{\'e}crits dans cette th{\`e}se.

\clearpage

\section*{Copyright notice}
The present thesis is copyrighted to Pierre de Buyl.

Most of the results and figures in chapter~\ref{chap:numerical} have appeared in de Buyl, Commun. Nonlin. Sci. Numer. Simulat. doi:10.1016/j.cnsns.2009.08.020 (2009).
Most of the results and figures in chapter~\ref{chap:fel} have appeared in de Buyl {\it et al}, Phys. Rev. ST Accel. Beams {\bf 12} 060704 (2009).

\section*{List of publications}

Some of the results presented in this thesis have been presented in the following publications~:
\begin{itemize}
\item {\it Out-of-equilibrium mean-field dynamics of a model for wave-particle interaction}\\
  P. de Buyl, D. Fanelli, R. Bachelard and G. De Ninno\\
  Physical Review Special Topics - Accelerators and Beams {\bf 12} 060704 (2009)
\item {\it Numerical resolution of the Vlasov equation for the Hamiltonian Mean-Field model}\\
  P. de Buyl\\
  Communications in Nonlinear Science and Numerical Simulation (2009)\\
  doi:10.1016/j.cnsns.2009.08.020
\item {\it Transition de phases hors-d'{\'e}quilibre dans le Laser {\`a} Electrons Libres}\\
P. de Buyl, R. Bachelard, M.-E. Couprie, G. De Ninno and D. Fanelli\\
``Comptes-Rendus de la 12$^{\textrm{{\`e}me}}$ Rencontre du Non-Lin{\'e}aire'', Non-Lin{\'e}aire Publications, 2009
\item {\it Deep saturation dynamics in a Free-Electron Laser}\\
  R. Bachelard, M.-E. Couprie, P. de Buyl, G. De Ninno and D. Fanelli\\
  Proceedings of the FEL09 Conference, Liverpool, 2009
\end{itemize}

% Local Variables:
% TeX-master: "main"
% End:

\pdfbookmark[2]{Table of contents}{Table of contents}\tableofcontents

\section*{Remerciements}
\addcontentsline{toc}{chapter}{Remerciements}

Au terme de 4 ans, 3 mois et quelques jours de travail comme doctorant, je peux enfin imprimer un gros paquet de papier avec des {\'e}quations et des beaux dessins!

La place d'honneur de ces remerciements va de droit {\`a} Pierre Gaspard, mon promoteur.
Ma passion pour la physique est assez {\'e}tendue, que ce soit pour la physique de tous les jours ou la physique th{\'e}orique.
Les cours donn{\'e}s par Pierre Gaspard ont {\'e}veill{\'e} mon int{\'e}r{\^e}t pour le domaine de la physique macroscopique, de mani{\`e}re plus pr{\'e}cise dans la dynamique non-lin{\'e}aire et la m{\'e}canique statistique.
Son enseignement est rigoureux et insiste sur le lien entre une description quantitative pr{\'e}cise et l'interpr{\'e}tation physique qui peut en d{\'e}couler.
Je souhaite mentionner {\'e}galement que sa capacit{\'e} {\`a} dessiner au tableau (en 2 dimensions) des sch{\'e}mas dans un espace de param{\`e}tres {\`a} 3 dimensions avec du relief m'a toujours impression{\'e}.

J'ai r{\'e}alis{\'e} mon m{\'e}moire de licence {\`a} Florence gr{\^a}ce {\`a} Pierre Gaspard, chez un de ses coll{\`e}gues dans le domaine de la m{\'e}canique statistique, Stefano Ruffo. Ce fut une exp{\'e}rience fort enrichissante et je suis reconnaissant au groupe de Florence pour son accueil. J'ai ensuite pu entamer un travail de doctorat {\`a} Bruxelles. % sous sa direction.
Je tiens {\`a} remercier Pierre Gaspard pour les opportunit{\'e}s qu'il m'a offertes et pour son soutien, plus particuli{\`e}rement dans les derni{\`e}res semaines tr{\`e}s rock'n roll de ma th{\`e}se.

Toutes ces ann{\'e}es d'{\'e}tudes auraient sembl{\'e} bien longues si je n'avais pu compter sur la particularit{\'e} souvent v{\'e}rifi{\'e}e des physiciens d'{\^e}tre fort sympathiques. Tout au long des ann{\'e}es, j'ai pu {\'e}tudier, travailler et faire des pauses avec Ariane, Quentin, Jean-R{\'e}my, Thomas, St{\'e}phane, Nassiba, Vincent, Nathan, Renaud, Sonia, Olivier, Jonathan, Julie, Julien, C{\'e}dric, Gilles, Ella-coloc, Ma{\"\i}t{\'e}, J{\'e}r{\^o}me, Nir, Cyril, Michael, Fran{\c c}ois, Audrey, S{\'e}verine, Giorgio, Jessica, Delphine, Laurence, Yasmina, Huguette (attention, des chimistes se cachent dans le tas, vers la fin!).
Une mention sp{\'e}ciale va {\`a} Ariane, ma bin{\^o}me en 1{\`e}re et 2{\`e}me candi (et plus tard mon t{\'e}moin!), et {\`a} Nathan, Nassiba et Vincent avec qui nous avons affront{\'e} la d{\'e}but de th{\`e}se unis, au milieu de la Plaine. Ils ont fini un peu avant, mais {\c c}a ne change pas grand chose\footnote{Voir V. Wens, PhD Thesis (2009)}.
Merci {\`a} Nathan d'avoir permis l'{\'e}coute de musique Folk dans le bureau 2.O5.102 toutes ces ann{\'e}es. Notre bureau fut un lieu d'amiti{\'e}, de nombreuses discussions scientifiques tout {\`a} fait ``correctes'' et de rock and roll.

Renaud a rapatri{\'e} son labo du Solbosch un peu plus tard. Sa pr{\'e}sence a la Plaine fut fortement appr{\'e}ci{\'e}e, ainsi que le partage de tous les aspects de l'universit{\'e}.

Lors de ma premi{\`e}re ann{\'e}e de th{\`e}se, {\`a} environ $30 s$ {\`a} pied se trouvait le bureau de ma soeur Sophie de Buyl. Elle a fait le tour du monde depuis, mais en suivant toujours mes progr{\`e}s, merci pour ton soutien. Merci {\`a} mes parents, {\`a} mon fr{\`e}re Martin et {\`a} ma famille et ma belle-famille pour leurs encouragements.
De fa{\c c}on tout {\`a} fait hors de propos, merci {\`a} tous les musiciens~: Adeline, Sophie, Arnaud, Jean, Luc, Marie-Paule, Ariane, Kim, Hugo, St{\'e}phane et Philou.

Chronologiquement, un de mes premiers contacts avec la science fut la venue de Pasquale Nardone expliquant des exp{\'e}riences {\`a} l'{\'e}cole primaire.
La suite se d{\'e}roule {\`a} l'{\'e}cole secondaire~: Philippe L{\'e}onard enseigne la physique en d{\'e}veloppant un sens de l'intuition tr{\`e}s fort, en lien avec la description math{\'e}matique, et donne {\`a} ses cours une dimension historique. Ce dernier point rend son enseignement m{\'e}morable et passionnant. Il a fort certainement orient{\'e} mon choix vers les sciences physiques. J'ai eu le grand plaisir de pouvoir travailler dans le mus{\'e}e de physique de l'ULB qu'il dirige dans le cadre de ma fonction d'assistant.

Les charges d'enseignement que j'ai assum{\'e}es au cours de ces ann{\'e}es de th{\`e}se sont une source d'inspiration p{\'e}dagogique et scientifique, ce pour quoi je souhaite remercier les {\'e}tudiants et leur int{\'e}r{\^e}t dans des choses diverses, allant de la bille qui tombe aux amplificateurs op{\'e}rationnels en passant par la thermodynamique et la m{\'e}canique quantique.
Je remercie {\'e}galement mes coll{\`e}gues d'enseignement avec qui les discussions furent nombreuses et int{\'e}ressantes~: Laura, Claire, Paola, Nicolas, Jonathan, Priscilla, Laurent, Kael, Paul, Michael, Dominique, Henri, Nicole, Fabian, Fabienne, Vincent, Michel, L{\'e}on, Yassin, Marc, Frank, Manu, Laetitia, Michel, Philippe, Barbara.

Deux coll{\`e}gues acad{\'e}miques ont {\'e}t{\'e} particuli{\`e}rement amicaux vis {\`a} vis d'un jeune coll{\`e}gue. Leurs pr{\'e}sence, leur soutien, leur franchise et leur exp{\'e}rience m'ont {\'e}t{\'e} indispensables. Malek et Francis, merci d'avoir {\'e}t{\'e} l{\`a}!

Dans le cadre de mes recherches, j'ai appr{\'e}ci{\'e} la pr{\'e}sence d'{\^e}tres humains dans ``le NO'', en particulier mais pas exclusivement au 5{\`e}me {\'e}tage: les fourmis\footnote{Pardon, je voulais dire ``les membres de l'Unit{\'e} d'Ecologie Sociale'' bien s{\^u}r.}, les chimistes dont la pr{\'e}sence matinale dans la salle caf{\'e} est fort appr{\'e}ci{\'e}e~: Anne, Jean-Christophe, Genevi{\`e}ve et Yannick (jadis illustre assistant). Les membres du nouvellement nomm{\'e} groupe ``Physique des Syst{\`e}mes Complexes et M{\'e}canique Statistique'' fournissent un environnement de recherche stimulant. Eric, Cem, Gr{\'e}gory, S{\'e}bastien, David, Jim, Vasilis, Anselmo, Jean-Pierre Boon, Gr{\'e}goire Nicolis, Daniel Kosov, Thomas Gilbert et plus r{\'e}cemment Massi. Les discussions avec Thomas sont toujours vives et enti{\`e}res! Je partage avec Jonathan un int{\'e}r{\^e}t excessif pour les librairies graphiques et num{\'e}riques.
Merci {\`a} Jean-Sabin d'avoir apport{\'e} le Canada ici tout pr{\`e}s.

Je remercie {\'e}galement les personnes avec qui j'ai eu la chance de travailler en dehors de l'ULB, au d{\'e}part de Florence principalement~: Duccio Fanelli, Romain Bachelard, Giovanni De Ninno, David Mukamel et Stefano Ruffo. Les collaborations qu'ils ont rendues possibles m'ont beaucoup aid{\'e} {\`a} d{\'e}marrer mon m{\'e}tier de chercheur.

Mon travail de chercheur a b{\'e}n{\'e}fici{\'e} d'un soutien amical de Malek (oui, oui, le m{\^e}me) qui m'a encourag{\'e} {\`a} plusieurs reprises et qui a m{\^e}me {\'e}t{\'e} jusqu'{\`a} m'autoriser {\`a} s{\'e}journer dans son bureau.

Je remercie pour leurs commentaires sur le texte de cette th{\`e}se Thomas Gilbert, Nathan Goldman, Sophie de Buyl et Pierre Gaspard.

Je remercie Stefano Ruffo, L{\'e}on Brenig, Michel Mareschal et Daniel Kosov d'avoir accept{\'e} de faire partie de mon jury.

Pour dire merci, pour parler de physique, j'ai trouv{\'e} bien des mots. Si vous voulez conna{\^\i}tre mon secret du bonheur, je vous r{\'e}pondrai {\it Sophie}. Merci pour tout ce bonheur et merci pour ton soutien ces derniers mois.

% Local Variables:
% TeX-master: "main"
% End:

\cleardoublepage
\chapter{Introduction}
\label{chap:intro}

\myabstract{The long-range character of the interaction modifies significantly the macroscopic behaviour of physical systems. We present general properties of such systems. %
The reasons motivating their investigations are observations in the context of galactic dynamics and in systems of charged particles. %
We specialize on the particular topic of the collisionless evolution in 1D models of systems with long-range interactions. %
The Vlasovian description of these systems has been the focus of recent interest, inducing the need for a numerical implementation.}{systems with long-range interactions, continuum limit.}

\section{Motivations}
\label{sec:intro-motivations}

%\nocite{chandrasekhar_principles_steldyn_1942,chandrasekhar_intro_stellar_structure_1939}
% Add celestial mechanics, context and physical motivation
Gravitational and electrostatic interactions are fundamental examples of systems with long-range interactions.
The long-range character of the interaction has a profound impact on the thermodynamical and dynamical features of these systems.
A treatment for neutral plasmas allows one to introduce a screening while for gravitational systems, non-neutral plasmas and beams of charged particles the unshielded interactions manifest long-range properties.

Despite its commonness in physics, the detailed study of long-range interacting systems has only recently attracted important interest from the statistical physics community, and has led to the organization of conferences and summer schools \cite{long-range-02,long-range-07,les-houches-2008}. The focus of the long-range community has been mainly devoted to toy models and equilibrium statistical mechanics in the microcanonical ensemble, bringing out peculiarities such as ensemble inequivalence, negative specific heat and non-additivity. These phenomena do not exist in systems with short-range interactions which, in contrast, are additive.

The aforementioned meetings brought together specialists from different fields, experimentalists and theoreticians alike, with the objective of bringing links between theory and experiments in the coming years. This perspective seems especially promising for trapped plasmas \cite{drewsen_et_al_assisi_2007}, dipolar interactions \cite{menotti_et_al_assisi_2007} or Free-Electron Lasers \cite{curbis_et_al_epjb_2007}.

Another trend is to focus on the dynamical phenomena found in systems with long-range interactions. More precisely, the slow relaxation to equilibrium, or the lack of relaxation, has attracted much interest. This feature offers challenges and motivations in the context of gravitational systems or wave-particle interactions. The transitory stage of the dynamics in which the system may remain trapped for long times before eventually reaching thermodynamical equilibrium is referred to as a Quasi-Stationary State (QSS) \cite{antoni_ruffo_1995,antoniazzi_et_al_pre_2007}. The existence of out-of-equilibrium phase transitions separating different dynamical regimes contributes to enrich the interest in dynamical studies  \cite{antoniazzi_et_al_prl_2007,antoniazzi_et_al_pre_2007,latora_et_al_prl_1998,de_buyl_et_al_prstab_2009}.

A precise description of the dynamical evolution can be achieved thanks to kinetic theory. Chavanis has dedicated a series of papers (\cite{chavanis_I} \nocite{chavanis_II,chavanis_III,chavanis_IV}to \cite{chavanis_V}) to the application of kinetic theory to systems with long-range interactions. In addition, a number of articles have focused on the Vlasov equation \cite{yamaguchi_et_al_physica_a_2004,barre_et_al_pre_2004,barre_et_al_physica_a_2006,antoniazzi_califano_prl}, which is adequate to describe systems where collisional effects are absent. This equation is also called {\it collisionless Boltzmann equation} in astrophysics. The Vlasov equation allowed the understanding of stability and dynamical properties, and is the basis for Lynden-Bell's theory, a statistical theory aimed at the prediction of collisionless evolution following the so-called {\it violent relaxation}. The theory was introduced in Ref. \cite{lynden-bell_1967}.

%A sign of the importance given to the Vlasov equation is the interpenetration of both fields taking place, for instance, at the Vlasovia workshop 2009.
%REFREF (Firpo Vlasovia 2009, Kaiser, Manos, Paskauskas, Turchi).

The purpose of this thesis is to study the dynamical properties of the Vlasov equation, especially in the process of relaxation. We consider the Hamiltonian Mean-Field model \cite{antoni_ruffo_1995} which can be viewed as a paradigm to study systems with long-range interactions and the Colson-Bonifacio model for the Free-Electron Laser \cite{bonifacio_et_al_pra_1986}, an example of wave-particle interaction. We study the Vlasov equation itself through direct numerical resolution, going beyond molecular dynamics simulations. We detail a numerical procedure to perform these computations. We finally use an analogy with a set of uncoupled pendula to refine our understanding of the dynamics.

\section{Examples of systems with long-range interactions}
\label{sec:intro-examples}

We give in this section a number of physical situations where the long-range nature of the interaction needs to be taken into account. A number of them derives from the fact that charged particles interact via a $1/r$ potential\footnote{In this introduction, $r$ is generically considered the distance between particles.}, and in some situations, only the mathematical description is effectively long ranged while the underlying physical system is not.

\paragraph{Charged particles.}

\begin{figure}[ht]
  \centering
  \includegraphics[width=1\linewidth]{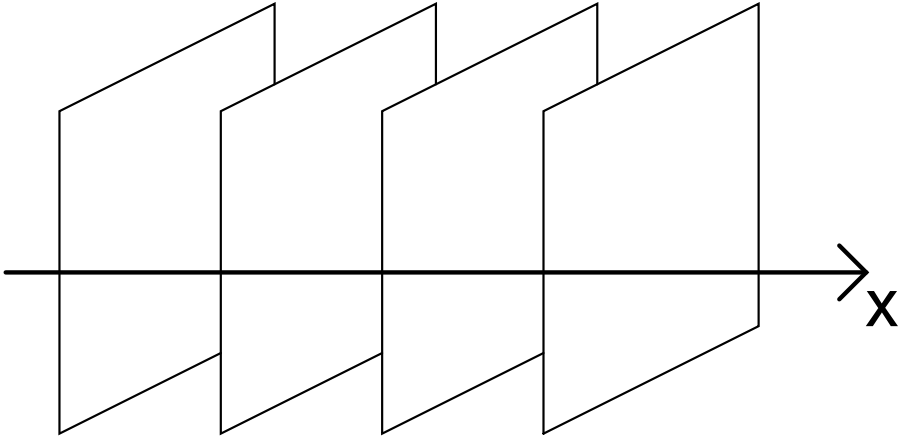}
  \caption{A system of parallel sheets. The position is given along the direction normal to the sheets.}
  \label{fig:intro-sheets}
\end{figure}

Given a density of particles $\rho$, the interaction potential $\Phi$ between particles under Coulombian interaction is given by the solution of Poisson's equation~:
\begin{equation}
  \Delta \Phi = - \rho
\end{equation}
where $\Delta$ is the Laplacian operator. The solution is an inter-particle potential $V(r)$ of $1/r$ in 3D, $-\log r$ in 2D and $-r$ in 1D. These potentials are considered long range~: the decay of $V(r)$ in function of $r$ is weak. They correspond to point particles in 3D, infinite charged parallel wires in 2D and sheets in 1D. The infinite charged wires are an adequate description for beams of charged particles, which we describe in a separate paragraph. The 1D system is equivalent to a set of parallel charged sheets (see Fig.~\ref{fig:intro-sheets}) and does not correspond to a physically observable system. It has served as a theoretical model to study plasma physics since a long time [see for instance Ref.~\cite{dawson_rmp}].

\paragraph{Models of gravitational interaction.}

The relaxation time of galaxies, i.e. the time after which we expect the galaxy to have reached thermodynamical equilibrium, can be as long as the age of the universe \cite{chandrasekhar_principles_steldyn_1942}. Galactic dynamics thus needs to be analyzed from a collisionless perspective. The 1D sheet model presented belows allows for a simplified study of the phenomenon. In this thesis, the topic of gravitational interaction motivates the use of the Hamiltonian Mean-Field model defined in chapter~\ref{chap:models}~: it can be obtained by retaining only the first Fourier mode of the 1D sheet model.

Given a density of particles $\rho$, the gravitational potential $\Phi$ between particles is given by the solution of Poisson's equation~:
\begin{equation}
  \label{eq:poisson}
  \Delta \phi = \rho
\end{equation}
where $\Delta$ is the Laplacian operator. The solution is an inter-particle potential $V(r)$ of $-1/r$ in 3D while in 1D the solution is $r$.

The 1D sheet model has been the subject of numerous numerical studies \cite{hohl_NASA_1968,luwel_et_al_relaxation_times_1984,luwel_severne_1985} motivated by its simplicity and the existence of exact (up to machine precision) algorithms to simulate its dynamics. It is usually called the gravitating sheets model, because the 1D version corresponds to a hypothetical system of parallel rigid sheets, described by their positions along the normal axis (see Fig~\ref{fig:intro-sheets}).

%Gravitational systems present peculiar thermodynamical and relaxational properties. A microcanonical treatment reveals 
% a peculiar particle distribution: most of the particles are located in the ``core'', while a number of more energetic ones orbit in a ``halo'' \cite{yamashiro_et_al_origin_of_core-halo}. This gives rise to a complicated description, and some modifications to statistical theories have been devised specifically to adapt to this core-halo distribution \cite{yamaguchi_pre_2008}.

\paragraph{Beams of charged particles.}

In a beam of charged particles (ions or electrons), the longitudinal velocity along the beam propagation direction (say $z$) can be considered a constant. The particles can then be described in the $x-y$ plane only, with a logarithmic repulsive interaction.

This system has been the subject of experimental work on vortex dynamics and turbulence \cite{driscoll_fine_experiments}, and is also a relevant description for charged particles circulating in an accelerator \cite{benedetti_et_al_physica_a_2006}.

\paragraph{Wave-particle interactions.}

Wave-particle interactions are useful to describe situations in which the inter-particle forces can be neglected in comparison with the mean-field coupling provided by an electromagnetic wave. While in this thesis we interpret this model as a description of a Free-Electron Laser device (see chapter~\ref{chap:fel} for details), it is of more general interest in plasma physics \cite{elskens_escande_book,del-castillo-negrete_procs_02}.

The coupling via a wave allows for effective long-range interactions and provides rich dynamical properties. The reduction of the wave to one component allows for a simplified treatment. The Colson-Bonifacio model that we investigate in this thesis is however relevant to describe features of realistic Free-Electron Laser devices \cite{bonifacio_et_al_nuovo_cimento_1990,bonifacio_et_al_pra_1986,curbis_et_al_epjb_2007}.

\paragraph{Other systems of interest.}

Two-dimensional flows can be considered effectively long-ranged. The vorticity field $\ddx{u_y} - \ddy{u_x}$ [where $u_x$ ($u_y$) is the velocity in the $x$ ($y$) direction] displays a logarithmic interaction. A statistical theory similar to Lynden-Bell's theory predicts the evolution of the flow \cite{chavanis_et_al_2D_vortices_1996}.

The dipolar interaction is also long-range, going as $1/r^3$ in 3D. The dipolar interaction is a candidate for experiments on long-range phenomena \cite{menotti_et_al_assisi_2007}.

We call ``Small system'' a system in which the range of the interaction is of the order of the size of the system. In that setup, each particle feels the presence of the complete set of particles in the system. Thermodynamics may be affected by this property and some features such as non-additivity or negative specific heat may appear \cite{chomaz_gulminelli_procs_02}.

\section{Thermodynamical behaviour of long-range interacting systems}
\label{sec:intro-thermo}

The range of the interaction potential defines a neighborhood for each particle. In order to model the properties of an ensemble of particles, we take into account the coupling of a particle to all of its neighbours. As an example to illustrate these properties, we take the case of the Ising model, the Hamiltonian of which is the following:
\begin{equation}
  \label{eq:ising_model}
  H = - J \sum_{<i,j>=1}^N S_i S_j
\end{equation}
where $J$ is a coupling constant, $S_i = -1,1$ the spin value at site $i$ and the brackets in the sum means that the sum is on nearest neighbours only. In the simple setup of a square lattice, a spin has 2 nearest neighbors in one dimension, 4 in two dimensions and 6 in three dimensions. It is clear that the interaction energy for a single site is bounded by its number of neighbours.

Extending the range of the interaction to infinity leads to mean-field models:
\begin{equation}
  \label{eq:ising_mean-field}
    H = -{J} \sum_{i\leq j=1}^N S_i S_j = -\frac{J}{2} \left( \sum_{i=1}^{N} S_i \right)^2 = -\frac{J}{2} N^2 M^2
\end{equation}
where the sum runs over all pairs $i,j$ and 
\begin{equation}
 M=\frac{1}{N}\sum_i S_i 
\end{equation}
is the magnetization.
The bound of the energy at each site is now growing as $N$ and the global energy scales as $N^2$. While $H$ can be made extensive by an appropriate rescaling, it cannot be made additive. 

\paragraph{Lack of additivity.} In systems with short-range interactions, the interactions between two subsystems take place at the interface between the two subsystems. The bulk energy then grows with the volume, while the interfacial interaction only grows in proportion to the surface. The interfacial term thus become negligible in the thermodynamic limit.

In long-range interacting systems, the interfacial interaction needs to take into account all particles and therefore grows in proportion to the volume. It cannot be neglected anymore when compared to the bulk energy. We illustrate that property using systems (\ref{eq:ising_model}) and (\ref{eq:ising_mean-field}).

\begin{figure}[ht]
  \centering
    \begin{minipage}[h]{0.5\linewidth}
      \includegraphics[width=\linewidth]{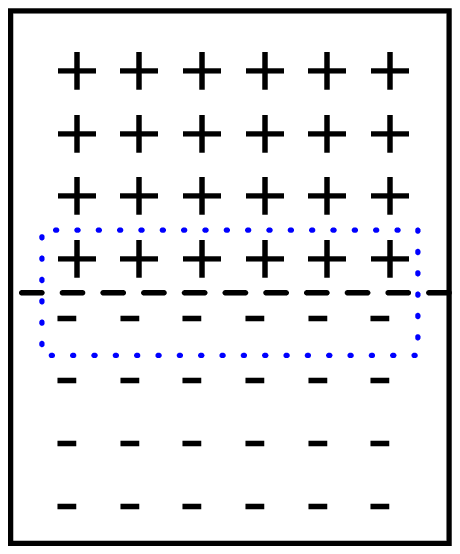}    
  \end{minipage}
  \hskip 0.02\linewidth
  \begin{minipage}[h]{0.47\linewidth}
  \caption{Additivity of a system of spins interacting with their nearest neighbours~: The division of the system into two parts is such that the overall energy is approximately the sum of the energy of the two parts considered separately. The dotted line indicates the region that contributes to the difference between $H^a+H^b$ and $H^{a+b}$, negligible with respect to the size of the bulk contribution.}
  \label{fig:NN_Ising}
  \end{minipage}
\end{figure}

Let us first consider the Hamiltonian~(\ref{eq:ising_model}) in two dimensions. If we divide the systems into two parts $a$ and $b$ whose spins are labeled respectively $S^a_i$ and $S^b_i$. When the subsystems are placed next to each other (see Fig.~\ref{fig:NN_Ising}), the total energy $H^{a+b}$ can be expressed as:
\begin{equation}
  \label{eq:NN_Ising_ab}
  H^{a+b} = H^a + H^b -J \sum_{i,j} S^a_i S^b_j 
\end{equation}
where the sum runs over nearest neighbours only (here along the boundary indicated in Fig.~\ref{fig:NN_Ising}).
The surface term scales as $\sqrt{N}$ and therefore becomes negligible compared to $H^{a+b}\propto N$ in the thermodynamic limit. Namely, we may write $H^{a+b}\approx H^a+H^b$.

\begin{figure}[ht]
  \centering
  \begin{minipage}[h]{0.5\linewidth}
  \includegraphics[width=\linewidth]{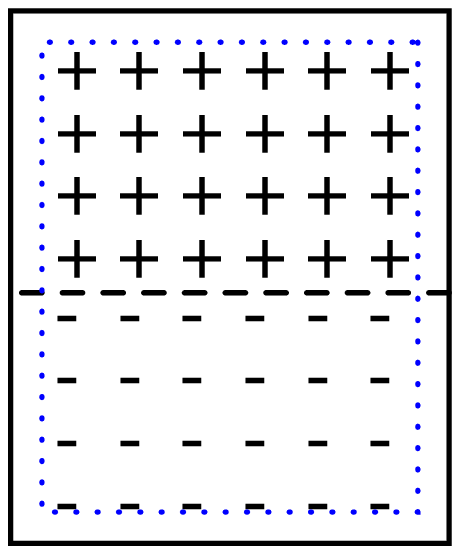}    
  \end{minipage}
  \hskip 0.02\linewidth
  \begin{minipage}[h]{0.47\linewidth}
    \caption{Non-additivity of a mean-field spin system. The division of the system into two parts cannot approximate the properties of the full system. The dotted line indicates the region that contributes to the difference between $H^a+H^b$ and $H^{a+b}$. In the case of long-range interaction, that region can be of the size of the complete system.}
    \label{fig:mean-field_Ising}
  \end{minipage}
\end{figure}

We now turn to the system~(\ref{eq:ising_mean-field}). Dividing it into two subsystems, we write the energy of the total system as:
\begin{equation}
  \label{eq:mean-field_Ising_ab}
  H^{a+b} = -\frac{J}{2} (\sum_{i=1}^{N^a} S^a_i + \sum_{i=1}^{N^b} S^b_i)^2 = H^a + H^b - J (\sum_{i=1}^{N^a} S^a_i) (\sum_{i=1}^{N^b} S^b_i)
\end{equation}
where the non-additive part cannot be neglected anymore (see Fig.~\ref{fig:mean-field_Ising}).

\paragraph{Ensemble inequivalence.} The usual equivalence between the microcanonical  and canonical ensembles can be proved on general grounds for systems satisfying the additivity property. In systems with long-range interactions, the equivalence of ensembles is no longer granted and needs to be checked individually for model under consideration. In the case of inequivalence, the adequate physical description depends on the constraints put on the system: isolated systems correspond to the microcanonical ensemble while systems in contact with a heat bath correspond to the canonical ensemble.

Ensemble equivalence holds for the models considered in this thesis. We refer the reader to Refs. \cite{barre_et_al_prl_2001,barre_phd,barre_et_al_j_stat_phys_2005,campa_et_al_phys_rep_2009} to find a discussion of ensemble inequivalence. Reference \cite{barre_et_al_prl_2001} investigates the phase diagram of the Blume-Emery-Griffiths model with infinite range interactions in which ensemble inequivalence is found.

\paragraph{Negative specific heat.} The positive character of the specific heat also follows from additivity. A negative specific heat can be found in the microcanonical study of systems with long-range interactions. The occurrence of negative specific heat implies the property of ensemble inequivalence~: it can be proved to be positive in the canonical ensemble \cite{barre_et_al_prl_2001,de_buyl_et_al_xy_2005}.
%REF thirring ?.

\section{Dynamical properties}
\label{sec:intro-dyn}

Numerical experiments on the Hamiltonian Mean-Field model (HMF) have led to the discovery of a particular dynamical phenomenon, the so-called quasi-stationary states (QSS). Starting from an initial condition, the system evolves in time towards a QSS before eventually reaching equilibrium. For a time which turns out to diverge when the number of particles is increased, the macroscopic quantities characterizing the system remain distinct from the value expected at equilibrium.
Situations of interest arise when the time within which we observe the system is such that equilibrium statistical mechanics does not provide a good prediction. It is consequently a challenge to understand these QSS and ultimately to predict their properties.

An unusual feature is that the emergence of dynamical properties in systems with long-range interactions greatly depends on the initial condition, while the initial condition is usually considered irrelevant in classical statistical mechanics.
It eventually leads to an {\it out-of-equilibrium phase transition}, a change in the macroscopic observable found when the system is kept away from thermodynamical equilibrium.

%Another model that has been the subject of much interest is the Gravitational Sheets Model where properties that also displays peculiar relaxation properties and a core-halo distribution.

\paragraph{Quasi-Stationary States (QSS).}

Quasi-Stationary States (QSS) are dynamical states in which the system displays features of equilibrium but that does not correspond to the equilibrium predicted by statistical mechanics in the following sense:
\begin{itemize}
\item The system eventually experiences a change in its macroscopic variable.
\item That change takes place at a time that increases when the number of particles considered increases.
\item The macroscopic quantities characterizing the system do not correspond to those predicted by statistical mechanics.
\end{itemize}
Their observation was reported in numerical experiments and have since attracted a lot of attention in the literature \cite{yamaguchi_et_al_physica_a_2004,barre_et_al_physica_a_2006,chavanis_qss_2006,antoniazzi_et_al_pre_2007}. The time spent in a QSS may be very large, and a prediction of that state is highly desirable.

The Vlasov equation allows one to cast a theoretical framework to understand the QSSs. The time of validity of the description by the Vlasov equation is investigated in \cite{jain_et_al_relaxation_times_2007} and discussed in relation with numerical simulations in \cite{antoniazzi_califano_prl}.

A statistical theory devised by Lynden-Bell and appropriate for the Vlasov equation \cite{lynden-bell_1967} has been used with success to predict the QSS for the Hamiltonian Mean-Field \cite{antoniazzi_et_al_pre_2007,antoniazzi_et_al_prl_2007}, and to describe them in the Free-Electron Laser \cite{barre_et_al_pre_2004,curbis_et_al_epjb_2007,de_buyl_et_al_prstab_2009}

Numerical simulations on gravitational systems report cases of slow relaxation \cite{luwel_et_al_relaxation_times_1984}.
Yamaguchi applied a slightly modified Lynden-Bell theory to the 1D sheet model and discussed the concept of QSS in that context \cite{yamaguchi_pre_2008}.

\paragraph{Core-Halo distribution.}

A core-halo distribution is a non-monotonous particle distribution. It has been first observed in the Gravitating Sheets Model \cite{yamashiro_et_al_origin_of_core-halo}, but can be found also in the Hamiltonian Mean-Field model. No statistical theory has been predicting these distributions which are understood as a purely dynamical phenomenon.

\paragraph{Out-of-equilibrium phase transition.}

An out-of-equilibrium phase transition is a dynamical phenomenon that can cause a system to evolve into radically different regimes, depending on the parameters of the initial condition. In the Hamiltonian Mean-Field model, for instance, such a transition is found numerically and can be predicted by Lynden-Bell's theory \cite{antoniazzi_et_al_prl_2007}.
Further analysis revealed that the transition is related to the occurrence of regular structures in phase space, from a one-cluster state to a two-clusters state \cite{bachelard_et_al_prl_2008}. In the Free-Electron Laser, a similar out-of-equilibrium phase transition exists, and the analysis of the associated phase space structures is expected to display a rich phenomenology \cite{de_buyl_et_al_prstab_2009}.

\paragraph{The Vlasovian approach.}

The description of the dynamical evolution by the Vlasov equation is valid in the thermodynamic limit $N\to\infty$. Its use helps the understanding of dynamical phenomena for systems with long-range interactions.
The description of the quasi-stationary states of the dynamics by Lynden-Bell's theory, which is based on the Vlasov equation, is a remarkable theoretical achievement. It has been to predict successfully the magnetization in the Hamiltonian Mean-Field model \cite{antoniazzi_et_al_prl_2007,antoniazzi_et_al_pre_2007} and the intensity in the free-electron laser \cite{barre_et_al_pre_2004,curbis_et_al_epjb_2007}.

The kinetic description the Vlasov equation provides does not take into account the number of particles $N$. However, the numerical simulations performed in the literature are almost exclusively of the $N$-body type. An exception is given by Ref.~\cite{antoniazzi_califano_prl}. The impact of the number of particles on the dynamics has been quantified, with special interest on the time scale for which the Vlasov equation and the $N$-body description coincide \cite{yamaguchi_et_al_physica_a_2004,jain_et_al_relaxation_times_2007}.

This thesis investigates the dynamical evolution of the Vlasov equation through direct numerical simulation in which the number of particles $N$ is irrelevant. This approach brings a new tool to the study of models with long-range interactions and helps to shed light on the detailed phase space dynamics \cite{de_buyl_numerical_hmf_2009}.

\paragraph{Relaxation scenario}

\begin{figure}[t]
  \centering
  \includegraphics{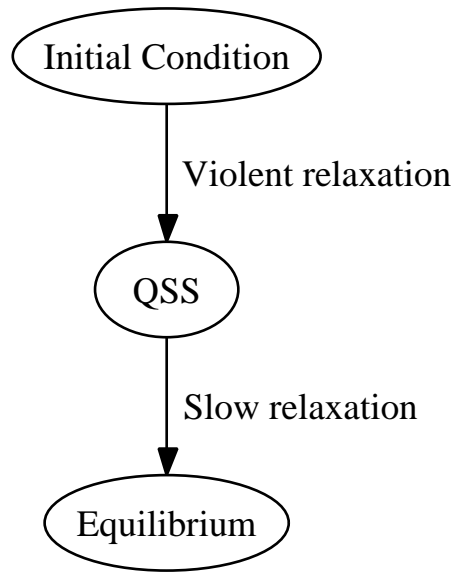}
  \caption{The relaxation scenario for Vlasov dynamics, as proposed in Ref.~\cite{barre_et_al_physica_a_2006}. An initial condition evolves through violent relaxation on a timescale of order $O(1)$. If the dynamics gets trapped into a quasi-stationary state (QSS), the time needed to reach equilibrium increases with the number of particles~: $t\approx N^\delta$ or $t\approx \log N$ \cite{jain_et_al_relaxation_times_2007}.}
  \label{fig:intro-relax}
\end{figure}

The authors of Ref.~\cite{barre_et_al_physica_a_2006} propose that the relaxation scenario in systems with long-range interactions follows the following steps~: a first stage is called {\it violent relaxation}, its time scale does not depend on the number of particles $N$ ; following violent relaxation, the system may get trapped in a {\it quasi-stationary state} (QSS) and eventually attains equilibrium after a time that depends on the number of particles.
This scenario is depicted in Fig.~\ref{fig:intro-relax}.
%The properties of the system in the QSS regime have been predicted shown to correspond to solutions of Lynden-Bell theory for the HMF model \cite{antoniazzi_et_al_prl_2007}.
The dependence on the number of particles present in the system is studied in Refs.\cite{yamaguchi_et_al_physica_a_2004} and \cite{jain_et_al_relaxation_times_2007}.

The use of direct simulations of the Vlasov equation allows one to work in the limit $N\to\infty$ and provides an useful addition to $N$-body simulations regarding the lifetime of the QSSs.
Vlasov simulations also offer the possibility to study the stability of arbitrary solution of Lynden-Bell's theory \cite{de_buyl_et_al_prstab_2009}.

\vfill

\pagebreak
 
\ 
\vfill

\pagebreak

\section{Outline of the thesis}
\label{sec:intro-outline}

\paragraph{Chapter \ref{chap:vlasov}} is a general introduction to the Vlasov equation and its properties.

\paragraph{Chapter \ref{chap:models}} gives recalls briefly the definition of the Hamiltonian Mean-Field model and presents some of its general properties. 

\paragraph{Chapter \ref{chap:numerical}} details the numerical procedure required to solve the Vlasov equation. A study of the Hamiltonian Mean-Field model is performed and the limitations of the method are discussed.
The case of the Free-Electron Laser is also given.

\paragraph{Chapter \ref{chap:fel}} introduces a model for the Free-Electron Laser, its associated Vlasov equation and presents the study of the out-of-equilibrium phase transition.

\paragraph{Chapter \ref{chap:hmf}} studies the relaxation in the Hamiltonian Mean-Field model on the basis of numerical simulations of the Vlasov equation. The analysis benefits of an analogy with fluid dynamics.

\paragraph{Chapter \ref{chap:pendulum}} concerns a set of uncoupled pendula. In that model, we derive an exact solution to the asymptotic evolution of the Vlasov dynamics under an ergodic hypothesis.

\paragraph{Chapter~\ref{chap:conclusions}} concludes this thesis and offers some perspectives to its results.
\\

The table of contents contains a short abstract of each chapter and constitutes a good starting point to the reading of this thesis.

% Local Variables:
% TeX-master: "main"
% End:

\cleardoublepage
\chapter{The Vlasov equation}
\label{chap:vlasov}

\myabstract{We introduce the Vlasov equation as the continuum limit in systems with long-range interactions. We discuss its derivation via the BBGKY hierarchy of kinetic theory and the conservation properties of Vlasov dynamics. We present Lynden-Bell's theory for the statistical mechanics of violent relaxation.}{kinetic description, filamentation, microscopic description, coarse-grained distribution function, violent relaxation.}

The dynamical evolution of particles under mutual interactions is described, within the range of validity of classical mechanics, by Newton's equation, ruling the evolution of each particle's position and velocity.
When the number of particles goes to infinity, it is impossible to follow the particles' coordinates individually.
Instead, we will be interested in the probability with which we find the set of particles at given positions and velocities.
The knowledge of that probability is sufficient to compute the average of macroscopic quantities.
The equation ruling the evolution of that probability is called the Liouville equation, and its resolution is as complicated as solving Newton's equations for the ensemble of individual trajectories, unless simplifying assumptions can be made. The objective of kinetic theory is to identify physically interesting approaches that allow such simplifications and to understand the statistical behaviour of dynamical phenomena~\cite{balescu_statistical_dynamics}.
For instance, we restrict our attention in this thesis to the probability of finding {\it one} particle at a given position and velocity, irrespective of the remaining particles.

%We also note briefly that the Vlasov equation is found to be an adequate semi-classical limit to Schro{\"e}dinger's equation \cite{feix_bertrand_universal_vlasov,bonifacio_et_al_quantum_lasing_2005} and is also analogous to Euler's equations for two-dimensional fluid dynamics \cite{chavanis_et_al_2D_vortices_1996}.

This chapter deals with one approach, namely the Vlasov equation, whose domain of validity is interesting for systems with long-range interactions which are considered in this thesis. This chapter is organized as follows: we discuss the phase space description in section~\ref{sec:vlasov-phasespace}, introduce the Vlasov equation in section~\ref{sec:vlasov-origin} and finally present Lynden-Bell's theory in section~\ref{sec:vlasov-LB}.

\section{The phase space description}
\label{sec:vlasov-phasespace}

We limit ourselves, as this is not a limitation for this thesis, to the study of periodic systems in which the position is labeled by an angle $\theta$ and is defined in the interval $[-\pi;\pi]$. The associated momentum is denoted $p$.

Given a Hamiltonian $H$ for particles with a kinetic energy $\frac{p_i^2}{2}$ and an inter-particle potential energy $V(r)$ (where $r$ is the distance between two particles), the evolution of $\theta$ and $p$ obeys the following equations:
\begin{eqnarray}
  \label{eq:hamilton_eq}
  H &=& \sum_i \frac{p_i^2}{2} + \frac{1}{2}\sum_{i\neq j} V(|\theta_i - \theta_j|) \cr
  \dot \theta_i &=& \frac{\partial H}{\partial p_i} = p_i \cr
  \dot p_i &=& -\frac{\partial H}{\partial \theta_i} = - \sum_{j\neq i} \frac{d}{d\theta_i}V(|\theta_i - \theta_j|)
\end{eqnarray}

\begin{figure}[h!]
  \centering
  \includegraphics[width=0.7\linewidth]{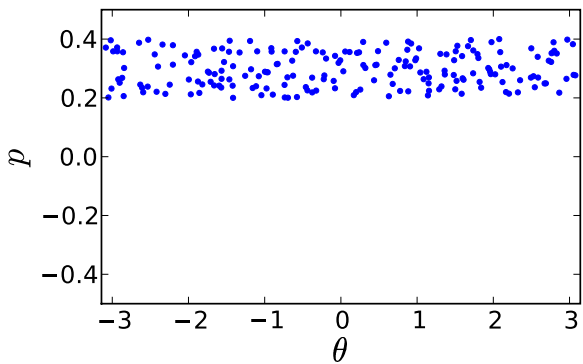}
  \caption{An example of distribution of 200 point particles in phase space. The particles are uniformly distributed and move with a positive average velocity of about $0.3$~.}
  \label{fig:uniform_beam}
\end{figure}

To display the state described by the system (\ref{eq:hamilton_eq}), we project all particles' positions and momenta in $\mu$-space, the space of single-particle coordinates and momenta; we will call it equivalently phase space in this thesis. Fig. \ref{fig:uniform_beam} gives an example of 200 particles uniformly distributed and moving to the right (i.e. with positive velocity).

The system can be described alternatively by a distribution with support is phase space: $f^{N}(\Theta, P)$ [here, $(\Theta, P)$ belongs to the $N$-particle phase space~: it represents the full set $\{(\theta_i,p_i)\}$ of positions and velocities]. $f^{N}(\Theta, P)$ depends on the time $t$, but this dependence is not explicitly written to simplify the notation. In the Vlasov regime (see paragraph \ref{sec:vlasov-origin}), we reduce the problem to the study of the one-particle distribution function~: $N f(\theta,p)\ d\theta\ dp$ is the number of particles having their position and momentum contained in a cell of volume $d\theta\ dp$ centered at $(\theta,p)$. The distribution function $f(\theta,p)$ is normalized to $1$:
\begin{equation}
  \int d\theta\ dp\ f(\theta,p) = 1
\end{equation}

\begin{figure}[h!]
  \centering
  \begin{minipage}{.48\linewidth}
    \includegraphics[width=\linewidth]{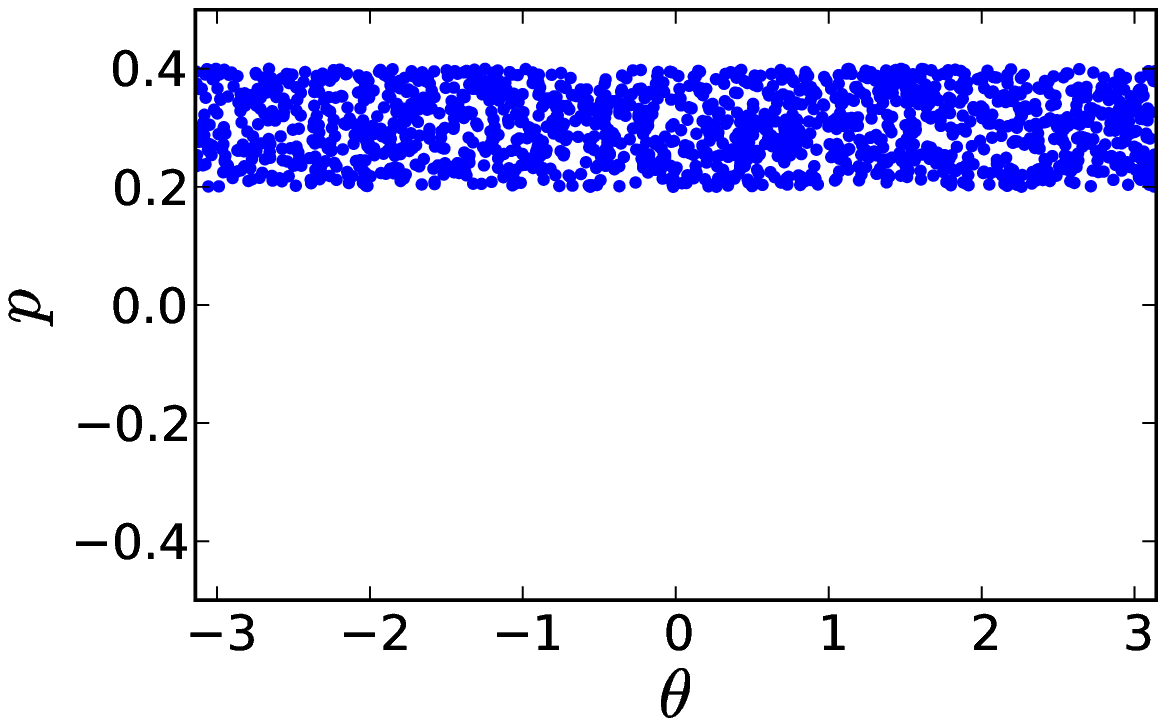}    
  \end{minipage}
  \hskip .03\linewidth
  \begin{minipage}{.48\linewidth}
    \includegraphics[width=\linewidth]{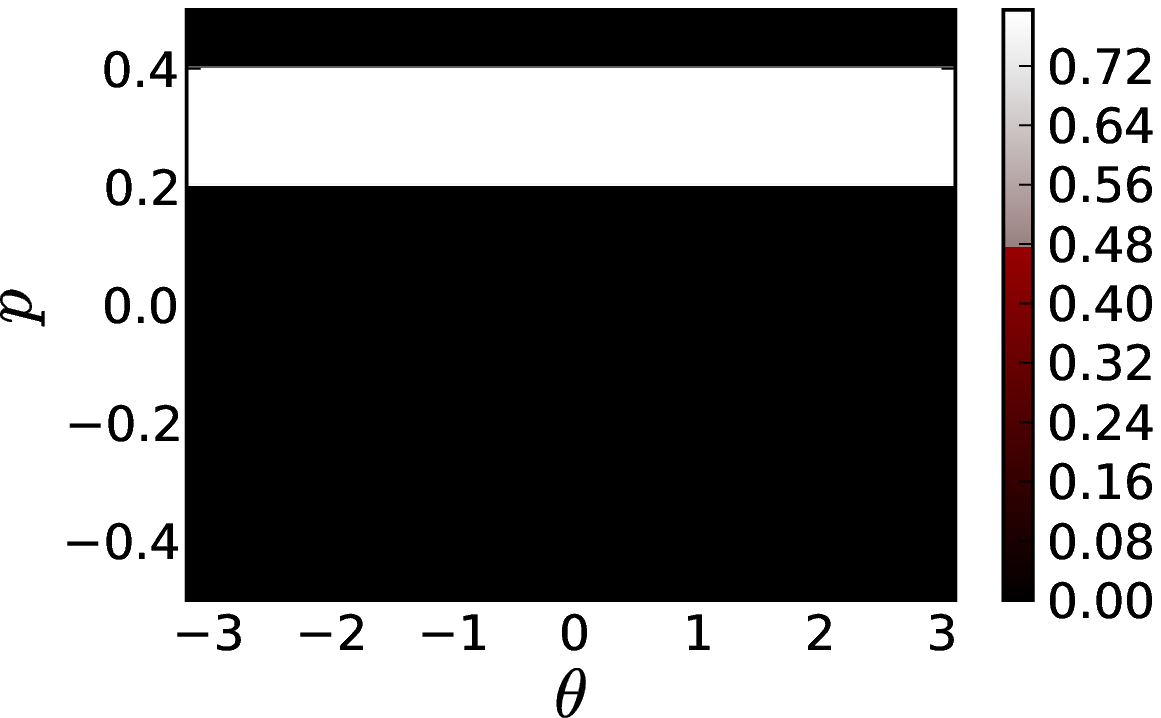}
  \end{minipage}
  \caption{(left panel) An example of distribution of 1500 point particles in phase space. The particles are uniformly distributed and move with a positive velocity of about $0.3$ .\label{fig:uniform_beam_large_n}}
    \caption{(right panel) A kinetic description of the situation depicted in Fig. \ref{fig:uniform_beam_large_n} with a distribution function. The color code on the right indicates the value of the distribution function $f$.\label{fig:uniform_beam_f}}
\end{figure}

We reproduce the example of Fig. \ref{fig:uniform_beam} with more particles in Fig. \ref{fig:uniform_beam_large_n}.
In the thermodynamic or kinetic limit, the distribution function $f(\theta,p)$ approaches a smooth distribution. We illustrate in Figs. \ref{fig:uniform_beam_large_n} and \ref{fig:uniform_beam_f} the change from the particles' point of view to the kinetic one.

\section{Kinetic equations}
\label{sec:vlasov-origin}

\subsection{Formalism}
\label{sec:vlasov-formalism}

We introduce the formalism of the reduced distribution functions (following \cite{balescu_statistical_dynamics}). Given the probability $F(\Theta, P) d\Theta dP$ that a system of $N$ particles finds itself with positions and velocities in the region of volume $d\Theta dP$ around $(\Theta,P)$, where $\Theta=\{\theta_1,\dots,\theta_N\}$ and $P=\{p_1,\dots,p_N\}$, we define the $s$-particles reduced distribution functions:
\begin{equation}
  \label{eq:s-part-RDF}
  f_s((\theta_1,p_1),\dots,(\theta_s,p_s)) = \frac{N!}{(N-s)!} \int d\theta_{s+1}dp_{s+1}..d\theta_Ndp_N\ F(\Theta,P)
\end{equation}
$F$ is implicitly dependent on the time $t$. We will not write down the explicit time dependence of $F$, and of the subsequent $f$'s and $g$'s defined in this section, unless needed for clarity. The evolution of $F$ is ruled by the Liouville equation:
\begin{equation}
  \label{eq:liouville}
  \ddt{} F(\Theta, P) = \mathcal{L} \ F(\Theta,P)
\end{equation}
where $\mathcal{L}$ is the Liouvillian operator, defined through a Poisson bracket:
\begin{equation}
  \mathcal{L}\ F = \sum_{j=1}^{N} \left\{ \frac{\partial H}{\partial \theta_j} \frac{\partial F}{\partial p_j} - \frac{\partial H}{\partial p_j} \frac{\partial F}{\partial \theta_j} \right\}
\end{equation}
Using the definition (\ref{eq:s-part-RDF}), and computing the evolution equations for the $f_s$'s leads to the BBGKY hierarchy of equations, equivalent to the Liouville equation \cite{balescu_statistical_dynamics}.

Under the assumption that no correlation exists between particles, we may write the uncorrelated one-particle distribution function as~:
\begin{equation}
  \label{eq:uncorr}
  f_s^{\mathrm{unc}}((\theta_1,p_1),\dots,(\theta_s,p_s)) \propto \prod_{j=1}^s f_1(\theta_j,p_j)   .
\end{equation}
Beyond that assumption, one may use the correlation functions $g_s$ to quantify the deviation from this ideal behaviour, $g_2$ being defined by the following relation:
\begin{equation}
  f_2((\theta_1,p_1),(\theta_2,p_2)) = f_1(\theta_1,p_1) f_1(\theta_2,p_2) + g_2((\theta_1,p_1),(\theta_2,p_2)) .
\end{equation}
Similar relations define $g_s$ for $s\geq 3$ in the {\it cluster representation}.

The evolution of $f_1$, deduced from the Liouville equation, is given by~:
\renewcommand{\arraystretch}{2}%
\begin{equation}
  \label{eq:one-p-liouville}
  \begin{array}{r l}
      \ddt{}f_1(\theta_1,p_1)& + p_1 \frac{\partial}{\partial\theta_1} f(\theta_1,p_1) = \\
      & \int d\theta_2\ dp_2\ \frac{\partial}{\partial \theta_1} V(|\theta_1-\theta_2|)\cdot \frac{\partial}{\partial p_1} f_1(\theta_1,p_1) f_1(\theta_2,p_2)\\
%      & \\
  & + \int d\theta_2\ dp_2\ \frac{\partial}{\partial \theta_1} V(|\theta_1-\theta_2|)\cdot (\frac{\partial}{\partial p_1} - \frac{\partial}{\partial p_2}) g_2((\theta_1,p_1), (\theta_2,p_2))
  \end{array}
\end{equation}

The meaning of the terms appearing in Eq.~(\ref{eq:one-p-liouville}) is the following:
\begin{itemize}
\item {\rule[-8pt]{0pt}{16pt} $ \mathrm{ADV} \equiv  p_1 \frac{\partial}{\partial\theta_1} f(\theta_1,p_1)$} is an advection term. It causes the displacement of molecules (or phase space elements in the kinetic description) according to their velocities. When only this term is present in the equation, the situation is called free streaming; each particle moves only at its constant velocity.
\item {\rule[-8pt]{0pt}{16pt} $ \mathrm{VLA} \equiv \int d\theta_2\ dp_2\ \frac{\partial}{\partial \theta_1} V(|\theta_1-\theta_2|)\cdot \frac{\partial}{\partial p_1} f_1(\theta_1,p_1) f_1(\theta_2,p_2)$} is the Vlasov term. Given the interaction potential $V$ and the distribution of particles in space, this term causes the particles to accelerate or decelerate.
\item {\rule[-8pt]{0pt}{16pt} $\mathrm{COL} \equiv \int d\theta_2\ dp_2\ \frac{\partial}{\partial \theta_1} V(|\theta_1-\theta_2|)\cdot (\frac{\partial}{\partial p_1} - \frac{\partial}{\partial p_2}) g_2((\theta_1,p_1), (\theta_2,p_2))$} is the collisional term. It causes the particles to accelerate or decelerate, as the Vlasov term, but depends on $g_2$, implying that information is needed from a superior term in the BBGKY hierarchy (absence of closure).
\end{itemize}
An external potential can easily be added to Eq.~(\ref{eq:one-p-liouville}). It takes the following form [in the right hand side of Eq.~(\ref{eq:one-p-liouville})]:
\begin{equation}
  \frac{dV^{\mathrm{ext}}}{d\theta} \ddp{f}
\end{equation}
where $V^\mathrm{ext}(\theta)$ is a potential that does not depend on $f$. The study of a kinetic equation containing only such a term is given in chapter~\ref{chap:pendulum}.

\subsection{Boltzmann's equation}
\label{sec:vlasov-boltzmann}

Assuming the following conditions:
\begin{enumerate}
\item A low density of particles ;
\item A short-range interaction ;
\item Weak inhomogeneity with respect to the range of the interaction ;
\item Molecular chaos~: the famous {\it Stosszahlansatz} ; colliding particles are considered uncorrelated ;
\end{enumerate}
we obtain the Boltzmann equation, the most famous kinetic equation. It can be derived on the basis of a gain and loss methodology. The variation of $f_1(\theta_1,p_1)$ is given, aside from the advection, by the number of colliding particles resulting in a position and velocity $(\theta_1,p_1)$ minus the number of particles undergoing collisions that make them leave the position $(\theta_1,p_1)$.

The Boltzmann equation takes the form:
\begin{equation}
  \label{eq:boltzmann_eq}
  \ddt{f_1} + p \ddth{f_1} = \textrm{G} - \textrm{L}
\end{equation}
where G represents the gain term and L the loss term, and G$-$L is the {\it collisional} term.

The right hand side of Eq.~(\ref{eq:boltzmann_eq}) is the driving force of the relaxation to equilibrium. It represents binary encounters in the system, events that cause two nearby particles to change their momenta, under the assumption of Boltzmann's equation.
An important property of Boltzmann's equation is the existence of an H-theorem proving the increase of entropy in the course of time and leading the system to a Boltzmann-Gibbs distribution.

\subsection{The Vlasov equation}
\label{sec:vlasov-vlasov}

In systems with long-range interactions, the aforementioned Vlasov term (see the term VLA in section~\ref{sec:vlasov-formalism}) becomes non-negligible.

Starting from Eq.~(\ref{eq:one-p-liouville}) and assuming that the interaction potential is weak $V(r)=O(\lambda)$, we establish the following estimation~:
\begin{equation}
  \label{eq:ordering}
  f(\theta,p) = O(\lambda^0) \textrm{  ;  } g_2 = O(\lambda^1)
\end{equation}
Keeping the terms of Eq.~(\ref{eq:one-p-liouville}) up to order $O(\lambda^1)$, we obtain the Vlasov equation:
\begin{equation}
  \ddt{f} + p \ddth{f} - \frac{dV}{d\theta} [f](\theta) \ddp{f} = 0
\end{equation}
where we have again dropped the $_1$ subscript. In this equation,
\begin{equation}
  V[f](\theta) = \int d\theta'\ dp'\ f(\theta',p') V(|\theta-\theta'|)
\end{equation}
is the self-consistently computed potential. As a result of the dependence of $V$ on $f$, the Vlasov equation acquires a nonlinear character.

Braun \& Hepp \cite{braun_hepp_1977} proved the weak convergence of the $N$-body dynamics to the Vlasov equation under the assumption of weak coupling.

\subsection{Properties of the Vlasov equation}
\label{sec:vlasov-properties}

The Vlasov equation possesses the form of an advection equation in phase space. As a result, the levels of $f(\theta,p)$ are conserved by the dynamics: the volume of phase space in which $f$ takes a value comprised between $\eta$ and $\eta + d\eta$ is constant. This property can be made explicit if we write the equation in a conservative form~:
\begin{equation}
  \frac{d}{dt} f = \ddt{} f + p \ddth{} f - \frac{dv}{d\theta} \ddp{} f = 0
\end{equation}
where $\frac{d}{dt}$ is the total time derivative. A major consequence is that an infinite number of quantities, the Casimirs, are conserved. For any function $s$, $\frac{d}{dt} C_s[f]=0$, where
\begin{equation}
  \label{eq:casimir}
  C_s[f] = \int d\theta\ dp\ s(f(\theta,p))
\end{equation}
Additionally, the preservation of all levels $\eta$ of $f$  implies that the maximum and the minimum of $f$ remain constant.

We define the $L_i$ norms, that we use in chapter~\ref{chap:numerical}.
\begin{equation}
  \label{eq:vlasov-l-norm}
  L_i[f] = \int d\theta\ dp\ \left(f(\theta,p)\right)^i~,
\end{equation}
where $i$ is a positive integer.

Consider the one-particle Hamiltonian defined by~:
\begin{equation}
  H_1(\theta,p) = \frac{p^2}{2} + V[f](\theta)
\end{equation}
Any distribution depending only on $H_1(\theta,p)$,  $f(\theta,p) \equiv f(H_1(\theta,p))$, is a stationary solution of the Vlasov equation. In particular, for homogeneous systems with no external force, this implies that any distribution of the momenta alone $f(\theta,p) = \varphi(p)$ is stationary. The stability of these solutions needs to be studied individually; stationary stable solutions are candidates for equilibrium in Vlasov dynamics while this is not the case for unstable stationary solutions.

\subsection{Filamentation}
\label{sec:vlasov-filamentation}

Even in the simplest cases, the dynamical evolution of the Vlasov equation displays filamentation. Filamentation is the thinning of the structure in phase space. We illustrate filamentation in the motion of free streaming, given by the Vlasov equation with no force term.
\begin{equation}
  \label{eq:vlasov-free-streaming}
  \ddt{f} + p \ddth{f} = 0
\end{equation}

\begin{figure}[ht]
%  \centering
  \begin{minipage}[h]{0.49\linewidth}
    \includegraphics[width=\linewidth]{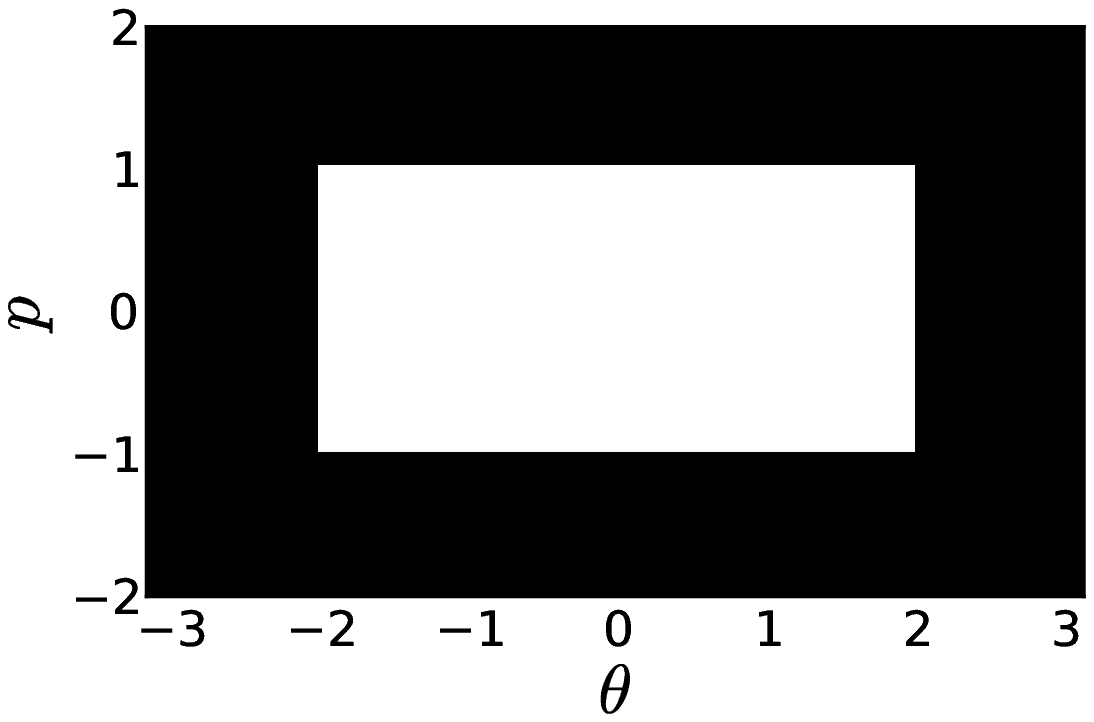}
  \end{minipage}
  \begin{minipage}[h]{0.49\linewidth}
    \includegraphics[width=\linewidth]{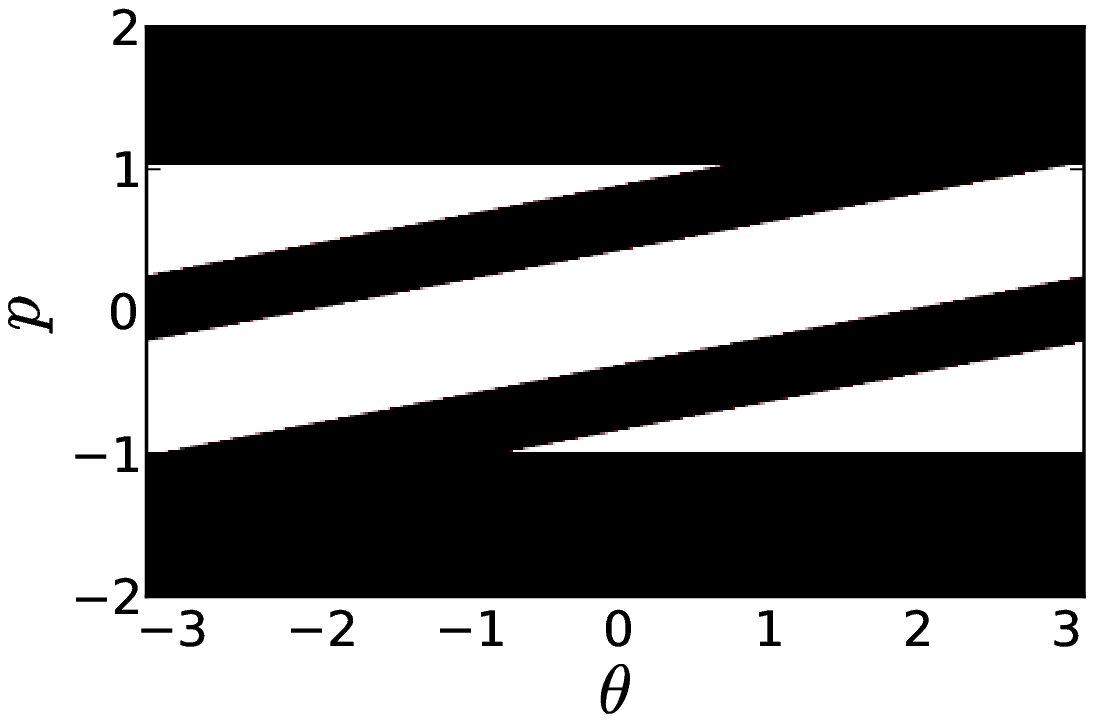}
  \end{minipage}
  \caption{(left): A waterbag initial condition.}
  \label{fig:vlasov-fil-IC}
  \caption{(right): Early evolution of the initial condition given in Fig.~\ref{fig:vlasov-fil-IC} by Eq.~\ref{eq:vlasov-free-streaming}. Each horizontal level moves at a constant speed.}
  \label{fig:vlasov-fil-1}
\end{figure}

\begin{figure}[ht]
  \centering
    \begin{minipage}[h]{0.49\linewidth}
      \includegraphics[width=\linewidth]{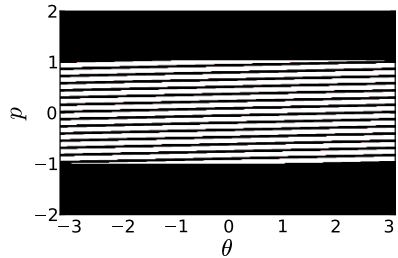}
  \end{minipage}
  \hskip 0.02\linewidth
  \begin{minipage}[h]{0.47\linewidth}
    \caption{Later evolution of the free-streaming.}
    \label{fig:vlasov-fil-2}
  \end{minipage}
\end{figure}

We display in Fig.~\ref{fig:vlasov-fil-IC} a waterbag initial condition\footnote{The waterbag initial condition is used here because of its frequent use in the following of this thesis.}. A waterbag is a distribution that is constant inside a region of phase space, it is defined as~:
\renewcommand{\arraystretch}{1}%
\begin{equation}
  \label{eq:vlasov-wb}
  f(\theta,p) = \left\{\begin{array}{l l}
      \frac{1}{4\Delta\theta \Delta p} & \mbox{if } |p|\leq \Delta p,\cr
          & \mbox{\ \ \ } |\theta|\leq \Delta\theta, \cr
      0 & \mbox{otherwise.}
  \end{array}\right.
\end{equation}
The evolution of an initial $f$ given by Eq.~\ref{eq:vlasov-wb} is portrayed in Figs.~\ref{fig:vlasov-fil-1} and \ref{fig:vlasov-fil-2}. Figure~\ref{fig:vlasov-fil-1} shows the free-streaming initial evolution. Then, in Fig.~\ref{fig:vlasov-fil-2}, we observe filaments. The width of the filaments and the space between them decreases in the course of time, and eventually becomes too small to allow a numerical description.

\section{The statistical theory of Lynden-Bell}
\label{sec:vlasov-LB}

A feature of some systems with long-range interactions is that the time spent without reaching thermodynamical equilibrium may be large. It has been reported to diverge when the number of particles increases in numerical observations \cite{barre_et_al_pre_2004,yamaguchi_et_al_physica_a_2004}. This result causes problems if one considers the thermodynamic limit $N\to\infty$ that is relevant for the Vlasov equation.
For the time scale considered (experimental or observational), equilibrium statistical mechanics provides no useful prediction. In the framework of Vlasov dynamics, Lynden-Bell proposed \cite{lynden-bell_1967} a statistical theory to predict the result of the so-called {\it violent relaxation} process.

Lynden-Bell's theory, and more generally the use of the Vlasov equation, allows to understand the properties of $N$-body systems if $N$ is large enough \cite{antoniazzi_et_al_pre_2007,jain_et_al_relaxation_times_2007}.

\subsection{Violent relaxation}
\label{sec:vlasov-violent}

Violent relaxation refers to the initial fast dynamical evolution of collisionless systems. It attracted special attention for the following reasons:
\begin{itemize}
\item Contrasting with the Boltzmann equation where the collision term provides explicit relaxation, the Vlasov equation does not describe {\it a priori} a relaxation process.
\item In collisionless systems, it appears that a fast evolution takes place rapidly after letting an initial condition evolve, followed by a slower process. This phenomenon has been largely discussed for galactic dynamics \cite{lynden-bell_1967,hohl_NASA_1968,severne_luwel_ass_1986,mineau_et_al_numerical_holes_1990}.
\end{itemize}

We give an example of such a dynamics for the Hamiltonian Mean-Field model, considered as a simplification of gravitational dynamics. A molecular dynamics in which we follow the magnetization $M$ in the course of time is performed (see chapter \ref{chap:numerical} or \ref{chap:hmf} for definitions). The initial magnetization is $M_0=0.2$ and the energy is $U=0.6$. In the initial evolution of $M$, we observe strong oscillations, followed by oscillations around $0.3$.
The run performed with $N=10^3$ particles then drifts towards a higher value.
When the number of particles is sufficiently high (in our illustration, $N=10^4$) $M$ remains near the value $0.3$. It does not experience relaxation apart from the initial evolution ($0 \leq t \lesssim 50$)
The fast initial evolution is termed ``violent relaxation''. In the limit where $N$ goes to infinity, this is the only source of relaxation.
\begin{figure}[ht]
  \centering
  \includegraphics[width=0.8\linewidth]{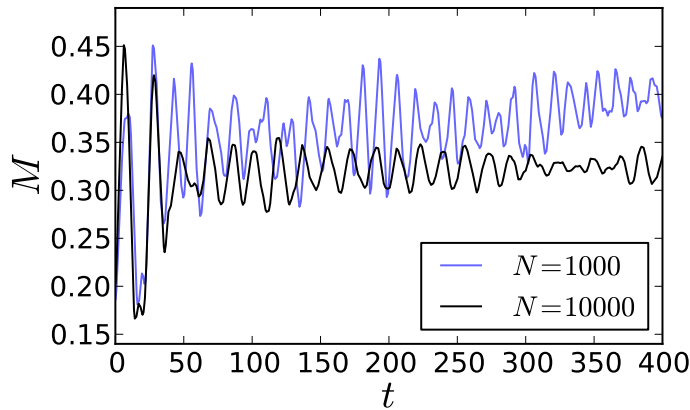}
  \caption{An example of violent relaxation in the Hamiltonian Mean-Field model. We display the magnetization $M$ in the course of time. After an initial value $M_0=0.2$, $M$ experiences strong oscillations and stabilizes around a mean value of about $0.3$ for $N=10^4$ while it drifts to a higher value for $N=10^3$. The initial regime is termed ``violent relaxation''. The simulation is performed at energy $U=0.6$, $N$ is indicated in the legend.}
  \label{fig:violent_HMF}
\end{figure}

In order to characterize the state effectively attained by the dynamics after the violent relaxation stage and to eventually predict its properties, Lynden-Bell proposed a ``Statistical mechanics of violent relaxation in stellar systems'' \cite{lynden-bell_1967}.

\subsection{Microscopic phase space dynamics}
\label{sec:vlasov-LB-microscopic}

\begin{figure}[ht]
  \centering
  \includegraphics[width=0.7\linewidth]{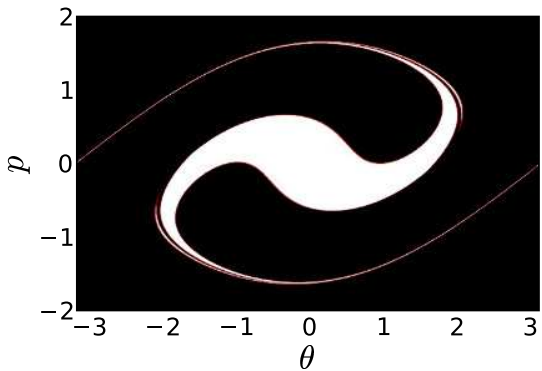}
  \caption{Phase space for the HMF model at time $t=8$, with a waterbag initial condition of energy $U=0.51$ and initial magnetization $M_0=0.05$.}
  \label{fig:HMF_initial}
\end{figure}

The Vlasov equation describes an incompressible fluid in phase space. An initial condition will be deformed, stretched, and filaments will form but the height and area of the waterbag will remain constant\footnote{The discussion of what happens in numerical simulations is postponed to chapter~\ref{chap:numerical}.}. Already at early times, for the Hamiltonian Mean-Field model, we observe a thinning of the contours. Figure \ref{fig:HMF_initial} displays such a thinning, leading to the formation of filaments.
The smallest scale eventually attained by the filaments decreases continuously, and it becomes impossible to follow that evolution analytically or numerically.

To clarify the discussion, we here define the meaning of the different scales of interest in the study of phase space dynamics.
\begin{enumerate}
\item Microscopic: The microscopic scale refers to the scale describing the fine filaments appearing in Vlasov dynamics. It is able to describe fully the ideal theoretical evolution of $f(\theta,p)$.
\item Mesoscopic: The mesoscopic scale is small with respect to the complete phase space, typically too small to be attained by numerical methods but larger than the filaments. From the mesoscopic scale, the filaments are averaged and not visible.
\item Macroscopic: The macroscopic scale is the one on which collective phenomena are observable. For the mean-field models studied in this thesis, the characteristic length of interest is typically $\pi$, in relation to the characteristic scale of the interaction potential.
\end{enumerate}

\begin{figure}[ht]
  \centering
  \includegraphics[width=0.95\linewidth]{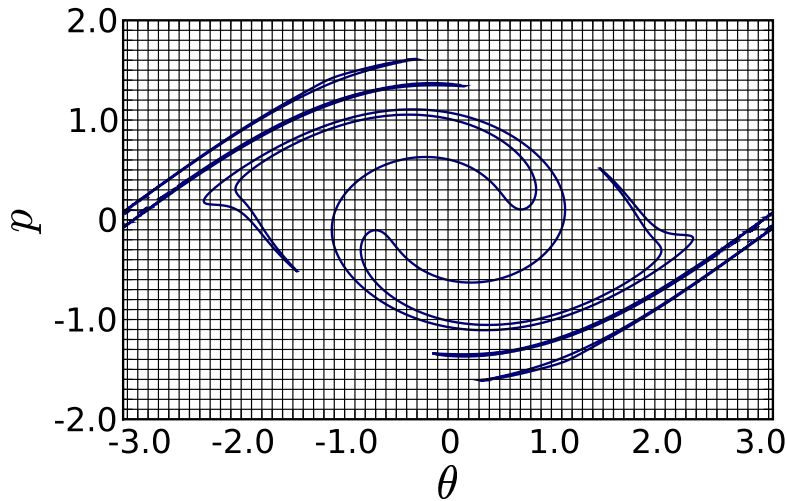}
  \caption{Illustration of the different scales of description in phase space.}
  \label{fig:HMF_grid}
\end{figure}
\begin{figure}[ht]
  \centering
  \includegraphics[width=0.95\linewidth]{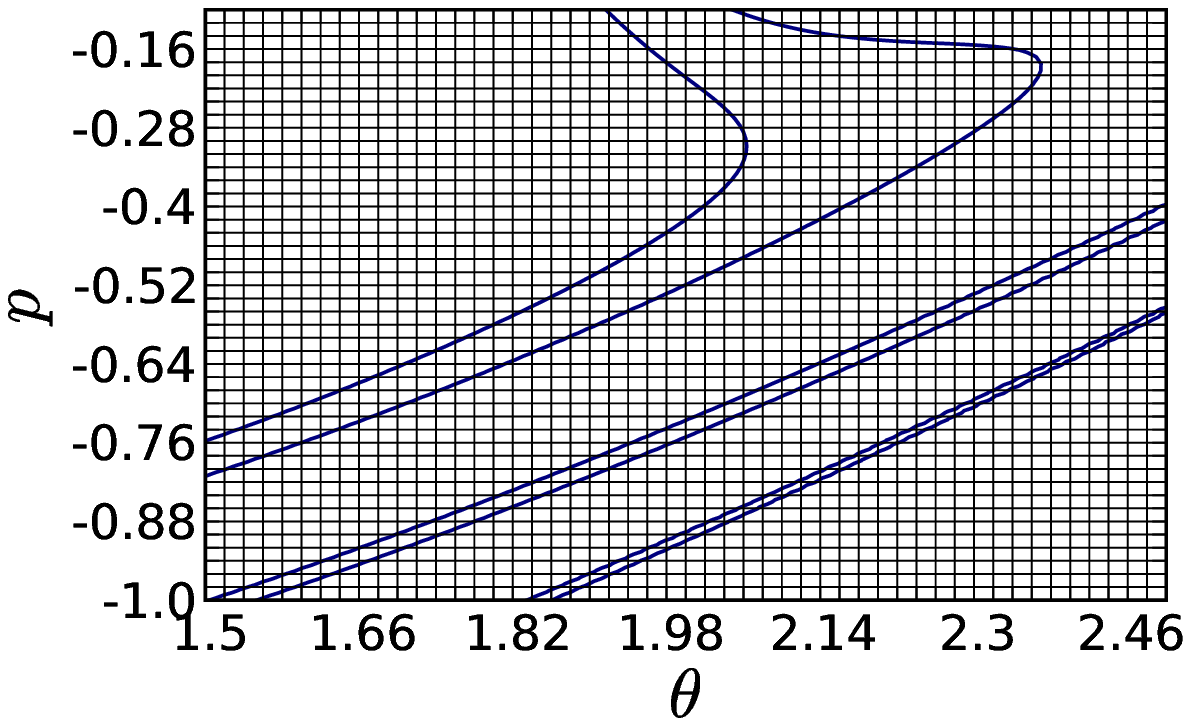}
  \caption{Zoom of Fig.~\ref{fig:HMF_grid}. The rightmost filament would need a grid finer than the one displayed in order to be correctly described. The scale below our ``observation'' scale is termed microscopic. Lynden-Bell used the analogy with the power of a microscope \cite{lynden-bell_1967}.}
  \label{fig:HMF_fine_grid}
\end{figure}

The initial condition we consider is a waterbag initial condition (see Fig.~\ref{fig:vlasov-fil-IC} for instance), in which $f$ takes only two possible values: $0$ or $f_0$. We display its evolution through Vlasov dynamics at time $t=8$ in Fig.~\ref{fig:HMF_initial}.
In Fig.~\ref{fig:HMF_initial}, filaments are seen to be a prominent characteristic feature in phase space.
We detail the analysis by displaying the contour lines of the waterbag in Fig.~\ref{fig:HMF_grid}.
A zoom on a particular region is displayed in Fig.~\ref{fig:HMF_fine_grid}, where we superimpose a grid to illustrate the concept of phase space cell.

The rightmost filament in Fig.~\ref{fig:HMF_fine_grid} is too thin to be described by the grid~: its width is smaller than a grid cell. It is therefore considered microscopic, while the leftmost filament is adequately described.

Figure~\ref{fig:HMF_fine_grid} also serves the purpose to illustrate the limitation of the numerical approach: if we consider the value to assign to grid points of the mesh, we can for instance assign to a grid region filled with a filament the area of occupied by the filament: for a grid point $(\theta_i, p_i)$, we would get:
\begin{equation}
  \label{eq:cell_average}
  f_{i,m} = \int_{\theta_i - \Delta\theta/2}^{\theta_i + \Delta\theta/2} \int_{p_m - \Delta p/2}^{p_m + \Delta p/2} d\theta\ dp\ f(\theta, p)
\end{equation}
which is a real number between $0$ and $f_0$.

Lynden-Bell's original idea is the following: the microscopic scale cannot be included in the predictive theory, and the prediction will give $\bar f(\theta,p)$, the coarse-grained one-particle probability distribution function, valid at the meso- and macroscopic scales.

\subsection{Lynden-Bell's entropy}
\label{sec:vlasov-LB-entropy}

We describe how to obtain Lynden-Bell's entropy from microscopic phase space dynamics. We refer the reader to the following references for more detailed explanations \cite{barre_phd,lynden-bell_1967,campa_et_al_phys_rep_2009}.

Lynden-Bell's theory predicts a coarse-grained distribution function $\bar f$. Whereas the microscopic distribution function $f$ can take only two values, $0$ or $f_0$, $\bar f$ can take any value in that interval. The process is the following:
\begin{enumerate}
\item We decompose phase space in microscopic cells of volume $\omega$.
\item We consider mesoscopic cells containing $\nu$ microscopic cells.
\item Considering the ``average value'' of $f$ in that cell, e.g. $$\bar f_i=\sum_{j=1}^\nu \frac{f^\mu_j}\nu = \frac{n_i f_0}\nu$$
  $f^\mu_j$ is the microscopic value of $f$ ($0$ of $f_0$), and $n_i$ is the number of microscopic cells having the value $f_0$.
\item In that mesoscopic cell, there are $\frac{\nu!}{(\nu-n_i)!}$ ways to place the elements.
\item There are $\frac{\mathcal{N}!}{\prod_i n_i!}$ to place the ${n_i}$'s in the mesoscopic cells, where $\mathcal{N}=\sum_i n_i$ is the total number of grid cells having the value $f_0$.
\end{enumerate}
The result is that the volume $W(\{n_i\})$ occupied by a given configuration is
\begin{equation}
  W(\{n_i\}) = \frac{\mathcal{N}!}{\prod_i n_i!} \times \prod_i \frac{\nu!}{(\nu-n_i)!} .
\end{equation}
Taking the logarithm, going to the continuum limit and using Stirling's formula leads to the Lynden-Bell entropy for a given $\bar f$:
\begin{equation}
  - \int d\theta\ dp\ \left[ \frac{\bar f}{f_0} \ln \frac{\bar f}{f_0} + \left(1-\frac{\bar f}{f_0}\right) \ln\left(1-\frac{\bar f}{f_0}\right) \right] .
\end{equation}
Maximizing this entropy under constraints (mass, energy, momentum) leads to the most probable $\bar f$.

\begin{figure}[ht]
  \centering
  \includegraphics[width=0.7\linewidth]{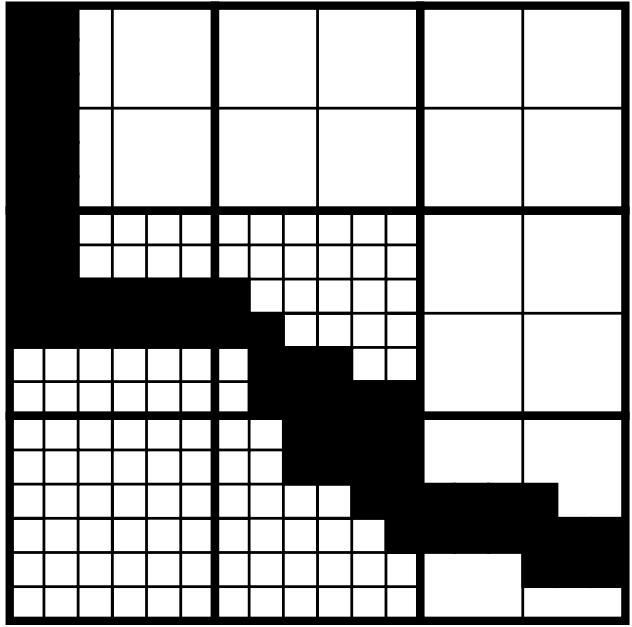}
  \caption{Example of subdivision of phase space into cells. The thick lines delimit mesoscopic cells and the small square represent microscopic cells, whose value is $0$ or $f_0$. From top-left to bottom right, horizontally, the occupation levels are: 12, 0, 0, 16, 11, 0, 0, 11 and 13.}
  \label{fig:LBcells}
\end{figure}

Fig.~\ref{fig:LBcells} illustrates the idea of dividing phase space into cells. $f(\theta,p)$ in the small (microscopic) cells takes only the values $0$ or $f_0$. If one considers a group of microscopic cells, the average value of $f(\theta,p)$ is bounded between $0$ and $f_0$.

% Local Variables:
% TeX-master: "main"
% End:

\cleardoublepage

\chapter{Definition of the Hamiltonian Mean-Field model}
\label{chap:models}

\myabstract{This chapter is dedicated to the definition of the Hamiltonian Mean-Field (HMF) model. We illustrate some of its properties with numerical simulations in the $N$-body context.}{Hamiltonian Mean-Field model, magnetization $M$.}

The Hamiltonian Mean-Field (HMF) model has been introduced in Refs.~\cite{inagaki_konishi_pasj_1993,ruffo_HMF_1994,antoni_ruffo_1995} [see also \cite{pichon_phd}].
It can be interpreted as either a set of ferromagnetic particles with infinite range coupling or as the simplification of a gravitational model in one dimension in which only the first term of the Fourier expansion of the interaction is kept.
The HMF model is meant to give a simplified framework in which it is possible to study features of the more realistic systems.

A phase transition is found for the HMF model in the microcanonical and the canonical ensembles, separating homogeneous and non-homogeneous situations \cite{antoni_ruffo_1995}. Molecular dynamics simulations in the microcanonical ensemble show a peculiar behaviour near the phase transition, a feature that gave rise to interest in the dynamical behaviour of the HMF model.

We give in this chapter a definition of the HMF model and illustrate some of its dynamical properties.

\section{Definition}
\label{sec:models-defs}

The HMF model is described by the Hamiltonian
\begin{equation}
  \label{eq:models-HMF-H}
  H = \sum_{i=1}^N \frac{p_i^2}{2} + \frac{J}{2N} \sum_{i,j=1}^N \left(1-\cos(\theta_i-\theta_j)\right)
\end{equation}
where $\theta_i$ is the position in $[-\pi;\pi[$ of the $i$th particle, $p_i$ its momentum, $N$ is the number of particles and $J$ is a coupling constant. We consider in this thesis the HMF model with $J=1$ and do not write this constant anymore.

The magnetization is commonly used to track the dynamical evolution of the system. It is defined as~:
\begin{equation}
  {\bf M} = (M_x,M_y) = \unsurn \sum_{i=1}^N \left( \cos\theta_i , \sin\theta_i \right)
\end{equation}
and we define $M=\sqrt{M_x^2+M_y^2}$\ . The magnetization also allows one to rewrite the Hamiltonian~(\ref{eq:models-HMF-H}) in a way explicitly extensive~:
\begin{equation}
  H = \sum_{i=1}^N \frac{p_i^2}{2} + \frac{N}{2} \left( 1 - M^2 \right) \quad.
\end{equation}
In the following, we make use exclusively of the energy per particle $U$~:
\begin{equation}
U=\frac{H}{N} \quad .
\end{equation}

The equations of motion for the particles are given by~:
\begin{equation}
  \label{eq:models-eom}
  \renewcommand{\arraystretch}{2}%
  \begin{array}{l l}
  \dot \theta_i &= p_i  ,\\
  \dot p_i &= - \unsurn \sum_{j=1}^N \sin\left(\theta_i - \theta_j \right) ,\\
           &= - M(t) \sin(\theta_i - \varphi(t)) ,
  \end{array}
\end{equation}
where $\varphi$ is defined so as to satisfy $M_x + i M_y = M e^{i\varphi}$ (here, $i$ is the complex number).
Equations \ref{eq:models-eom} are similar to the equations of motion for a pendulum in which the field intensity and direction are dependent on the time.

\section{Dynamical evolution}

We comment on the properties of the $N$-body dynamics for the HMF model. We focus on the case of waterbag initial conditions that are of relevance in this thesis. In the waterbag initial condition, particles are distributed uniformly in phase space according to the following conditions:
\begin{eqnarray}
  | \theta_i | \leq \Delta\theta & , \textrm{ for all $i$} \\
  | p_i | \leq \Delta p & , \textrm{ for all $i$\ .}
\end{eqnarray}
One can characterize the initial distribution equivalently by the energy $U=\frac{\Delta p^2}{6} + \frac{1}{2} \left( 1 - M_0^2 \right)$ and the initial magnetization $M_0=\frac{\sin\Delta\theta}{\Delta\theta}$ .

\begin{figure}[ht]
  \centering
  \includegraphics[width=.8\linewidth]{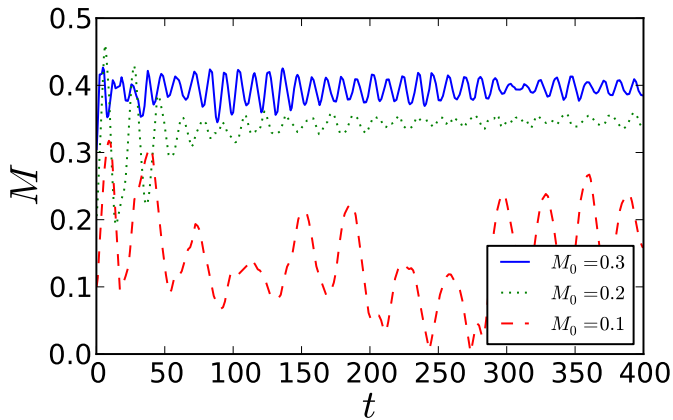}
  \caption{Evolution of the magnetization $M$ in the HMF model. $U=0.6$ and $N=10^4$. Depending on the initial magnetization $M_0$, we observe significantly different values for $M$.}
  \label{fig:models-HMF-M0}
\end{figure}

\paragraph{Dependence on the initial magnetization $M_0$.}
Statistical mechanics predicts for a given energy $U$ a unique value of the magnetization $M$ \cite{antoni_ruffo_1995}.
In contrast, the dynamical evolution of the HMF model depends also on the value given to $M_0$. This is illustrated in Fig.~\ref{fig:models-HMF-M0} where different evolutions are observed depending on the value of $M_0$. A phase diagram separating the parameter space $(U,M_0)$ in regions with $M=0$ and $M\neq 0$ can be computed with the help of Lynden-Bell's theory \cite{antoniazzi_et_al_prl_2007}. This is an {\it out-of-equilibrium} phase transition, a feature of some systems with long-range interaction.

\begin{figure}[ht]
  \centering
  \includegraphics[width=.8\linewidth]{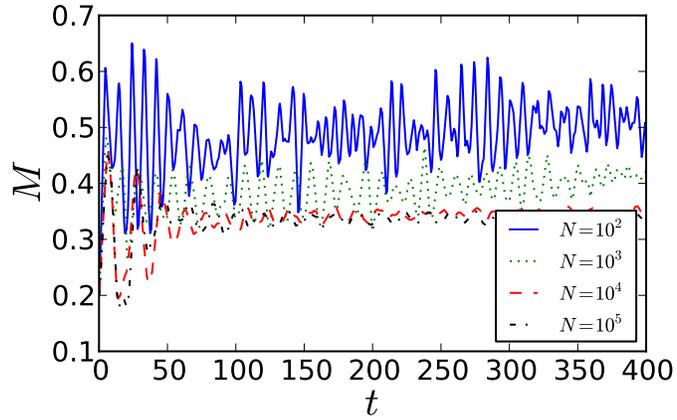}
  \caption{Evolution of the magnetization in the HMF model. $U=0.6$ and $M_0=0.2$ . We observe a qualitatively different evolution depending on the number of particles in the simulation.}
  \label{fig:models-HMF-N-body}
\end{figure}

\paragraph{Dependence on the number of particles $N$.}
The evolution of the magnetization $M$ shows a strong dependence on the number of particles $N$. After strong oscillations in $M$, the simulations with $N=10^4$ and $N=10^5$ particles in Fig.~\ref{fig:models-HMF-N-body} stabilizes at a value of about $0.35$, which is not the value predicted by statistical mechanics. The simulations with a lower number of particles evolve towards a higher value of $M$.

\bigskip
The strong dependence of the values attained by $M$ on the initial magnetization $M_0$ and on the number of particles $N$ represents a conceptual challenge for statistical mechanics and nonlinear dynamics.

% Local Variables:
% TeX-master: "main"
% End:

\cleardoublepage

\chapter{Numerical resolution of the Vlasov equation}
\label{chap:numerical}

\myabstract{The nonlinear dynamics resulting from the Vlasov equation is not tractable analytically and one needs to resort to a numerical treatment. We introduce the semi-Lagrangian method, its numerical properties and its application to the Hamiltonian Mean-Field model. We also discuss the case of a model for the Free-Electron Laser.}{advection equation, numerical mesh, time splitting, $L_i$ norm.}

The numerical resolution of the Vlasov equation remains an active research domain, despite its foundations being laid as early as in 1976 \cite{cheng_knorr_1976}. This method is a fundamental tool for the understanding of plasma physics and is used, for instance, to study the behaviour of tokamaks.
The major limitations is the resolution (number of grid points) used to describe the evolution of the one-particle distribution function (1-PDF). The 1-PDF is an infinite dimensional object and its representation on a computer, whatever the method chosen, is imperfect.

The literature on the numerical resolution of the Vlasov equation covers mostly the Vlasov-Poisson system of equations describing plasmas in various configurations \cite{shoucri_eulerian_codes_2008}. Refinements include relativistic descriptions \cite{sonnendrucker_et_al_semi-lag_1999}, parallel computations \cite{crouseilles_patches_2007} and the use of wavelets \cite{gutnic_et_al_wavelets_cpc_2004}, for instance.

In this thesis, we use the numerical resolution of the Vlasov equation to study one-dimensional models in which peculiar dynamical phenomena have been discovered \cite{long-range-02,long-range-07}. The Vlasovian approach has played a considerable role in their understanding \cite{yamaguchi_et_al_physica_a_2004,antoniazzi_et_al_prl_2007,antoniazzi_califano_prl}; meanwhile, the performed simulations have remained almost only of the molecular dynamics type.
The study of direct Vlasov simulations is an essential addition to this approach and requests a detailed analysis of their properties.

Conversely, taking profit of the relative simplicity of our mean-field models and of the growing interest for their Vlasov dynamics in the literature, we compare our simulation to recent studies on the Vlasov-Poisson system that are concerned with the limitation of the numerical scheme. These comparisons aim at making the Hamiltonian Mean-Field model a test-case for Vlasov algorithms.

This chapter is structured as follows: we review the problem of the numerical resolution of the Vlasov equation in section \ref{sec:num-overview}. We then proceed to detail the resolution of the 1D advection equation that forms the basis of the complete algorithm in section \ref{sec:num-adv}. We present the algorithm for the complete 1D Vlasov problem, following the work of Cheng and Knorr \cite{cheng_knorr_1976} and Sonnendr{\"u}cker {\it et al} \cite{sonnendrucker_et_al_semi-lag_1999} in section \ref{sec:num-semil} and detail its application to the Hamiltonian Mean-Field model (section \ref{sec:num-HMF}) and to the Free-Electron Laser (section \ref{sec:num-FEL}).

\section{Overview of the problem}
\label{sec:num-overview}

In order to study the evolution of $f(\theta,p ; t)$ \footnote{The notation and description of physical quantities is discussed in chapter \ref{chap:vlasov}.}, we need to choose a method to represent and store this distribution function on a computer. Several points of view exist, which we briefly recall:
\begin{enumerate}
\item Lagrangian Methods~:\\
  $f$ is here computed with the knowledge of a set of particles. These particles represent themselves a number of microscopic elements and are evolved at each time step. This methodology is called particle-in-cell (PIC).
\item Eulerian Methods~:\\
  The computation is based on $f$ directly. The function is stored on the computer, and serves as the basis for the computation of the force and of the macroscopic quantities. A number of methods exist to store $f$ (regular Cartesian grid, irregular grid, a Fourier basis, wavelets), and the implementation of the algorithm is tailored to the storage method.
\end{enumerate}

The main advantage of the PIC method is its ability to treat the full 6D problem in terms of storage capacity [see for instance Ref.~\cite{hockney_eastwood_book}]. Its major drawback is to  suffer from an important noise level arising from the sampling in phase space. Another fundamental difference is that, in the case of Eulerian methods, one needs to run a single simulation by physical situation. Since there is no random generation of particles corresponding to the macroscopic situation, the averaging over several realisations is not needed.

Given the modest computational cost in storage as well as in CPU-time for 1D systems, we can afford the choice of an Eulerian method. The highest resolution that we will use, $8192\times 8192$, requires $\approx 540$MB of memory for a copy of $f$ in double precision, making the computation accessible, memory-wise, to single-CPU computers. The use, in our algorithm of $f$, a copy of $f$ and the storage of the second derivative needed by the spline interpolation elevates the memory cost by a factor of three.

In order to assess the quality of the numerical resolution, we make use of $L_i$ norms, which are ideally conserved.
These conserved quantities are not expected to hold exactly (up to machine precision) in the numerical resolution. We will compute the following quantities:
\begin{itemize}
\item $L_1[f]$ corresponds to the normalization of $f$.
\item Deviation of $L_2[f]$ indicates the presence of numerical dissipation.
\item The energy $U[f]$ is also a constant.
\end{itemize}

The limitation of the resolution to describe $f$ during the development of filaments comes from the fact that the ideal $f$ needs a microscopic phase space definition (see section \ref{sec:vlasov-LB-microscopic} for a definition), which is impossible. The problem of filamentation is detailed in section~\ref{sec:vlasov-filamentation}.

A rough criterion based on the free-streaming situation gives an estimation of the time during which the numerical simulation is able to follow the development of the filaments. We recall the steps given by Canosa {\it et al} \cite{canosa_gazdag_fromm_recurrence}. The explicit solution to the free-streaming Vlasov equation is written in terms of a Fourier decomposition, for an initial condition perturbed at wavenumber $k$:
\begin{equation}
  \label{eq:fourier-free-streaming}
  f(\theta, p ; t) = \exp\left(i k \theta\right)\exp\left(-ikpt\right) \quad .
\end{equation}
Equation~(\ref{eq:fourier-free-streaming}), specialized for a discrete set of momenta $p_m = m \Dp$, reads~:
\begin{equation}
  \label{eq:fourier-free-streaming-Dp}
  f(\theta, m\Delta p ; t) = \exp\left(i k \theta\right)\exp\left(-ik m \Delta p\ t\right) \quad .
\end{equation}
Equation~(\ref{eq:fourier-free-streaming}) indicates that for any given $m$, the distribution function comes back to its initial value at a time $t_m = \frac{2\pi}{k m \Delta p}$.
Considering the resolution for a set of values $m$, one finds that after a time~:
\begin{equation}
  \label{eq:numerical-T_R}
  T_R = \frac{2\pi}{k \Delta p} \quad ,
\end{equation}
the complete distribution function $f$ experiences a recurrence, clearly indicating a limitation of the resolution procedure. $T_R$ is called the recurrence time.

\section{The advection equation}
\label{sec:num-adv}

To solve the Vlasov equation, we make use of the time-split algorithm introduced in the work of Cheng and Knorr \cite{cheng_knorr_1976}. Only a one-dimensional advection problem needs to be solved in this situation.

The 1D advection with a constant velocity field reads:
\begin{equation}
  \label{eq:1d-adv}
  \ddt{f(x;t)} + c \ddx{f(x;t)} = 0
\end{equation}
Considering the problem on a fixed grid, a simple way to solve numerically this problem is a finite-difference explicit scheme. We note $f^s_i$ the value of $f(x_i ; s \Delta t)$ where $x_i = x_0 + i \Delta x$, $i\in [1;N_x]$, and $\Delta t$ is the time step.

The finite-difference version of Eq. (\ref{eq:1d-adv}) is:
\begin{equation}
  \frac{f^{s+1}_{i} - f^s_{i}}{\Delta t} + c\ \frac{f^s_{i}-f^s_{i-1}}{\Delta x} = 0 ,
\end{equation}
giving an explicit formula for $f^{s+1}_i$.

This formulation allows for a simple computation of $f$ at time $s+1$ given $f$ at time $s$. It is however strongly dissipative. Dissipation in this situation refers to an artifact of the algorithm acting as a diffusion operator. A simple finite-difference scheme as this one can be used only in situation where a physical diffusion operator is present, as in Fourier's heat equation. Another important property of this scheme is the bound on the value of the time step, the so-called Courant-Friedrichs-Lewy condition [see for instance Ref.~\cite{zwillinger_handbook_3rd}]
\begin{equation}
   c \frac{\Delta t}{\Delta x} < 1 \quad .
\end{equation}

Let us introduce the method of the characteristics that will be used further on. This algorithm is based on the fact that, for Eq.~(\ref{eq:1d-adv}), the value of $f(x ; t+\Delta t)$ correspond to the value of $f(x^\ast ; t)$ where $x^\ast$ is located at the foot of the characteristic curve. We focus on the 1D problem, but the presentation is valid in higher dimensions with the same notations.

$x^\ast$ is computed \cite{zwillinger_handbook_3rd} according to~:
\begin{equation}
  x^\ast = x + \int_t^{t-\Delta t} dt\ c = x - c \Delta t
\end{equation}
allowing to write an explicit solution to Eq.~(\ref{eq:1d-adv}) on a grid as
\begin{equation}
  \label{eq:1d-semi-lag}
  f_i^{s+1} = f^s(x_i - c \Delta t)
\end{equation}

In general, $x^\ast$ does not lie on the storage mesh, meaning that we need an interpolated value of $f$ to proceed. The choice of the interpolation method has a great impact on the numerical properties of the resolution, as detailed in several articles \cite{filbet_et_al_conserv_schemes_2001,arber_vann_critical_compar_2002}.

In the rest of this work, we use the cubic spline method \cite{NR_in_f90} that gives a good compromise between numerical dissipation, computational time and complexity.

\section{The semi-Lagrangian method for the Vlasov equation}
\label{sec:num-semil}

Solving the Vlasov equation in 1D amounts to solve a 2D partial differential equation, one for the position and one for the velocity. Cheng and Knorr have already proposed in 1976 a simplification via a time-splitting algorithm \cite{cheng_knorr_1976}. Their work laid the foundation for further works.

Cheng and Knorr introduced the following discretization, accurate to second order:
\begin{equation}
  \label{eq:CandK}
  f^{s+1}(\theta,p) = f^s( \theta- \Delta t (p - \frac{1}{2} F^\ast(\bar \theta)\Delta t) , p - \Delta t F^\ast(\bar \theta))
\end{equation}
where $\bar \theta = \theta - p\Delta t /2$ and $F^\ast$ is the force field computed at half a time step. We use the notation $f^s(\theta,p) = f(\theta,p ; s\ \Delta{}t)$.

A practical implementation to compute Eq. (\ref{eq:CandK}) on a numerical mesh is the following :

\begin{tabular}{l l l}
  1. & Advection in the & $f^\ast(\theta_i,p_m) = f^s(\theta_i-p_m\Delta t/2, p_m)$\cr
     & $\theta$-direction, 1/2 time step & \cr
  2. & Computation of the force field & \cr
     & for $f^\ast$ & \cr
  3. & Advection in the & $f^{\ast\ast}(\theta_i,p_m) = f^\ast(\theta_i, p_m - F^\ast(\theta_i) \Delta t)$\cr
     & $p$-direction, 1 time step & \cr
  4. & Advection in the & $f^{s+1}(\theta_i,p_m) = f^{\ast\ast}(\theta_i-p_m\Delta t/2, p_m)$\cr
     & $\theta$-direction, 1/2 time step & 
\end{tabular}

\begin{figure}[ht]
  \centering
  \includegraphics[width=\linewidth]{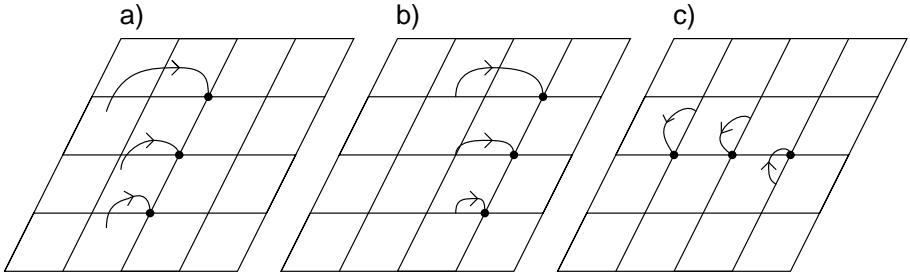}
  \caption{The semi-Lagrangian algorithm: a) illustrates the computation of the value of $f$ at the grid points indicated by the dots from the feet of the characteristic curves (the start of the arrow). b) and c) represent steps 1 and 3 of the time-split algorithm.}
  \label{fig:grid_all}
\end{figure}

Each of these steps is performed for the whole grid before going to the next one. A sketch of these steps is given in Fig. \ref{fig:grid_all}. This method, involving a fixed mesh and trajectories along the characteristics backwards in time is said to be semi-Lagrangian. It is used in fluid dynamics, e.g. for weather forecasts (see \cite{staniforth_cote_1991}).

In the framework of the Vlasov equation, especially the Poisson-Vlasov system, it was introduced by Sonnendr{\"u}cker {\it et al} \cite{sonnendrucker_et_al_semi-lag_1999}, also associated with cubic splines. Let us mention that the semi-Lagrangian method is symplectic if one neglects the approximation present in the interpolation step. Watanabe and Sugama proposed a generalization of symplectic algorithms for the Vlasov equation and presented the result up to sixth order in time in Ref.~\cite{watanabe_sugama_symplectic_vlasov_2004}.

\section{The study of the Hamiltonian Mean-Field model}
\label{sec:num-HMF}

We discuss in this section the computations related to the properties of the simulation for an initial Gaussian homogeneous distribution, then we check the stability criterion of Ref. \cite{yamaguchi_et_al_physica_a_2004}.
References~\cite{galeotti_califano_prl_2005,califano_galeotti_pop_2006} propose original analyses for the numerical resolution for the Vlasov-Poisson system of equations of relevance for plasma physics. We comment our results in the same spirit.

\subsection{The Vlasov equation for the HMF model}
\label{sec:num-vlasov-hmf}

The Vlasov equation for the Hamiltonian Mean-Field (HMF) model is~:
\begin{eqnarray}
  \label{eq:num-vlasov}
  \frac{\partial f}{\partial t} &+& p \frac{\partial f}{\partial \theta} - \frac{dV[f]}{d\theta} \frac{\partial f}{\partial p} = 0 \cr
    & & \cr
  V[f](\theta) &=& 1 - M_x[f] \cos\theta - M_y[f] \sin\theta \cr
  M_x[f] &=& \int d\theta dp\ f \cos\theta, \cr
  M_x[f] &=& \int d\theta dp\ f \sin\theta \quad .
\end{eqnarray}
Equations~(\ref{eq:num-vlasov}) preserve the total energy $U$~:
\begin{equation}
  \label{eq:num-U}
  U(t)[f] = \int d\theta\ dp\ f(\theta,p ; t) \left( \frac{p^2}{2} + \undemi \left( 1 - M_x[f]\cos\theta - M_y[f]\sin\theta \right) \right)
\end{equation}
and the total momentum $P$~:
\begin{equation}
  \label{eq:num-P}
  P(t)[f] = \int d\theta\ dp\ f(\theta,p ; t) \left( p \right) .
\end{equation}

\subsection{Comparison of Vlasov and molecular dynamics}
\label{sec:numerical-check}

We compare the dynamics of the Vlasov resolution with $N$-body simulations.
The Vlasov equation is a good description of the system up to a validity time that depends on the number of particles. Braun \& Hepp \cite{braun_hepp_1977} prove the convergence to the Vlasov equation for weakly coupled systems. Jain \etal{}  \cite{jain_et_al_relaxation_times_2007} detail the validity time in the case of the HMF for homogeneous initial conditions and quasi-stationary states.

In this paragraph, we compare a Vlasov simulation with a sligthly perturbed initial condition (\ref{eq:IC_G}) with $N$-body simulations setup in the same fashion, but with no perturbation. The instability is triggered by fluctuations of order $1/\sqrt{N}$  in the magnetization. The energy is set to $U=0.51$. For the Vlasov simulation, the recurrence time is $T_R\approx 357$, which is longer than the simulation.
\begin{figure}[ht]
  \centering
  \begin{minipage}[h]{0.49\linewidth}
    \includegraphics[width=\linewidth]{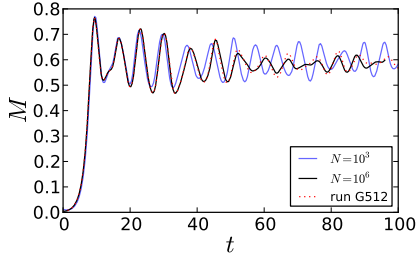}
    \caption{Comparison of the magnetization $M$ in Vlasov and molecular dynamics simulations for the HMF model. Run G512 is described in section~\ref{sec:prop}, the energy is $U=0.51$.}
    \label{fig:numerical_check}
  \end{minipage}
  \begin{minipage}[h]{0.49\linewidth}
    \includegraphics[width=\linewidth]{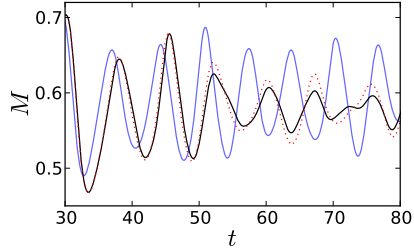}
    \caption{Zoom on Fig.~\ref{fig:numerical_check}, same legend. The simulation with $N=10^6$ stays close to the Vlasov simulation for a longer time than the one with $N=10^3$.}
  \end{minipage}
\end{figure}

The initial dynamics is similar for all simulations, but after $t\approx 30$, the run with $N=1000$ particles clearly separates from the two other runs, confirming that the Vlasov dynamics corresponds to a high number of particles.

Let us remark that the use of the energy per particle $U=H/N$ and of the average magnetization ${\bf M} = \unsurn \sum_i\left(\cos\theta_i,\sin\theta_i\right)$ allow to cast the results for various values of $N$ in a simple manner.

\subsection{Properties of the numerical solution}
\label{sec:prop}

We run simulations starting with a Gaussian profile, slightly perturbed~:
\begin{equation}
  \label{eq:IC_G}
  f(\theta, p) = \sqrt{\frac{\beta}{2\pi}} e^{-\beta p^2/2} \left( 1 + \epsilon \sin\theta \right)
\end{equation}
at an energy of $U=0.51$, $\epsilon=10^{-4}$. $\beta$ is computed from $U$, for an homogeneous state~: $\beta=\undemi\frac{1}{U-1/2}$. At $U=0.51$, the Gaussian profile is unstable; $M$ grows, then oscillates around a mean value to which it eventually relaxes.

The following parameters are used : $\Delta t = 0.1$, size of the box in the $p$-direction $[-4.5:4.5]$. The grid size is varied from $N_\theta=N_p=64$ (G64) to $N_\theta=N_p=512$ (G512); we call these runs G64, G128, G256 and G512.

We detail in Table \ref{tab:cons} the conservation properties of the algorithm. The conservation of $U$ could be improved by decreasing $\Delta t$, implying more computational time. This is the sole constraint on $\Delta t$ as the semi-Lagrangian method has no Courant condition.

\begin{table}[ht]
  \centering
  \begin{tabular}{l | r | r}
    run & $ \max \frac{L_1(t)-L_1(0)}{L_1(0)} $ & $ \max \frac{U(t)-U(0)}{U(0)}$ \cr
    \hline\hline
    & & \cr
    G64  & $1.6\ 10^{-4}$ & $3.2\ 10^{-4}$ \cr
    G128 & $2.1\ 10^{-5}$ & $2.0\ 10^{-4}$ \cr
    G256 & $7.2\ 10^{-6}$ & $2.0\ 10^{-4}$ \cr
    G512 & $2.8\ 10^{-6}$ & $2.0\ 10^{-4}$
  \end{tabular}
  \caption{Conservation properties for different grid sizes. Increasing the grid size allows for a better conservation of $L_1$. The conservation of $U$ depends also on the time step (second order scheme) and could be improved by decreasing $\Delta t$.}
  \label{tab:cons}
\end{table}

\begin{figure}[ht]
  \centering
  \includegraphics[width=.7\linewidth]{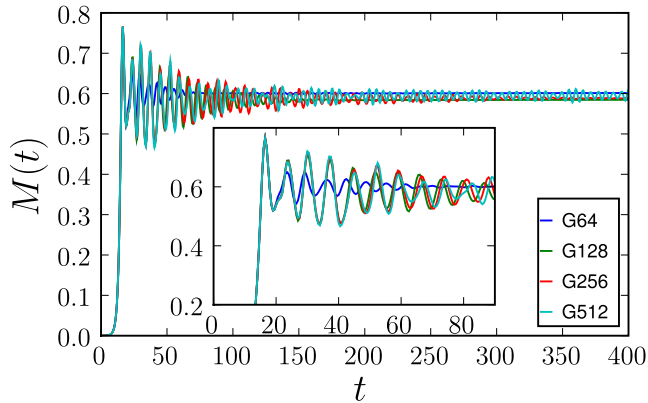}
  \caption{Evolution of $M$ for the Gaussian initial condition given by Eq.~(\ref{eq:IC_G}) ($U=0.51$,$\epsilon=10^{-4}$). A magnification of the same curve is presented in the inset.}
  \label{fig:G_M}    
\end{figure}
\begin{figure}[ht]
  \centering
  \includegraphics[width=.7\linewidth]{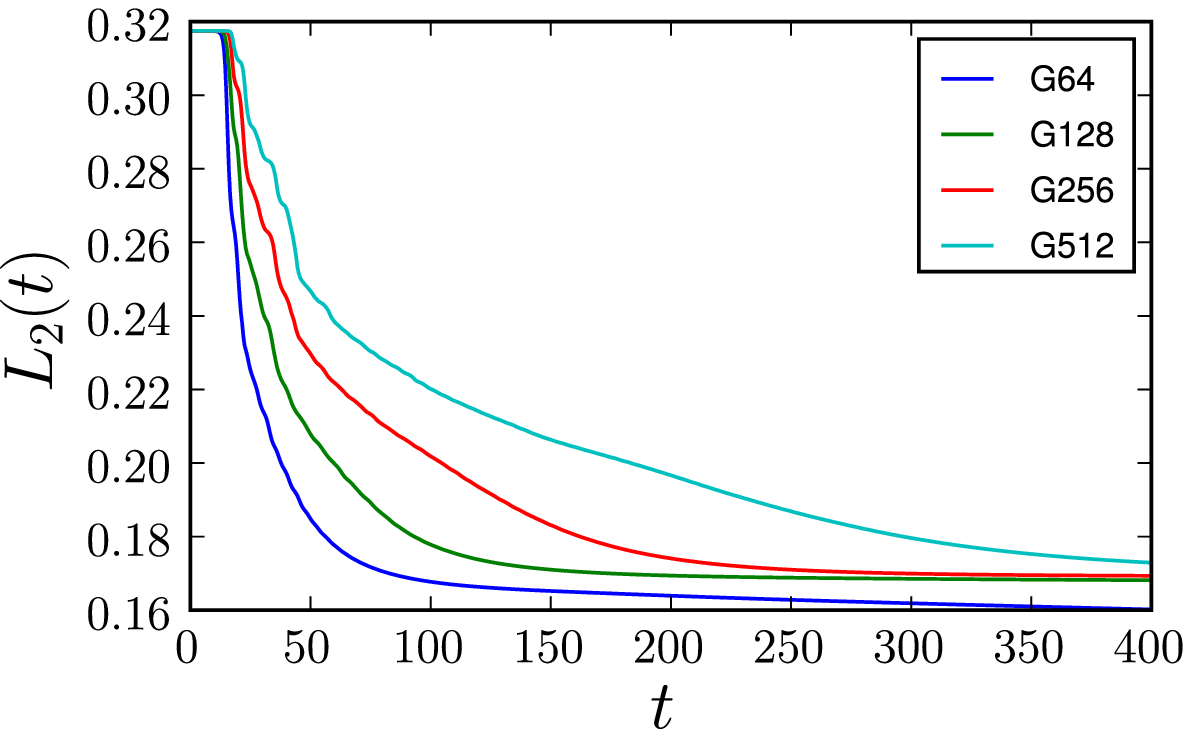}
  \caption{Evolution of $L_2$, same runs as Fig.~\ref{fig:G_M}.}
  \label{fig:G_I2}    
\end{figure}

$M$ displays a similar behaviour for all values of the resolution until $t\approx 60$, except for the lowest resolution run G64; $M$ has been checked to correspond to $N$-body dynamics with $N=10^6$ for short times (data shown in section~\ref{sec:numerical-check}).
$M$ grows up to a saturation value, identical for all runs, then oscillations take place and eventually damp to a stable value.

\begin{figure}[ht]
  \centering
  \includegraphics[width=\linewidth]{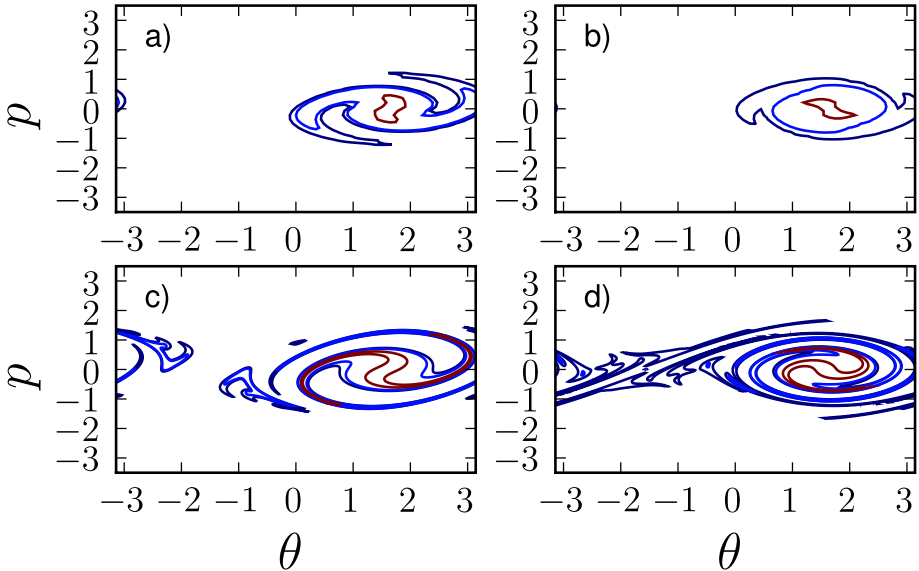}
  \caption{Phase space contour lines~: a) and b) depict the run G64 at times 32 and 64. c) and d) depict the from run G512 at the same times. While a) still presents the correct shape as compared to c), b) has lost most of the spiral shape present in d).}
  \label{fig:G_f}
\end{figure}

The damping of the oscillations is strongest in G64, they disappear at $t\approx70$. This damping is present in the other runs as well, a phenomenon that diminishes when the grid size is increased. This fact is to be put in correlation with the evolution of $L_2$ (Fig.~\ref{fig:G_I2})~: The decrease of $L_2$ means that the small scales are not well described anymore by the numerical procedure. This decrease is the strongest in G64 where $L_2$ does not reach a plateau and diminishes continuously.
The shape of $f$, a vortex rotating in phase space, is giving the oscillations of $M$. When this structure is smoothed out by the lost of the small scales, see Fig.~\ref{fig:G_f}, $M$ tends to a constant value. The run G512 displays a filamentary structure and oscillations in $M$ up to the end of the simulation.

The absence of oscillations in $M$ are directly linked to the fact that $f$ becomes
invariant under rotations in phase space in the process of numerical dissipation. Macroscopic quantities (such as $M$) cannot vary in time anymore, except for the small effects of the ongoing dissipation (see the continuous decrease of $L_2$ for run G64 in Fig.~\ref{fig:G_I2}). This phenomenon may differ significantly from the ideal evolution of $f$ and $M$ : in a number of situations, oscillations persist due to the presence of coherent structures; this is only observed in run G512.

These observations suggest a limit in time of the validity of the simulations. In the case of the free streaming, the evolution of $f$ leads to a recurrence after a time $T_R = 2\pi/(k\Delta{}v)$, called adequately ``recurrence time'' ($k$ is the mode number of the initial perturbation and $\Delta v$ is the grid spacing in velocity space, see section) at which $f$ recovers its initial value \cite{canosa_gazdag_fromm_recurrence}.
In situations close to free streaming, for instance in the linear Landau damping, $T_R$ is clearly identified in the evolution of the first mode of the electric field \cite{canosa_gazdag_fromm_recurrence}. In other situations, one can only use $T_R$ as a rough estimate and cannot identify its effects as sharply. It however agrees with observations made otherwise observing $\mathbf{M}(t)$ or cuts in $f$. For the run G64 the recurrence time $T_R$ is approximately $45$, discarding the simulation in Fig.~\ref{fig:G_fcut_64} where $t = 64$, agreeing with the observed discrepancy; for the run G512, we are close to the simulation time~: $T_R\approx 357$.

Asymptotically, all runs tend to a similar value of $M$, which we characterize with the standard deviation of $M$, $\sigma_M$, and the mean value across simulations, $\bar M$~: $\sigma_M/\bar M = 1.1\%$, or $0.7\%$ if we leave G64 aside.

\subsection{Stability of the initial condition}

The stability of homogeneous initial condition has been studied in Ref.~\cite{yamaguchi_et_al_physica_a_2004}. The authors of Ref.~\cite{yamaguchi_et_al_physica_a_2004} devise a criterion of formal stability for any homogeneous initial condition.
We consider a distribution function $f_0$ of the form~:
\begin{equation}
  \label{eq:num-hom-f0}
  f_0(\theta,p) = \frac{1}{2\pi}\ \varphi_0(p) \quad .
\end{equation}
Given a fixed point of the form given in Eq.~(\ref{eq:num-hom-f0}) of a conserved functional $F[f]$,
\begin{equation}
  \label{eq:functional}
  F[f] = C_s[f] - \beta H_V[f] - \mu \int f(\theta,p)\ d\theta\ dp\ ,
\end{equation}
where $C_s$ is a Casimir (depending on a {\it concave} function $s$) and $H_V$ the energy functional, one computes its variation $\Delta{}F[\delta f] = F[f_0+\delta f] - F[f_0]$. The sign of the second order term of $\Delta F$ determines the formal stability of $f_0$. $s$ disappears from the computation and criterion, depending only of $f_0$, becomes~:
\begin{equation}
I[f_0] = 1+\pi \int_{-\infty}^{+\infty} \frac{\partial_p f_0(\theta,p)}{p} dp > 0 \Leftrightarrow f_0 \textrm{ is formally stable.}
\end{equation}
The criterion defines, for a given family of $f_0$, a critical energy $U_c^\ast$ below which the initial condition is unstable.

We perform a check for the waterbag and Gaussian initial profiles. The waterbag is the following profile, for a given $U$:
\begin{equation}
  \label{eq:IC_WB}
  f(\theta,p) = \left\{
    \begin{array}{r l}
      \frac{1}{4 \pi \sqrt{6U}} \left(1+\epsilon\sin\theta\right), & p \leq \sqrt{6U}\cr
      0,                        & \mbox{else.}
    \end{array}\right.
\end{equation}
where we have again applied a small perturbation of order $\epsilon$ to trigger the instability. $\epsilon$ is set to $10^{-4}$ for all runs.

\begin{figure}[ht]
  \centering
  \includegraphics[width=.7\linewidth]{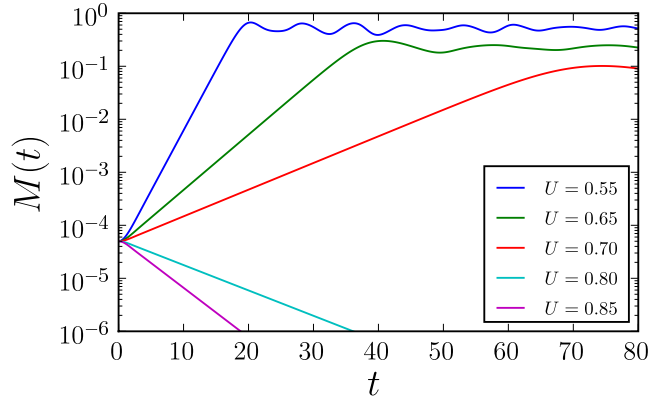}
  \caption{$M(t)$ for initial Gaussian profiles. $M$ grows if $U$ is below $U_c^\ast=3/4$.}
  \label{fig:SG_M}
\end{figure}

\begin{figure}[ht]
  \centering
  \includegraphics[width=.7\linewidth]{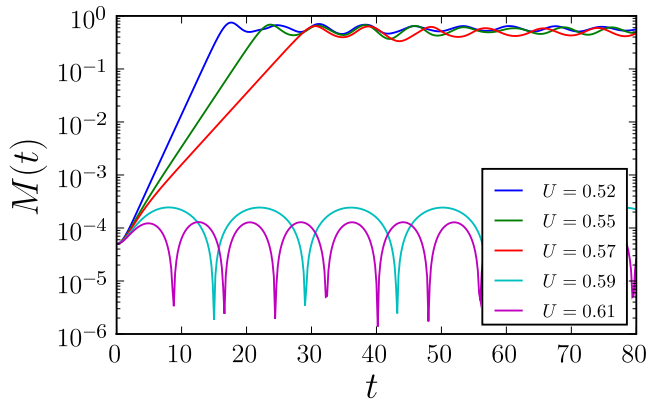}
  \caption{$M(t)$ for initial waterbag profiles. $M$ grows if $U$ is below $U_c^\ast=7/12\approx 0.58$. The slope before saturation of the three first runs is used to compute the exponential growth factor $\lambda$.}
  \label{fig:SW_M}
\end{figure}

Starting from both side of the critical energy ($U_c^\ast=7/12$ for the waterbag and $U_c^\ast=3/4$ for the Gaussian), we observe in Figs. \ref{fig:SG_M} and \ref{fig:SW_M} the change in stability. For values of $U$ above $U_c^\ast$, $M$ stays at the same order of magnitude or decreases.


In the case of the waterbag, Yamaguchi \etal{} \cite{yamaguchi_et_al_physica_a_2004} compute the exponential growth rate of $M(t)$. We compare that prediction with the simulation results of Fig. \ref{fig:SW_M} [fitting $\log M(t)$] and found a very good agreement. The exponential growth rate depends on the energy $U$~: $\lambda = \sqrt{6\left(U_c^\ast -U\right)}$. The theoretical and numerically computed values are given in couples $(\lambda_\textrm{th},\lambda_\textrm{num})$ for $U=0.52,0.55 \textrm{ and } 0.57$ respectively : $(0.61,0.61)$, $(0.43,0.43)$ and $(0.32,0.31)$.

\subsection{Effect of the numerical grid size}

\begin{figure}[ht]
  \centering
  \includegraphics[width=.7\linewidth]{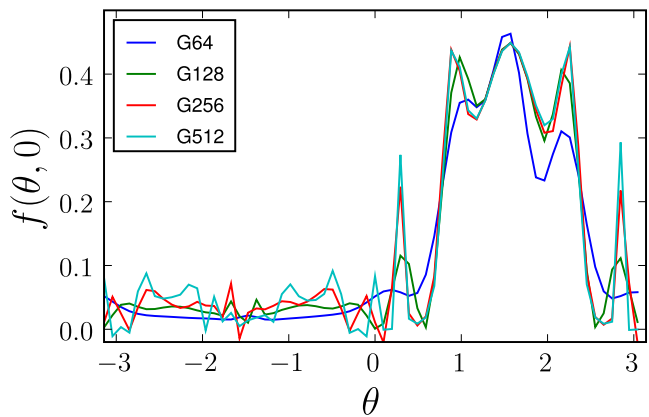}
  \caption{Cut in the function $f$ for $p^\ast=0$ at time $t=64$. Same runs as in Fig.~\ref{fig:G_M}. The locations of the peaks is similar in all runs. We observe that the heights of the peaks in run G64 stand apart.}
  \label{fig:G_fcut_64}
\end{figure}

\begin{figure}[ht]
  \centering
  \includegraphics[width=.7\linewidth]{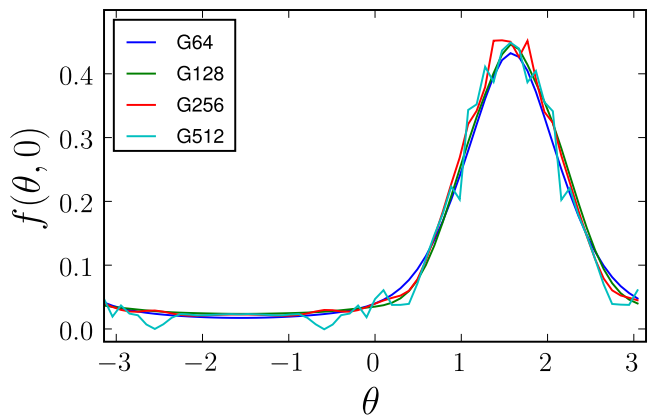}
  \caption{Cut in the function $f$ for $p^\ast=0$ at time $t=400$. Same runs as in Fig.~\ref{fig:G_M}. The location of the maximum is the same in this asymptotic state, and the value of the maximum is very similar for all runs.}
  \label{fig:G_fcut_400}
\end{figure}

The results of section~\ref{sec:prop} have considered different grid sizes for the same problem; $f(\theta,p)$ was inspected visually, but most of the analysis focused on macroscopic quantities : $L_1$, $L_2$, $U$ and $M$. Galeotti, Califano and Mangeney \cite{galeotti_califano_prl_2005,califano_galeotti_pop_2006} compared cuts in $f$, i.e. $f(\theta,p^\ast)$ where $p^\ast$ is fixed or $f(\theta^\ast,p)$ where $\theta^\ast$ is fixed, after smoothing of all runs to the lowest accuracy. This method explores with more precision the structure of $f$. For runs G64 to G512, we display $f(\theta,0)$ in Figs. \ref{fig:G_fcut_64} and \ref{fig:G_fcut_400} at two times~: $t=64$ when runs G128 to G512 still have a similar $M$, and at $t=400$ in order to discuss the asymptotic evolution.


At $t=64$, although the values for $L_2$ are different, which may imply a different evolution of $f$, we find peaks in $f(\theta,0)$ at identical locations. This indicates that $f$ behaves similarly. Indeed, the force on the particles (or phase space elements) only depends on $\mathbf{M}$. Thus, as long as $\mathbf{M}$ is identical for the runs, so is the force term.
What this means is that an equal ${\bf M}(t)$ gives a similar $f(\theta,p;t)$, up to the accuracy of a given grid on which $f$ is represented; this claim is supported by Fig.~\ref{fig:G_fcut_64} where the runs G128-G512 agree very well, in correlation with the agreement of $\mathbf{M}(t)$ shown in Fig. \ref{fig:G_M} up to $t\approx 64$.

At $t=400$, G64 is completely smooth, and the other runs still show a small filamentary structure. The location of the maximum is the same, a fact that contrasts with most results of Ref.~\cite{califano_galeotti_pop_2006} for a Vlasov plasma, and that indicates a similar asymptotic state.
Regarding the position of the final vortex, the initial dynamics of the runs G64-G512 forms only one vortex, whose location is stable, immediately after the growth of the initial perturbation of the initial condition (\ref{eq:IC_G}); an identical position is found in Ref.~\cite{califano_galeotti_pop_2006} in a particular case (final runs D,E and F).
On the other hand, the agreement in the cut of $f$ for the runs G128-G512 in Fig.~\ref{fig:G_fcut_64} is better in our case and reflects the similarity in $\mathbf{M}(t)$, as discussed in the previous paragraph.

We learn from these observations, in the case of the HMF model, the fact that one observable, $M$, describes not only the macroscopic evolution, but also monitors the ``quality'' of $f$ (i.e. the macroscopic evolution {\it and} the fact that $f$ has conserved a shape which is not invariant under rotation in phase space\footnote{See discussion on the oscillations in paragraph \ref{sec:prop}}; We point out that $M$ is not meant to replace the more complete information contained in $f$). The quantity $M$ is similar to the energy spectrum for a Vlasov plasma, the energy spectrum of the HMF model containing only one mode. The development of Vlasov codes may benefit from the simplicity of using the single quantity $M$. This is especially useful in the process of comparison between different algorithm that is often carried out in numerical analysis (for Vlasov plasmas, see for instance Ref. \cite{arber_vann_critical_compar_2002}).

\subsection{Conclusions}
\label{sec:conclu}

We presented in this chapter a detailed study of Vlasov numerical simulations for the Hamiltonian Mean-Field (HMF) model, that have been compared to previous work on the HMF model and to Vlasov results in plasma physics. We gave evidence that the mean-field order parameter $M$ of the HMF model is sufficient to compare different runs in terms of phase space evolution as well as dissipative (i.e. damping) properties.

Illustration was given of the ``non-Vlasov'' limit of the numerical simulation which we need to keep in mind to analyze simulations results carefully, as stated recently by other authors \cite{galeotti_califano_prl_2005,califano_galeotti_pop_2006,antoniazzi_califano_prl}.

Finally, we hope that cross research between long-range systems obeying a Vlasov equation and Vlasov plasmas may benefit to both fields by providing the first with a new tool and the latter with new insight on the numerical method.

\section{Application to the Free-Electron Laser}
\label{sec:num-FEL}

The Vlasov equation for the Colson-Bonifacio model for the Free-Electron Laser (FEL) reads (see chapter~\ref{chap:fel}):
\begin{eqnarray}
  \label{eq:numerical-vlasovwave}
  \frac{\partial f}{\partial \zb} &=& -p \frac{\partial f}{\partial \theta} + 2 (A_x\cos\theta-A_y\sin\theta) \frac{\partial f}{\partial p},\\
  \frac{\partial A_x}{\partial \bar z} &=& \int d\theta dp\ f \cos\theta, \label{eq:numerical-Ax}\\
  \frac{\partial A_y}{\partial \bar z} &=& -\int d\theta dp\ f \sin\theta. \label{eq:numerical-Ay}
\end{eqnarray}
These equations are very similar to the ones of the HMF model, but the mean-field term is not computed self-consistently. $A_x$ and $A_y$ are evolved through Eqs.~(\ref{eq:numerical-Ax}) and (\ref{eq:numerical-Ay}). An initial condition for $f$ is accompanied by an initial value for $A_x$ and $A_y$. When the initial $f$ is homogeneous, almost vanishing value for $A_x(0)$ and $A_y(0)$ correspond to self-amplified spontaneous emission (SASE). We will set $A_x(0)=10^{-4}$ and $A_y(0)=0$ in the following simulations, in order to trigger the instability.

The quantities $b_x[f]$ and $b_y[f]$ allow to compute the bunching of the electrons $b=\sqrt{b_x^2+b_y^2}$:
\begin{equation}
  b_x[f] = \int d\theta dp\ f \cos\theta \textrm{\quad ; \quad} b_y[f] = \int d\theta dp\ f \sin\theta
\end{equation}

We do not perform an in-depth study as for the HMF model, but present the general properties of simulations of interest for chapter~\ref{chap:fel}.

\subsection{Time splitting}
\label{sec:numerical-fel-time}

In order to integrate the system (\ref{eq:numerical-vlasovwave}), (\ref{eq:numerical-Ax}), (\ref{eq:numerical-Ay}), a time splitting needs to be defined. After testing several possibilities, we have settled for the following:
\begin{enumerate}
\item Advection in the $\theta$-direction, 1/2 time step
\item Computation of $b_x^\ast$ and $b_y^\ast$
\item Advection in the $p$-direction, 1 time step
\item Advection in the $\theta$-direction, 1/2 time step
\item Computation of $A_x^{s+1}=A_x^s + \frac{1}{2} (b_x^\ast + b_x)$ and $A_y^{s+1} = A_y^s - \frac{1}{2} (b_y^\ast + b_y)$
\end{enumerate}
This time splitting scheme avoids energy drifts, which is an important property for the situations encountered in paragraph~\ref{sec:fel-struct}.

\subsection{Properties}

\begin{figure}[ht]
  \centering
  \includegraphics[width=0.8\linewidth]{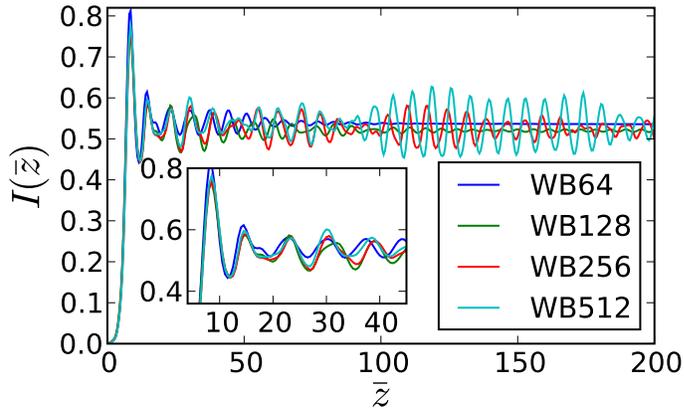}
  \caption{Evolution of $I={A_x^2+A_y^2}$ as a function of $\bar z$ in Vlasov simulations for the Free-Electron Laser.}
  \label{fig:numerical_FEL_I}
\end{figure}

Figure~\ref{fig:numerical_FEL_I} displays the evolution of $I=A_x^2+A_y^2$ as a function of $\bar z$ for a waterbag initial condition with $b_0=0.05$ and energy $U=0.10$, for grid sizes going from $N_\theta=N_p=64$ (WB64) to $N_\theta=N_p=512$ (WB512). As in the case of the HMF model, the initial dynamics is similar, and we observe a growing disagreement between the runs. The average final intensity is similar for all grid sizes.

The oscillations in $I(\bar z)$ follow a similar ordering as for the HMF model, indicating that the run with the largest grid size run is able to describe details that are too fine for the other runs. The added precision makes a qualitative difference in the behaviour of $I$.

\begin{figure}[ht]
  \centering
  \includegraphics[width=0.8\linewidth]{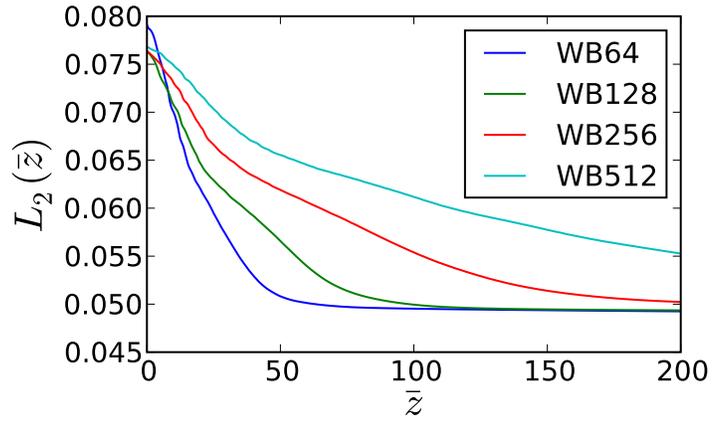}
  \caption{Evolution of the $L_2$ norm for the FEL. Same runs as in Fig.~\ref{fig:numerical_FEL_I}.}
  \label{fig:numerical_FEL_L2}
\end{figure}

Table~\ref{tab:fel-cons} presents the conservation properties of the simulations and Fig.~\ref{fig:numerical_FEL_L2} displays the evolution of the $L_2$ norm. The normalization is well preserved and the energy conservation does not improve significantly with the resolution.
\begin{table}[ht]
  \centering
  \begin{tabular}{l | r | r}
    run & $ \max \frac{L_1(t)-L_1(0)}{L_1(0)} $ & $ \max \frac{U(t)-U(0)}{U(0)}$ \cr
    \hline\hline
    & & \cr
    WB64  & $1.2\ 10^{-3}$ & $2.3\ 10^{-2}$ \cr
    WB128 & $6.8\ 10^{-5}$ & $3.8\ 10^{-4}$ \cr
    WB256 & $1.0\ 10^{-5}$ & $3.9\ 10^{-4}$ \cr
    WB512 & $4.9\ 10^{-6}$ & $4.3\ 10^{-4}$
  \end{tabular}
  \caption{Conservation properties for different grid sizes. Increasing the grid size allows for a better conservation of $L_1$. The conservation of $U$ depends also on the time step $\Delta t$ and could be improved by decreasing $\Delta t$.}
  \label{tab:fel-cons}
\end{table}

% Local Variables:
% TeX-master: "main"
% End:

\cleardoublepage

\chapter[Out-of-equilibrium dynamics for the free-electron laser]{Out-of-equilibrium dynamics in the Colson-Bonifacio model for the free-electron laser}
\label{chap:fel}

\myabstract{The Free-Electron Laser is a device in which a coherent beam of light is produced from a beam of electrons. A simple model for the wave-particle interaction describes the Free-Electron Laser and manifests a complex nonlinear dynamics. The Vlasovian formulation is given and we perform the corresponding simulations. In the context of the Free-Electron Laser, we describe the out-of-equilibrium phase transition taking place between lasing and non-lasing phases. We interpret the transition with the help of Lynden-Bell's theory and comment on the structures appearing in phase space.}{Colson-Bonifacio model, wave-particle interaction, out-of-equilibrium phase transition.}

%The free-electron laser is an advanced light source, with an output spectrum ranging from the X-rays to UV light. The physical process taking place is the emission of electromagnetic radiation by relativistic electrons. The resulting wave is coherent, making the device usable as a light source for the investigation of matter. A one-dimensional model describing the behaviour of the electron beam has been proposed by Colson REFREF and Bonifacio REFREF and then cast in an Hamiltonian setting \cite{bonifacio_et_al_opt_commun_1987}. The formulation is similar to the single wave model \cite{OWM_pof_1971,del-castillo-negrete_physica_a_2000}, and our results hold in that framework.

%\section{Introduction}
%\label{sec:fel-intro}

Mean-field models have been widely studied as paradigmatic representatives of the important class of systems subject to long-range coupling.
In the simplest scenario, $N$ particles are made to interact in one dimension, subject to a varying field which is self-consistently sensitive to the individual trajectories. A global network of connections is hence driving the dynamics of every constituting element, as it certainly happens for more realistic settings where e.g. gravity or unscreened Coulomb interactions are at play.

Free-electron lasers (FELs) are lasing devices consisting of a relativistic beam of charged particles, interacting with a co-propagating electromagnetic wave. The interaction is assisted by the static and periodic magnetic field generated by an undulator. FELs admit a mean-field description in term of the so--called Colson-Bonifacio model~\cite{colson_phys_lett_a_1976,bonifacio_et_al_opt_commun_1987}, which captures the essence of the collective wave-particle dynamics.

The FEL displays quasi-stationary states (QSS) \cite{barre_et_al_pre_2004} and is known to possess a threshold in energy above which the intensity drops to zero.
These QSSs bear an extraordinary conceptual importance as they potentially corresponds to the solely experimentally accessible regimes, for instance in plasma physics.

The QSS in the Hamiltonian Mean-Field model (HMF) have been explained by resorting to a maximum entropy principle, pioneered by Lynden-Bell in astrophysical context, and fully justified from first principles \cite{lynden-bell_1967}.

The Lynden-Bell protocol, also termed violent relaxation theory, has already proved effective in predicting the saturate intensity of a FEL \cite{curbis_et_al_epjb_2007,barre_et_al_pre_2004}.
However, no detailed study has been carried out for the FEL case aiming at unravelling the possible existence of out-of-equilibrium transitions of the type mentioned above. Are these transitions ubiquitous in mean field dynamics and, in this case, can we provide a consistent interpretative framework for their emergence? This chapter is dedicated to answering such questions, and  makes reference to the specific FEL setting. We also stress that the Colson-Bonifacio model of FEL dynamics can be regarded as a general formulation for all those applications where the complex interplay between particles and fields is well known to be central, e.g. electrostatic instabilities in plasma physics~\cite{elskens_escande_book}.  As a closing remark, we notice that, on the practical implication side, by disposing of reliable predictive tools on the system evolution, one can aim at guiding the system towards different experimental regimes.

This chapter is structured as follows: Section~\ref{sec:fel-model} casts the model and its Vlasov formulation, whose dynamics is studied in section~\ref{sec:ICandevo}. The Lynden-Bell theory is formulated in section~\ref{sec:fel-lb} and forms the basis of the analysis performed in section~\ref{sec:lb0andlba} and summarized in section~\ref{sec:fel-summary}.
The structures appearing in phase space are discussed in more detail in section~\ref{sec:fel-struct}.

\section{The FEL model}
\label{sec:fel-model}

The Colson-Bonifacio model for the FEL dynamics describes the coupled evolution of the electrons with a co-propagating wave \cite{bonifacio_et_al_nuovo_cimento_1990}. The equations read:
\begin{equation}
  \left\{\begin{array}{l l l}
      \frac{d\theta_j}{d\bar{z}} &=& p_j, \cr
      & &\cr
      \frac{dp_j}{d\bar{z}} &=& - \mathbf{A} e^{i\theta_j} - \mathbf{A^{\!\ast}} e^{-i\theta_j}, \cr
      & &\cr
      \frac{d\mathbf{A}}{d\bar{z}} &=& \frac{1}{N} \sum_j e^{-i\theta_j},
    \end{array}\right.
    \label{bonifacio}
\end{equation}
where $\theta_j$ stands for the particle phase with respect to that of the optical wave, $p_j$ being its conjugate normalized momentum.
The complex quantity $\mathbf{A} = A_x + i A_y$ represents the transverse field and $N$ the number of electrons composing the electron bunch. We assume here that electrons are perfectly resonant with the ponderomotive field generated by the optical wave and by the undulator. Should that condition not be met, an additional term quantifying the deviation from this ideal behaviour can be added to the above description. We make use of that possibility in sections~\ref{sec:fel-detuned} and \ref{sec:fel-poincare}.

In Eqs.~(\ref{bonifacio}), $\zb$ labels the longitudinal position along the undulator and it effectively plays the role of time.
The intensity of the laser field is $I=A_x^2+A_y^2$. As it can be seen from the last of Eqs. (\ref{bonifacio}), the bunching term, $b=\frac{1}{N} \sum_j e^{-i\theta_j}$, is the source of wave amplification.  The bunching quantifies the degree of localization of the electrons in the generalized space of their associated phases.
The above discrete system of equations admits a Hamiltonian formulation to which we shall make reference as
the $N$-body model. In the $N \rightarrow \infty$ limit, the system  (\ref{bonifacio}) converges to the following Vlasov-wave
set of equations \cite{barre_et_al_pre_2004}: 
\begin{eqnarray}
  \label{eq:vlasovwave}
  \frac{\partial f}{\partial \zb} &=& -p \frac{\partial f}{\partial \theta} + 2 (A_x\cos\theta-A_y\sin\theta) \frac{\partial f}{\partial p},\cr
  \frac{\partial A_x}{\partial \bar z} &=& \int d\theta dp\ f \cos\theta, \cr
  \frac{\partial A_y}{\partial \bar z} &=& -\int d\theta dp\ f \sin\theta.
\end{eqnarray}

Eqs. (\ref{eq:vlasovwave}) can be simulated numerically, allowing us to monitor the evolution of the phase space distribution function $f(\theta,p)$ along the $\zb$ axis. The numerical algorithm is given in chapter~\ref{chap:numerical}, with details regarding the FEL in section~\ref{sec:num-FEL}.

The results of the numerical integration are also checked versus $N$-body simulations and shown to return a perfect matching on relatively short time scale, for large enough values of $N$. On longer times, finite--$N$ corrections do matter. The discrete system is in turn sensitive to intrinsic granularity effects, stemming from the intimate finiteness of the simulated medium, and progressively migrate from the Vlasov state towards the deputed equilibrium configuration. When increasing its size, the system spends progressively more time in the Vlasov-like, out-of-equilibrium regime.
Formally, in the $N \rightarrow \infty$ limit, it never reaches equilibrium, being permanently trapped in the QSS.

As previously anticipated, our study is hence ultimately concerned with the emergence of QSSs, in a context where particles and waves evolve self-consistently. We shall be particularly interested in elucidating the occurrence of out-of-equilibrium phase transitions via dedicated numerical simulations, and in substantiating our claims analytically.
This study allows us to virtually extend the conclusion of Ref.~\cite{antoniazzi_et_al_prl_2007} to a broad spectrum of potentially relevant applications, beyond the specific case under inspection. Among other, it is again worth mentioning plasma physics: A formulation equivalent to model (\ref{eq:vlasovwave}) is in fact often invoked, when studying the collective effects of beam-plasma dynamics \cite{elskens_escande_book}.

\section{On the initial conditions and their subsequent dynamical evolution}
\label{sec:ICandevo}

Let us turn to discussing our results, as obtained via numerical integration of (\ref{eq:vlasovwave}). In order to make contact with the investigations reported in \cite{antoniazzi_et_al_prl_2007}, we shall employ in the following a two-dimensional water--bag initial condition in phase space, which can be seen 
as a rough approximation of a smooth Gaussian profile. A (rectangular) waterbag is formally parametrized by two quantities, namely the semi-width of the spanned 
interval in phase, $\Dth$, and its analogue in the momentum direction, $\Dp$.  
The corresponding expression for $f$ can be cast in the form (see also Fig. \ref{fig:lbwb} top-left):
\begin{equation}
  \label{eq:wbic}
  f(\theta,p) = \left\{\begin{array}{l l}
      f_0 & \mbox{if } |p|\leq \Delta p,\cr
          & \mbox{\ \ \ } |\theta|\leq \Delta\theta, \cr
      0 & \mbox{otherwise.}
  \end{array}\right.
\end{equation}
The initial conditions can be also characterized by defining 
\begin{equation}
\label{eq:bunch}
\left\{\begin{array}{r l}
  b_0= & \frac{\sin\Dth}{\Dth} , \cr
  \epsilon = &\frac{\Dp^2}{6},
\end{array}\right.
\end{equation}
where $b_0$ is the initial bunching, and $\epsilon$ the initial average kinetic energy per particle. Notice that we access all possible values of the bunching $b_0 \in [0,1]$ by properly tuning $\Dth$, and all positive energies 
$\epsilon$ by varying $\Dp$. Here, we limit our discussion 
to the case of vanishing initial optical field, $I_0 \simeq 0$, the relevant parameter space being therefore solely bound to the plane ($b_0$, $\epsilon$). 

This choice of initial parameters allows for a rich out-of-equilibrium dynamics. We insist nonetheless on the fact that we are specializing on a given bi-dimensional subset, and deliberately ignore the third, in principle available, direction of the reference parameter space. Quantifying the role of such an additional degree of freedom ultimately amounts to investigate the so-called seeded configuration \cite{doyuran_et_al_prl_2001,de_ninno_et_al_prl_2008} and will be the subject of future work.

\begin{figure}[ht]
  \centering
  \includegraphics[width=3.3in]{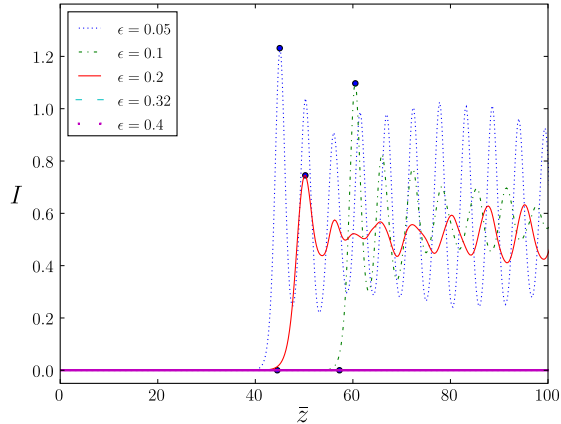}
  \caption{(Color online) As a result of the numerical integration of Eqs. (\ref{eq:vlasovwave}), the evolution of $I$ as a function of $\zb$ is reported for different choices of the energy 
  $\epsilon$ ($\epsilon = 0.05,0.1,0.2,0.3,0.4$); $b_0=0$. Symbols pinpoint the   
  position of the first peak in the intensity time series, thus returning an indication on the 
  saturation time. Notice that for $\epsilon > \epsilon_c=0.315$, 
  the peak is found for $I \ll 1$ : the corresponding initial conditions are hence stable, and no instability develops. }
  \label{fig:b0_0_wb}
\end{figure}

Let us start by discussing the simplest scenario, where the initial beam of particles is uniformly distributed over $[-\pi;\pi]$. From a physical point of view, this amounts to specialize to the case of Self-Amplified Spontaneous Emission (SASE, \cite{bonifacio_et_al_opt_commun_1984,brinkmann_xfel_2006}), where $b_0=0$ and no seed is applied externally. Such a choice was also considered by  
Barr{\'e} {\it et al.} \cite{barre_et_al_pre_2004} and Curbis {\it et al.} \cite{curbis_et_al_epjb_2007},  
where the dependence of the system evolution on the energy was numerically monitored within the  
$N$-body discrete viewpoint. Interestingly, $b_0=0$ is a stationary solution of the Vlasov system: a local perturbative calculation can hence be straightforwardly 
implemented so as to investigate its inherent stability \cite{bonifacio_et_al_nuovo_cimento_1990}; the calculations are detailed in appendix \ref{sec:stabilitywb}. For $\epsilon < 0.315$, an instability occurs: both 
the wave intensity and the bunching factor rapidly grow, before relaxing towards an oscillating plateau. The average value of $I$ reached in the oscillating regime is called the saturated intensity $\Imean$. This behaviour is displayed in Fig. \ref{fig:b0_0_wb}, where the simulations with $\epsilon > 0.315$ do not show an amplification of $I$. 
This is a well--known property, indeed correctly reproduced by our numerical simulations, and which first signals the existence of phase 
transitions, of the type depicted in \cite{antoniazzi_et_al_prl_2007}.
%Let us point here that this critical energy is linked to the stability of our homegeneous initial condition and that when we will consider non-homogeneous ones, the initial behaviour will be non-stationary and a non-linear evolution will follow.
To further corroborate our guess on the $b_0=0$ behaviour, we turn to measuring the saturated intensity $\Imean$ as function of the energy $\epsilon$, where $\Imean$ stands for the mean of $I$ after saturation. It is here computed during four oscillations of $I$.

\begin{figure}[ht]
  \centering
  \includegraphics[width=3.3in]{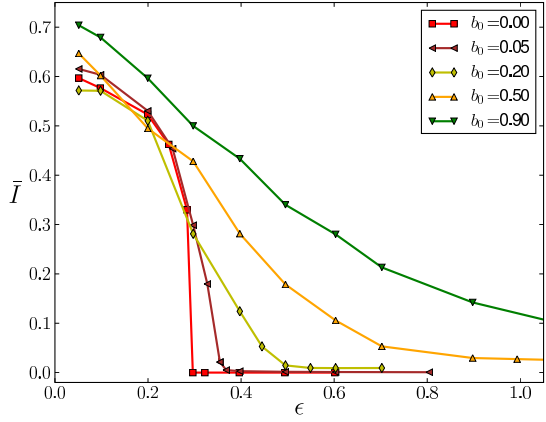}
  \caption{(Color online) Saturated intensity, $\bar{I}$, vs. $\epsilon$ for different choices of the initial bunching  $b_0 = 0.0,\ 0.05,\ 0.20,\ 0.50\ \mbox{and } 0.90$.}
  \label{fig:b0_all_I_of_e}
\end{figure}

As shown in Fig.  \ref{fig:b0_all_I_of_e}, $\Imean$ rapidly shrinks, 
when increasing the energy $\epsilon$, until a critical value is reached where a sudden transition to $\Imean \simeq 0$ is observed, bearing 
the  characteristic of a first order phase transition. This is a further point of contact with the analysis carried out in \cite{antoniazzi_et_al_prl_2007}
for the HMF toy model.

\begin{figure}[ht]
  \centering
  \includegraphics[width=3.3in]{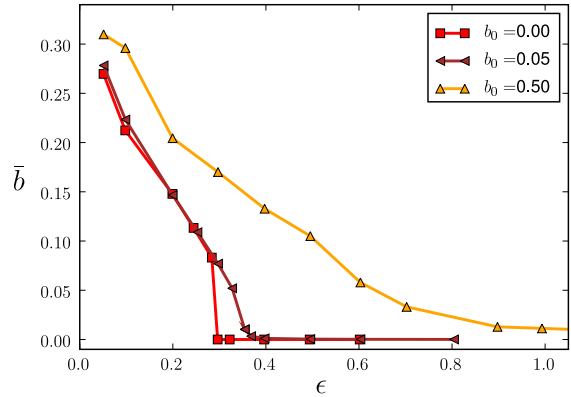}
  \caption{(Color online) Same as Fig. \ref{fig:b0_all_I_of_e}, for the saturated bunching, $\bar b$.}
  \label{fig:b0_all_b_of_e}
\end{figure}

Motivated by these findings, and to push the analogy with the HMF setting, we consider bunched initial distributions. From a physical point of view, this choice is relevant to the case of FELs working in the so-called harmonic generation regime \cite{doyuran_et_al_prl_2001,de_ninno_et_al_prl_2008}.
The positions of the particles are here initially assigned in order to uniformly span a limited portion of the allowed support, symmetric with respect to the origin, controlling the associated bunching via Eq. (\ref{eq:bunch}). 

The inhomogeneous ($b_0 \ne 0$) distribution in phase space is by nature non--stationary. The non-zero bunching amplifies the wave that rapidly acquires a non-zero value rapidly and, thanks to its initial velocity profile, the waterbag changes its shape with no possibility to remain stationary.
The Vlasov dynamics can however smooth it out into a homogeneous distribution 
($b = 0$, $I = 0$), possibly not of the waterbag type, or evolve to a bunched situation. The saturated mean-field average intensity 
$\Imean$ vs. the energy parameter $\epsilon$ is depicted in Fig.~\ref{fig:b0_all_I_of_e}, 
showing the newly collected data for different values of $b_0>0$ to the reference profile relative to $b_0=0$. In all cases the intensity is shown to decrease, 
as the energy increases. Importantly, for small values of $b_0$, an abrupt transition is observed, which can be naively interpreted as of the
first--order type. For larger values of $b_0$, the observed transition becomes smoother, such as for a second--order one. A substantially 
identical scenario holds for the bunching, which evolves towards an asymptotic plateau $\bmean$, also sensitive to the $\epsilon$ and 
$b_0$ parameters, see Fig. \ref{fig:b0_all_b_of_e}.  This scenario points towards a unifying picture on the 
emergence of out-of-equilibrium phase transitions within the considered class of mean-field Hamiltonian model. As previously anticipated, 
the Lynden-Bell theory of violent relaxation was successfully applied to the HMF problem, allowing one to gain a comprehensive 
understanding on the out-of-equilibrium phase transition issue, including a rather accurate characterization of the associated transition order. In the 
following section we set down to apply the Lynden-Bell argument to the present case, benchmarking the theory to numerical experiments.

\begin{figure}[ht]
  \centering
  \includegraphics[width=0.9\linewidth]{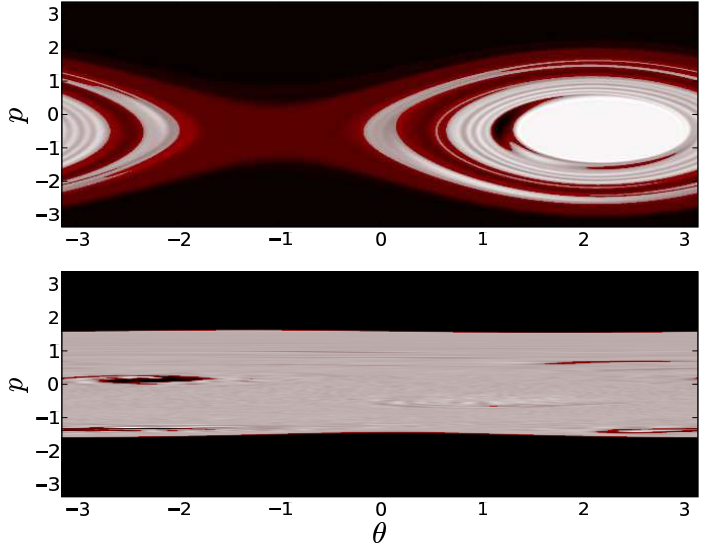}
  \caption{(Color online) Phase space density $f(\theta,p)$ for $b_0=0.05$ and $\epsilon = 0.2 \mbox{ (top) and } 0.4 \mbox{ (bottom)}$.}
  \label{fig:PhaseSpace}
\end{figure}

Before ending this section, we report observations of phase--space structures resulting from 
the Vlasov-based simulations. Two phase--space portraits are enclosed in Fig. \ref{fig:PhaseSpace}, and refer to different values 
of the energy $\epsilon$, respectively below (upper panel) and above (lower panel) the critical transition energy relative
to the selected (fixed) $b_0$ amount. When the system evolves towards a state at $\Imean \ne 0$, then $f(\theta,p)$  shows  a  
large resonance.
At variance, in the opposite regime, a hole-resonance dipole structure is observed, see also \cite{antoniazzi_et_al_wave-particle_2005}.  This observation seems to suggest that the out-of-equilibrium phase transition materializes via a bifurcation of invariant structures, an observation that has recently been made for the HMF model \cite{bachelard_et_al_prl_2008}.
These findings will be further discussed in section~\ref{sec:fel-struct}.
%, but their introduction is made here because they contribute to understand the lack of success of Lynden-Bell's theory in describing the phase transition.

\section{On the violent relaxation theory}
\label{sec:fel-lb}

In his work on self-gravitating systems, Lynden-Bell suggested \cite{lynden-bell_1967} that the collisionless dynamics governed by the Vlasov
equation tends to maximize a fermionic entropy. The latter is obtained from the classical definition, where the counting of the microscopic 
states, compatible with a given macroscopic configuration, results from a combinatorial calculation, and is sensitive to the underlying Vlasov dynamics.  The method was successfully employed    
in the study of the HMF model \cite{antoniazzi_et_al_prl_2007,chavanis_hmf_epjb_2006,antoniazzi_et_al_prl_2007} and also applied to predict the quasi-stationary amplitude of the FEL wave~\cite{barre_et_al_pre_2004,curbis_et_al_epjb_2007}. In these works, however, the analysis just focused on the unstable regime 
($\Imean \ne 0$): no attempt was in fact made to reconcile it, with the high energy homogeneous state, via the phenomenon of out-of-equilibrium phase transitions.  

In the following we shall review the main steps of the
derivation of the violent relaxation theory, applied to the FEL setting. Starting from a waterbag, 
the entropy to be maximized (see section~\ref{sec:vlasov-LB-entropy} for details) is:
\begin{equation}
  \label{eq:LBentro}
  s(\bar f) = - \int dp\ d\theta\ \left[ \frac{\bar f}{f_0}\ln\frac{\bar f}{f_0} + \left(1-\frac{\bar f}{f_0}\right)\ln\left(1-\frac{\bar f}{f_0}\right)  \right],
\end{equation}
where $f_0$  is specified in (\ref{eq:wbic}) and $\bar f$ is the coarse-grained distribution function.
The maximization problem is subject to constraints on normalization, total momentum and energy.

Following Barr{\'e} {\it et al.} \cite{barre_et_al_pre_2004,barre_phd}, maximizing the functional  (\ref{eq:LBentro}), results in the following set of equations
\begin{eqnarray}
  \label{eq:systemofeq}
    f_0 \frac{x}{\sqrt{\beta}} \int d\theta\ \zeta F_0(\zeta x) &=& 1, \cr
    f_0 \frac{x}{\sqrt{\beta}} \int d\theta\ \sin\theta\ \zeta F_0(\zeta x) &=&  A^3, \cr
    f_0 \frac{x}{2 \beta^{1.5}} \int d\theta\ \zeta F_2(\zeta x) &=& \epsilon + \frac{3}{2} A^4,
\end{eqnarray}
where $\zeta  = \exp\left(-2 A \beta \sin\theta \right)$. The functions $F_0$ and $F_2$ are defined as follows:
\begin{equation}
  \label{eq:fel-f0}
F_0(y)=\int_{-\infty}^\infty \frac{e^{-\frac{v^2}{2}} dv}{1+y\ e^{-\frac{v^2}{2}}}  
\end{equation}
\begin{equation}
  \label{eq:fel-f2}
  F_2(y)=\int_{-\infty}^\infty \frac{v^2 e^{-\frac{v^2}{2}} dv}{1+y\ e^{-\frac{v^2}{2}}}
\end{equation}

$\beta$ and $x$ are (rescaled) Lagrange multipliers and
ultimately stem from the conservation of mass, momentum and energy.
$A$, $\beta$ and $x$ are calculated by solving  Eqs. (\ref{eq:systemofeq}) 
numerically via a Newton-Raphson method~\cite{NR_in_f90}.  The resulting (real) value of $A$ is expected to return an estimate of the laser intensity at (Vlasov) saturation, $\Imean = A^2$, while $f(\theta,p)$ is:
\begin{equation}
  f(\theta,p) = f_0 \frac{1}{1+x\ e^{\beta (p^2/2 + 2 A \sin\theta + A^2p + A^4/2)}}.
\end{equation}

A useful simplification occurs for $A=0$ (namely $\Imean=0$) and the optimization problem (\ref{eq:systemofeq}) reduces to:
\begin{equation}
  \label{eq:oneeq}
  x=\sqrt{12 \frac{F_2(x)}{F_0(x)^3} \frac{\Dth}{\pi}}.
\end{equation}

\begin{figure}[ht]
  \centering
  \includegraphics[width=\linewidth]{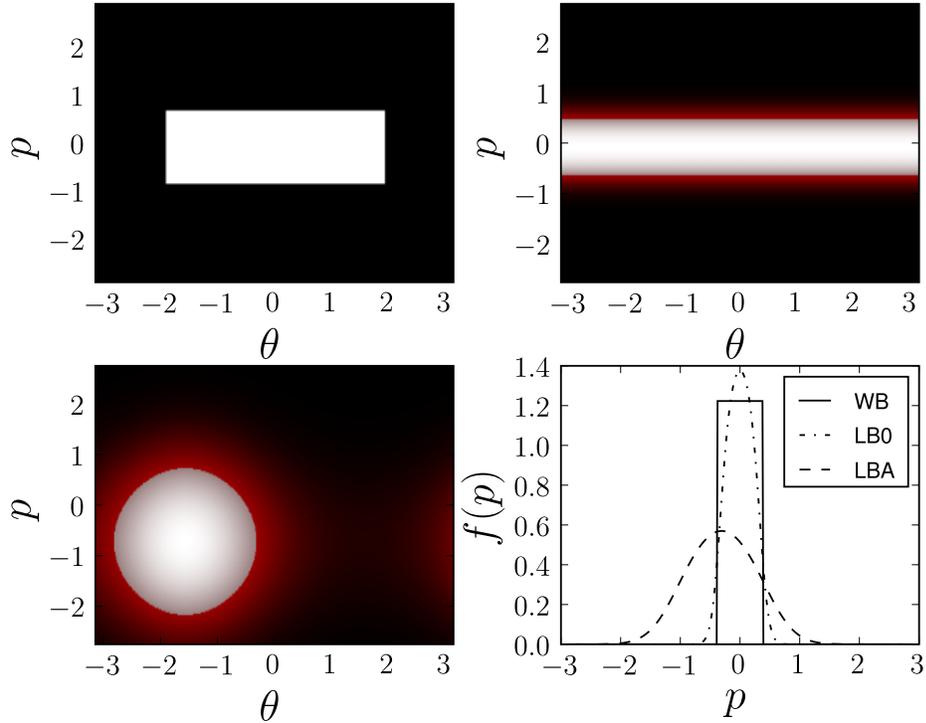}
  \caption{(Color online) Examples of $f(\theta,p)$: the waterbag (top-left), the LB0 solution (top-right), the LBA solution (bottom-left) and their corresponding velocity distribution function integrated over space.}
  \label{fig:lbwb}
\end{figure}

There exists a value of $x$ which solves the above equation for any choice of $\Dth$. The homogeneous state is a stationary 
solution of the Lynden-Bell entropy and thus a potentially attractive state of the Vlasov dynamics. 
Additional inhomogeneous solutions ($A \ne 0$, or, equivalently, $\Imean \ne 0$)  might however emerge from investigating the full 
system (\ref{eq:systemofeq}). The homogeneous and inhomogeneous solutions will be referred to as to LB0 and LBA, 
respectively (see Fig. \ref{fig:lbwb}). The forthcoming discussion will focus on how to discriminate between the two, and eventually predict the asymptotic fate of the system. 

\noindent

\section{Interpreting the out-of-equilibrium transition as a dynamical switch between two solutions of the theory}
\label{sec:lb0andlba}

\begin{figure}[ht]
  \centering
  \includegraphics[width=3.3in]{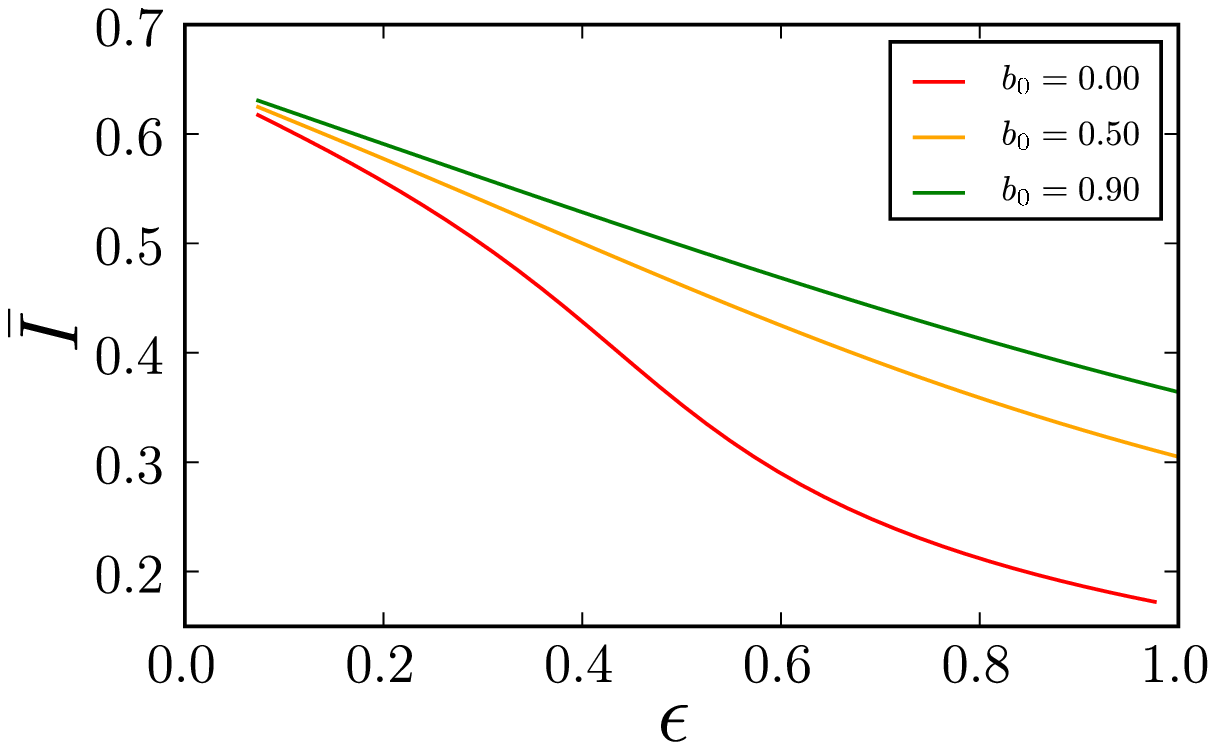}
  \caption{(Color online) The saturated intensity $\Imean$ for the (inhomogeneous) LBA solution of system (\ref{eq:systemofeq}) plotted as 
  function of the energy $\epsilon$. Different curves refer to distinct values of the initial bunching $b_0$ (see legend).
  }
  \label{fig:LBA_A}
\end{figure}

\begin{figure}[ht]
  \centering
  \includegraphics[width=0.9\linewidth]{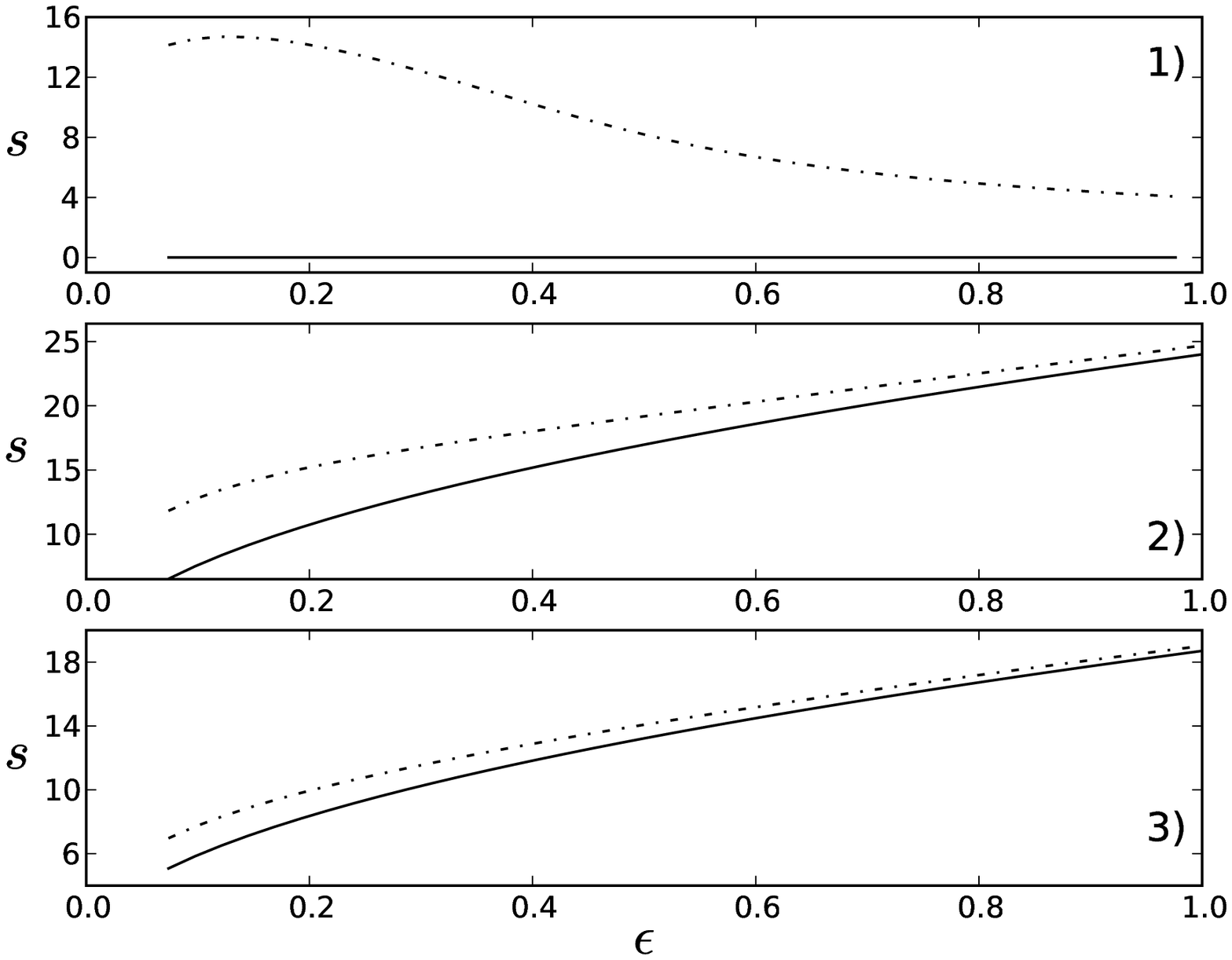}
  \caption{The Lynden-Bell entropy calculated respectively for the LB0 (plain line) and LBA (dash-dotted line) solution. Here 
  $b_0 = 0.00,\ 0.50,\mbox{ and } 0.90$ (resp. panels 1,2 and 3).}
  \label{fig:LB_entropy}
\end{figure}

Let us focus first on the LBA solution. In Fig. \ref{fig:LBA_A}, we report  the value of $\Imean$ as it follows from Eqs.~(\ref{eq:systemofeq}), for different choices of the energy and initial bunching.
The predicted intensity is shown to decrease when the energy gets larger but no transition is observed, in contradiction with the 
results of our numerical simulations. As previously stressed, the homogeneous LB0 state is also solution of the optimization problem
(\ref{eq:systemofeq}) and could in principle prevail over the former. To shed light on this issue, we 
calculated the entropy values $S_A$ and $S_0$, associated to LBA and LB0, respectively. 
Results of the computations are shown in Fig. \ref{fig:LB_entropy}, where the dependence on the energy $\epsilon$
is monitored for various choices of $b_0$. Surprisingly, and at odds with what happens for the HMF model \cite{antoniazzi_et_al_prl_2007}, 
$S_A$ is always larger than $S_0$. The two curves do not cross each other and the LBA configuration is entropically favored. Let us note that for higher values of $\epsilon$, $\Imean$ decreases and the two solutions get close to each other, also from the point of view of the entropy, without crossing.

The observed transition could possibly stem from a purely dynamical mechanism, 
and this would justify the discrepancy between the simulation output and the statistical prediction.
More specifically, we here argue that, depending on the selected initial conditions, 
the system explores a local basin of attraction and struggles to find its way to the deputed, 
global maximum of the entropy. To clarify this point, we focus on a single numerical simulation, assuming the system to be initialized in a LB0 state. 
The dynamics can progressively take the system towards the LBA configuration, leading to the maximization of the Lynden-Bell entropy.  
The opposite is not possible, and a simulation started in the LBA state will certainly not evolve to the LB0. However, dynamical effects might be also 
at play and interfere with the ideal situation here schematized, by virtually blocking the system in the neighborhood of an initially assigned 
LB0 configuration. Is this the correct scenario? And how to explain the observed transitions that instead relate to the waterbag initial condition? These
issues are addressed in the following, where the stability of LB0 and LBA is investigated via direct Vlasov simulations.

\begin{figure}[ht]
  \centering
  \includegraphics[width=3.3in]{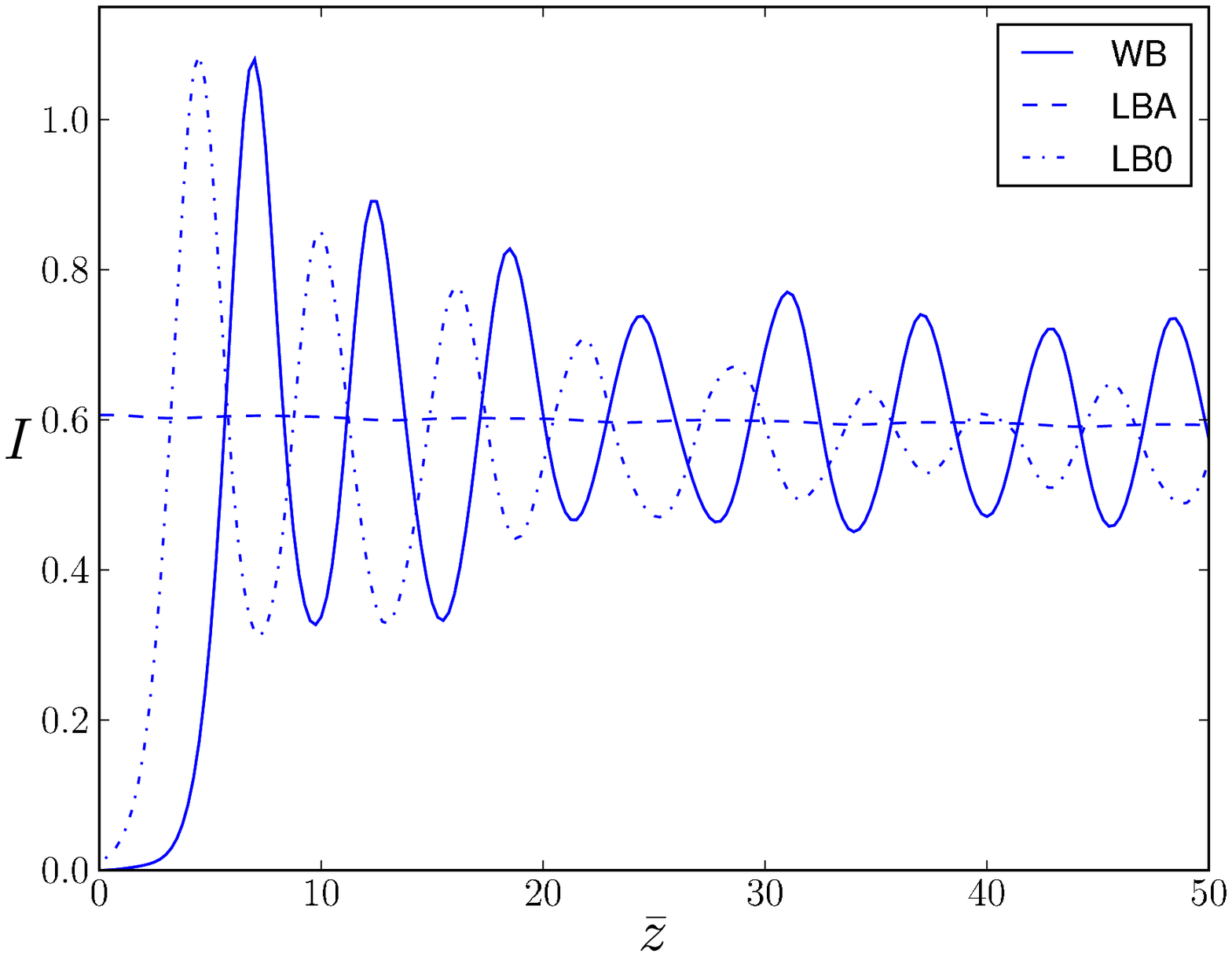}
  \caption{(Color online) The intensity $I$ as a function of time, for three different 
  choices of initial condition: WB (plain line), LBA (dashed line) and 
  LB0 (dotted line). All conditions refer to $b_0=0.05$,$\epsilon=0.10$}
  \label{fig:FEL_LBA_stable}
\end{figure}

In Fig.~\ref{fig:FEL_LBA_stable}, the dynamical evolution of the intensity $I$ is depicted, for three different classes 
of initial conditions, relative to the same choice of $\epsilon$ and $b_0$. The LBA is indeed stable, no deviation from the initial configuration being observed as an effect of the Vlasov dynamics. Conversely, the LB0 condition proves unstable, and the intensity converges towards an oscillating plateau. 
Interestingly, the LB0 and waterbag (WB) evolutions  are qualitatively similar, and, moreover, display the same average 
asymptotic value for the intensity $I$. Even more important, the asymptotic value corresponds to the LBA (maximum entropy) solution.

\begin{figure}[ht]
  \centering
  \includegraphics[width=3.3in]{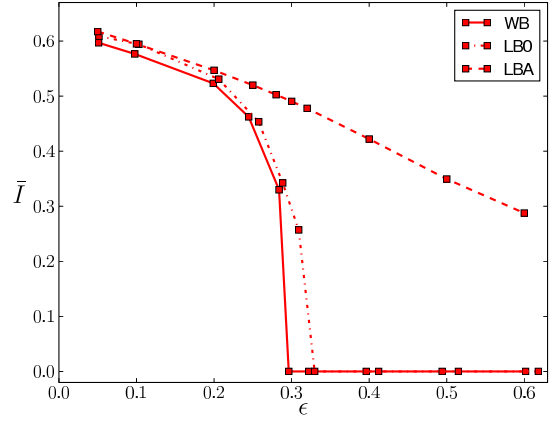}
  \caption{(Color online) The saturated intensity $\Imean$ is plotted as function of the energy $\epsilon$, for $b_0=0$.  WB, LB0 and LBA initial 
  conditions are considered, see legend.}
  \label{fig:FEL_comp_0.00}
\end{figure}

\begin{figure}[ht]
  \centering
  \includegraphics[width=3.3in]{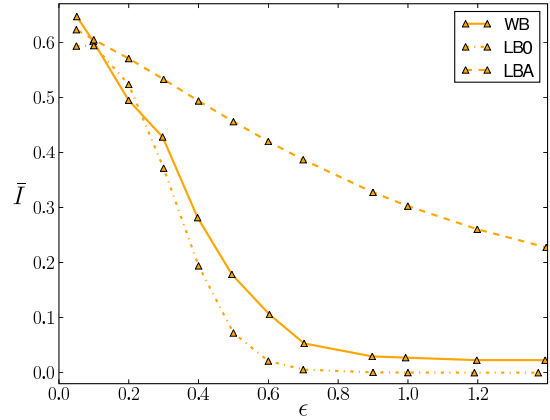}
  \caption{(Color online) The saturated intensity $\Imean$ is plotted as function of the energy $\epsilon$, for $b_0=0.50$.  WB, LB0 and LBA initial 
  conditions are considered, see legend.}
  \label{fig:FEL_comp_0.50}
\end{figure}

To further elucidate the analogies between LB0 and WB, and also clarify the issue of stability, we performed an extensive campaign of 
simulations, aimed at generalizing the results of Fig.~\ref{fig:FEL_LBA_stable}. The results of 
the investigations are reported in Figs.~\ref{fig:FEL_comp_0.00} and \ref{fig:FEL_comp_0.50}, where $\Imean$ is depicted versus 
$\epsilon$, for two selected $b_0$ amounts. The LBA initial condition is always stable under 
Vlasov dynamics. It is in fact a global maximum of the Lynden-Bell entropy, which, in this respect, proves adequate to describe the system 
at hand. The observed LB0 evolution is by far more complex. For large values of the energy, LB0 is stable. The stability is eventually lost 
when reducing the energy parameter: A transition materializes and the LB0 evolves towards the LBA state. In the vicinity of 
the transition, the LB0 initial condition approaches an asymptotic configuration, which slightly differs from the LBA one, and
possibly results from a balance between opposing dynamical strengths. The WB evolution mimics that of LB0, the two curves returning within a pretty close 
correspondence. In practice, during a short transient, an initially bunched WB expands (almost) ballistically, 
the particles being essentially transported by their own initial velocities, so visiting the whole interval $[-\pi;\pi]$. The obtained 
distribution can be approximated by a homogeneous LB0 state (the field has not yet developed, its intensity being effectively negligible), 
which in turn explains the observed correspondence. We shall however emphasize 
that this mechanism applies to relatively small $b_0$ amounts. This fact is testified in Fig.~\ref{fig:FEL_comp_0.50}: 
The agreement between LB0 and WB evolution is shown to worsen, when compared to that of Fig.~\ref{fig:FEL_comp_0.00}.
This is understood as follows: Starting from a high degree of bunching, the induced field opposes the natural ballistic contribution, 
by further enhancing the tendency to form a coherent clump of particles. 

In summary, our calculation returns two stationary points of the Lynden-Bell entropy. The first, which we termed LBA, corresponds to 
a inhomogeneous (laser on) configuration and it is a global maximum of the entropy. The second, labelled LB0, is homogeneous (laser off).
For sufficiently large energies, the system can be locally trapped in the vicinity of the LB0. This happens if the system is initiated close enough to a LB0 state, as e.g. in the case of WB with moderate $b_0$ values.
For smaller $\epsilon$, the LB0 loses stability and the system departs towards the entropically favored LBA state.
%Having detected no additional stationary points of the fermionic entropy, other than LB0 and LBA, we interpret the observed transition as a change of stabiity of LB0 which is a saddle point.
%Of all direction of evolution in $f(\theta,p)$-space, the one giving rise to the observed instability (eventually leading to LBA) becomes stable and the system is trapped near LB0, at least for the duration of our numerical experiments.

Having detected no additional stationary points of the fermionic entropy, other than LB0 and LBA, we interpret LB0 and LBA as a saddle point and a global maximum, respectively.
%While LB0 could in principle also be a global minimum, this hypothesis is invalidated by the fact that there exist (at least) a $f$ with a lower entropy.
%We give such a $f$ which, while not a stationary point of Lynden-Bell's entropy, is a possible stationary solution of system (\ref{eq:vlasovwave}) in the coarse-grained point of view implied by the theory~:
While LB0 could also be in principle a global minimum, this hypothesis is invalidated by the fact that there exists (at least) a function $f$, compatible with the system dynamics, which
yields a lower entropy value.  Consider in fact~:
\begin{equation}
  \label{eq:fLowerS}
  f(\theta,p) = \left\{\begin{array}{l l}
      f_0 \times \frac{\Dth}{\pi} & \mbox{if } |p|\leq \Delta p,\cr
      0 & \mbox{otherwise.}
  \end{array}\right.  
\end{equation}

This is not a stationary point of Lynden-Bell's entropy but represents one of the admissible stationary solutions of the dynamical system (\ref{eq:vlasovwave}), in its coarse-grained perspective as implied by the theory.
The Lynden-Bell entropy associated to Eq. (\ref{eq:fLowerS}) can be straightforwardly computed via Eq. (\ref{eq:LBentro}) and it is found to be always smaller than $S_0$, the value associated to the LB0 configuration, for all $b_0$ and $\epsilon$. Based on the above, and as previously anticipated,  we can exclude the possibility for LB0 to be a global minimum. Following our deductive reasoning, we hence suggest that LB0 is instead a saddle-point and the observed transition is consequently interpreted to stem from a local modification of LB0 stability properties or morphological characteristics (e.g. width/flatness of the stability basin). A detailed analytical characterization of LB0 stability is at present missing and could eventually help clarifying the underlying scenario.

%Lynden-Bell's entropy associated with Eq. (\ref{eq:fLowerS}) can be computed using Eq. (\ref{eq:LBentro}); it always falls below $S_0$, the value associated with LB0, for all values of $b_0$ and $\epsilon$. We are thus left with LB0 as a saddle point; we suspect a change in its shape (width of the stability basin) or stability at the transition, but analytical knowledge is missing on that point.

\section{Summary for the out-of-equilibrium phase transition}
\label{sec:fel-summary}

Using the case of a FEL as a paradigmatic example, we described the out-of-equilibrium dynamics of a mean--field model for wave-particle interaction. Our numerical investigation moves from the Vlasov version of the model, which rigorously applies to 
the continuous limit and is believed to constitute the correct interpretative framework to elaborate on the QSS peculiarities. Working within this context, and assuming a specific class of initial conditions, we identify a switch between different macroscopic regimes.  Such a transition is ultimately controlled by the nominal 
energy value and by the initial particles bunching, in qualitative agreement with what was previously observed for the HMF model in Ref.~\cite{antoniazzi_et_al_prl_2007}.

The Lynden-Bell violent relaxation theory is developed with reference to the FEL setting to quantitatively substantiate our findings. 
We numerically characterized the stability of the stationary points of the fermionic entropy functional.
A sudden change in their characteristic is related to the occurrence of the observed transition, 
supporting the adequacy of the violent-relaxation theory to describe the states reached by the dynamical system.

As a final comment, and beside stressing the unifying picture that is here brought forward, we 
emphasize that the transition here predicted can be in principle observed in real devices.
We regard this as a suggestion for future experiments 
which could eventually result in a direct proof on the existence of QSS for wave--particle systems.

\section{Structures in phase space}
\label{sec:fel-struct}

We come back in this section to the phenomenon of phase space coherent structures observed in section~\ref{sec:ICandevo}. The study of phase space topology is common in dynamical systems, and has already been the subject of investigations in mean-field models: Bachelard \etal{} \cite{bachelard_et_al_prl_2008} show the occurrence of regular structures in the kinetic limit for the Hamiltonian Mean-Field ; del-Castillo-Negrete \cite{del-castillo-negrete_procs_02} finds a coherent dipole in the wave-particle interactions through numerical investigations.

These structures prevent an exploration of the complete phase space for a given energy, preventing the statistical theory of Lynden-Bell to match the results of simulation. This reason has been invoked in the HMF model \cite{antoniazzi_califano_prl,antoniazzi_et_al_prl_2007}, the FEL (see section~\ref{sec:lb0andlba}), non-neutral plasmas \cite{levin_et_al_plasmas_prl_2008} or gravitational systems \cite{mineau_et_al_numerical_holes_1990,yamaguchi_pre_2008}.

In the present section, we aim at describing these features via numerical Vlasov dynamics.
Thereafter, we present the perspective of integrating Poincar{\'e} sections of particles evolving under a field $\left(A_x(t),A_y(t)\right)$ prescribed by a numerical integration of the Vlasov equation.

Ultimately, we would like to obtain a reduced model for the FEL describing these different structures, which would allow a better theoretical understanding.

Whereas we restricted ourselves in the present chapter to the resonant situation (see section~\ref{sec:fel-model}), we make use in this section of the detuning parameter $\delta$ that quantifies the deviation from this ideal behaviour. Varying $\delta$ unveils a rich phenomenology in the dynamical evolution.

The Vlasov-wave equations~(\ref{eq:vlasovwave}) become, with this addition~:
\begin{eqnarray}
  \label{eq:vlasovwave_det}
  \frac{\partial f}{\partial \zb} &=& -p \frac{\partial f}{\partial \theta} + 2 (A_x\cos\theta-A_y\sin\theta) \frac{\partial f}{\partial p},\cr
  \frac{\partial A_x}{\partial \bar z} &=&  - \delta A_y + \int d\theta dp\ f \cos\theta, \cr
  \frac{\partial A_y}{\partial \bar z} &=& \delta A_x    - \int d\theta dp\ f \sin\theta,
\end{eqnarray}
and the corresponding energy is
\begin{equation}
  \label{eq:fel_det_U}
  U[f]({\bf A}) = \int d\theta\ dp\ f(\theta,p) \left( \frac{p^2}{2} + 2 \left( A_x \sin\theta + A_y\cos\theta\right) \right) - \delta \left(A_x^2 + A_y^2\right) \quad .
\end{equation}

\subsection{The resonant case}
\label{sec:fel-resonant}

Setting $\delta = 0$ in the FEL model, the initial condition of the waterbag type is defined uniquely by $U$ and $b_0$.

\begin{figure}[ht]
  \centering
  \includegraphics[width=0.8\linewidth]{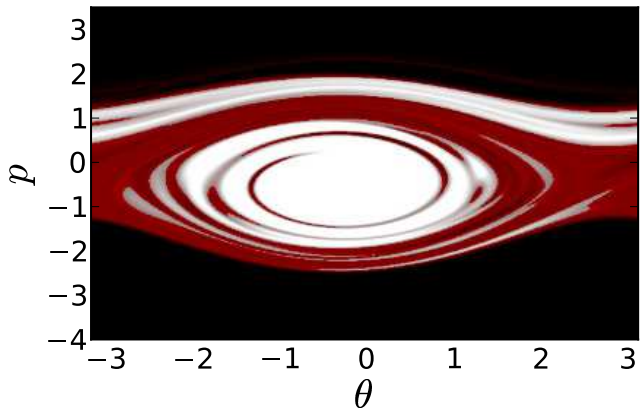}
  \caption{The resonant macroparticle in the FEL. $U=0.4$ and $b_0=0.2$.}
  \label{fig:reson_a}
\end{figure}

The most common occurring structure is the macroparticle, leading to a high value of the field. We depict it in Fig.~\ref{fig:reson_a}.

\begin{figure}[ht]
  \centering
  \includegraphics[width=0.8\linewidth]{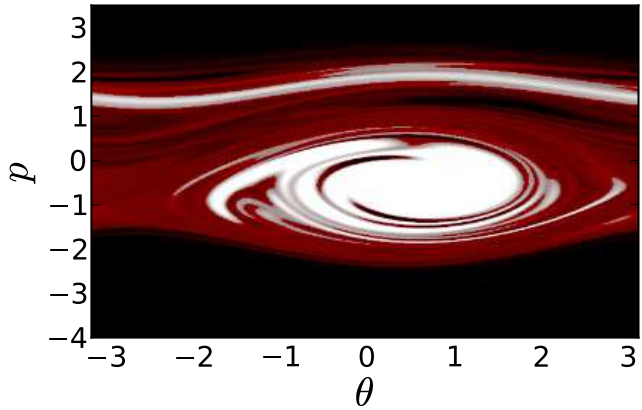}
  \caption{A resonant macroparticle and a depression. $U=0.6$ and $b_0=0.5$.}
  \label{fig:reson_c}
\end{figure}

When the macroparticle becomes less pronounced, we observe in addition a depression (``hole'') in $f$, travelling in the opposite direction as the macroparticle. This situation is represented in Fig.~\ref{fig:reson_c}.

\begin{figure}[ht]
  \centering
  \includegraphics[width=0.8\linewidth]{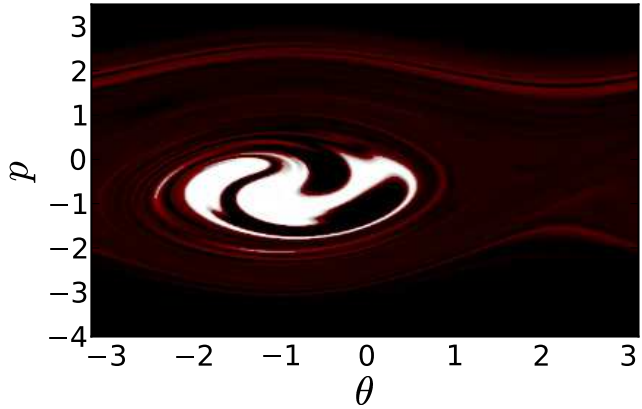}
  \caption{An inhomogeneous macroparticle. $U=0.6$ and $b_0=0.9$.}
  \label{fig:reson_d}
\end{figure}

In other situations, the macroparticle is not homogeneous and presents an internal structure. See for instance Fig.~\ref{fig:reson_d}.

These situations cannot be properly described by Lynden-Bell theory~: the theory predicts a distribution function in phase space that depends on the energy of the fluid element only, with a monotonously decreasing dependence (an illustration of the result for the theory can be found in Fig.~\ref{fig:lbwb}).

Indeed, the parameters given in Figs.~\ref{fig:reson_c} and \ref{fig:reson_d} fall in the region of disagreement between the theory and the simulations. For these parameters, the structure present in phase space prevent the possible description of the state attained by the dynamics by a solution of Lynden-Bell theory.
The results of the investigation for $b_0=0.5$, of application for Fig.~\ref{fig:reson_c} where $U=0.6$ and $b_0=0.5$, can be found in Fig.~\ref{fig:FEL_comp_0.50}. The evolution of the initial waterbag leads to a value for the intensity far below the one predicted (LBA solution).
One may also observe in Fig.~\ref{fig:reson_c} that a part of the initial waterbag behaves similarly to orbits outside of the separatrix of an equivalent pendulum and therefore spread uniformly on the interval $[-\pi ; \pi]$. This fraction of the phase space does not contribute to the bunching, diminishing the amplification of the wave.

\subsection{The detuned case}
\label{sec:fel-detuned}

The study performed in sections~\ref{sec:fel-model} through \ref{sec:lb0andlba} is restricted by the resonant condition ($\delta=0$ in the model). The phenomenology of dynamical regimes is however richer if we relax that condition. We present, as an illustration, some numerical results of interest because of the intrinsic difference they show with respect to the resonant case.

\begin{figure}[ht]
  \centering
  \includegraphics[width=0.8\linewidth]{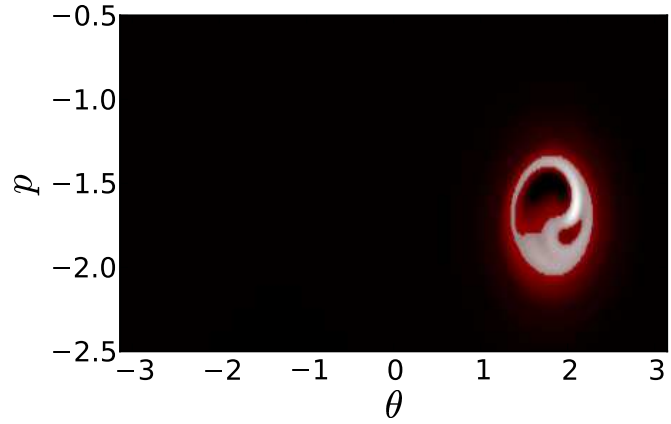}
  \caption{A dipolar structure in the detuned case. $U=4.2\ 10^{-4}$, $b_0=0$ and $\delta=-1.40$.}
  \label{fig:det_m140}
\end{figure}
\begin{figure}[ht]
  \centering
  \includegraphics[width=0.8\linewidth]{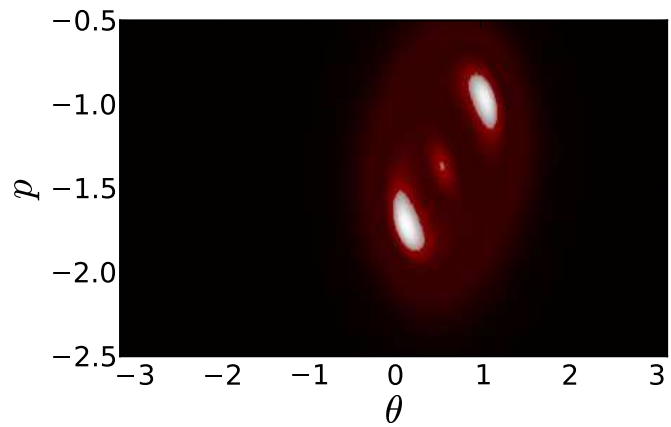}
  \caption{A tripolar structure in the detuned case. $U=4.2\ 10^{-4}$, $b_0=0$ and $\delta=-1.05$.}
  \label{fig:det_m105}
\end{figure}

Figures~\ref{fig:det_m140} and \ref{fig:det_m105} display a dipolar structure and a tripolar structure respectively. Similarly to what happens in section~\ref{sec:fel-resonant}, these results cannot be described by a solution of Lynden-Bell's theory.

\subsection{Poincar{\'e} sections}
\label{sec:fel-poincare}

A common technique to investigate nonlinear dynamics is the Poincar{\'e} section.
It consists in observing the variables of a full dynamical system at discrete times in a reduced space. The times at which we collect these position are defined in terms of the variables of the complete system.

We will use a criterion based on ${\bf A}(t)$ to collect data in the FEL. This technique has been applied recently to the HMF model, in a $N$-body setting, in order to understand the collective behavior in the thermodynamic limit.

We present in this paragraph a technique that allows a quantitative improvement for the study of phase space.
We consider a system of particles subjected to an external forcing ${\bf A}(t) = \left( A_x(t), A_y(t) \right)$ where the forcing is taken as the time series resulting from a Vlasov simulation.
The positions and velocities of a set of particles are collected at each passage of $I=A_x^2+A_y^2$ through its mean value.

The use of Poincar{\'e} sections in the study of the HMF and FEL models has already been used in a full $N$-body setting \cite{morita_kaneko_prl_2006,bachelard_et_al_prl_2008}. Both papers indicate that the dependence on the number of particles $N$ is significant. Our approach is original and removes the dependence on $N$ from the start. It allows the use of Poincar{\'e} sections in a new context, making its power of interpretation available to understand the dynamical properties of the Vlasov equation.

\begin{figure}[ht]
  \centering
  \includegraphics[width=0.75\linewidth]{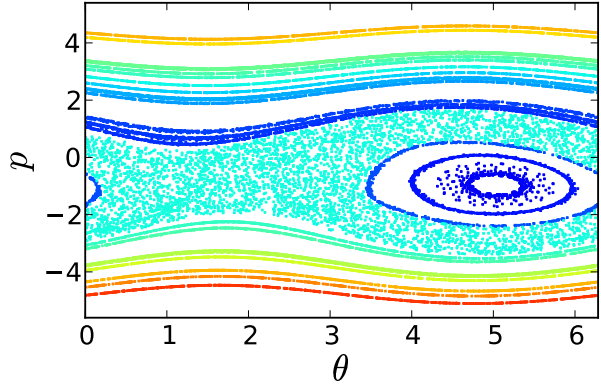}
  \caption{Poincar{\'e} section for the FEL. The particles are subjected to the force field of a Vlasov simulation, for parameters $b_0=0$, $U=4.2\ 10^{-4}$ and $\delta=-0.5$. The color code is arbitrary. Phase space displays an elliptic fixed point at $(\theta,p)\approx (5,-1)$.}
  \label{fig:fel-sec-a}
\end{figure}

\begin{figure}[ht]
  \centering
  \includegraphics[width=0.75\linewidth]{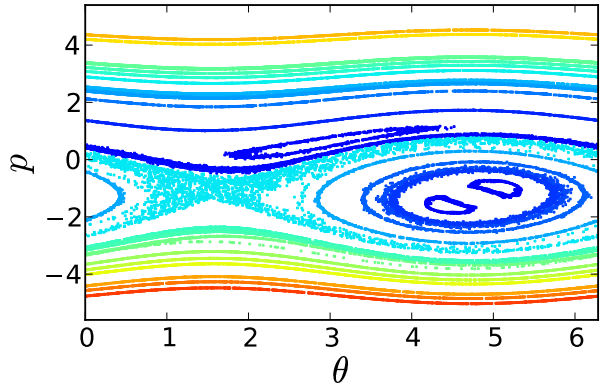}
  \caption{Poincar{\'e} section for the FEL. The particles are subjected to the force field of a Vlasov simulation, for parameters $b_0=0$, $U=4.2\ 10^{-4}$ and $\delta=-1.0$. The color code is arbitrary. Phase space displays an orbit of period 2 at $(\theta,p)\approx (5,-1)$.}
  \label{fig:fel-sec-b}
\end{figure}

The Poincar{\'e} section displays structures with significant details and an analysis of the trajectories allows us to identify periodic orbits.
Figures~\ref{fig:fel-sec-a} and \ref{fig:fel-sec-b} display two qualitatively different regimes, indicating a bifurcation between different phase space structures.

Figure~\ref{fig:fel-sec-a} contains an elliptic-like structure, similar to the macroparticle observed earlier. Figure~\ref{fig:fel-sec-b} on the other hand shows a period-2 orbit~: around $(\theta,p)\approx (5,-1)$, a single particle finds itself successively near $(4.5,-2)$ and $(5.1,-0.8)$.

A wide range of behaviors are found in the system while varying $\delta$, including orbits of higher periods.
Work is in progress to extend the results towards the understanding of the bifurcation at the origin of the period-2 and higher period orbits.

%\section{Summary}

% Local Variables:
% TeX-master: "main"
% End:

\cleardoublepage

\chapter{Relaxation in the Hamiltonian Mean-Field model}
\label{chap:hmf}

\myabstract{Vlasov dynamics exhibits incomplete relaxation in a number of situations. We propose an analogy with the mixing in fluids and apply it to the study of Vlasov dynamics in phase space. We find that stretching and folding structures appear in the evolution of the Hamiltonian Mean-Field model. In order to quantify the development of these structures, we compute the contour of the fluid in phase space through high accuracy simulations.}{mixing ; stretching and folding.}

The statistical mechanics of systems with long-range interactions is known for simple models with good detail (see for instance \cite{campa_et_al_phys_rep_2009}). The approach to equilibrium remains however a challenging aspect in these systems.

For instance, given appropriate initial conditions, a system may not reach equilibrium in the thermodynamic limit.
It is instead trapped in a so-called quasi-stationary state (QSS) whose lifetime grows with the number of particles \cite{antoni_ruffo_1995,latora_et_al_prl_1998}.

A most remarkable achievement is the prediction of the state of the system in these QSSs thanks to Lynden-Bell's theory \cite{lynden-bell_1967,antoniazzi_et_al_prl_2007}. Lynden-Bell's theory bases itself on the microscopic properties of Vlasov dynamics and on the assumption that there is a mechanism leading the system to its most probable state.
The framework of the Vlasov equation leads to a better understanding of the dynamical properties of systems with long-range interactions \cite{yamaguchi_et_al_physica_a_2004,barre_et_al_physica_a_2006,jain_et_al_relaxation_times_2007,antoniazzi_califano_prl}.

%Relaxation in models with long-range interactions has been the subject of much interest in the case of gravitational models in the form of theoretical and numerical studies \cite{luwel_severne_1985,funato_et_al_not_relaxation_1992}.
%It should also be noted that Lynden-Bell's theory fails when the dynamics conserves certain structures in phase space \cite{antoniazzi_califano_prl,levin_et_al_plasmas_prl_2008,levin_et_al_gravit_pre_2008}.

It is argued \cite{antoniazzi_califano_prl,levin_et_al_plasmas_prl_2008,levin_et_al_gravit_pre_2008} that the discrepancies of Lynden-Bell's theory 
to successfully predict the outcome of a system for some ranges of initial conditions are due to purely dynamical effects.

In order to detail the relaxation process taking place in phase space, we propose an analogy with fluid mixing. Ottino \cite{ottino_book_1989} gives a detailed introduction to the dynamical theory underlying mixing in fluids.

Taking the Hamiltonian Mean-Field (HMF) model as an example, we compute properties of the dynamics from the phase space fluid point of view.
We identify regions in which no dynamical evolution takes place, whereas the contour of the distribution function experiences substantial deformations in phase space.

We would like to add that the results presented in this chapter would greatly benefit to the understanding of relaxation in the gravitational sheet model. Extensive studies of this model exist in the literature \cite{luwel_severne_1985,funato_et_al_not_relaxation_1992,yamaguchi_pre_2008} and the approach that we present would complete adequately the dynamical considerations, especially those of Ref.~\cite{funato_et_al_not_relaxation_1992} in which a separation of the initial waterbag into distinct regions is performed in phase space.

This chapter is structured as follows: section~\ref{sec:hmf-defs} defines the HMF model and the initial conditions that we consider, section~\ref{sec:hmf-mix-fluids} introduces the analogy between fluid dynamics and phase space dynamics, section~\ref{sec:hmf-pendulum} details results for the simplified case of the pendulum and section~\ref{sec:hmf-struct} presents the results for the HMF model.

\section{Dynamical evolution of the HMF model}
\label{sec:hmf-defs}

The Vlasov equation describing the HMF is:
\begin{eqnarray}
  \label{eq:hmf-vlasov}
  \frac{\partial f}{\partial t} &+& p \frac{\partial f}{\partial \theta} - \frac{dV[f]}{d\theta} \frac{\partial f}{\partial p} = 0 \cr
    & & \cr
  V[f](\theta) &=& 1 - M_x[f] \cos\theta - M_y[f] \sin\theta
\end{eqnarray}
where ${\bf M} = (M_x,M_y)$ is defined by:
\begin{equation}
  M_x[f] = \int d\theta dp\ f \cos\theta \quad \textrm{ ; } \quad  M_x[f] = \int d\theta dp\ f \sin\theta \quad \textrm{ ,}
\end{equation}
and the magnetization $M$ is the modulus of the vector $({\bf M})$.
Equation~\ref{eq:hmf-vlasov} is solved numerically with the code presented in chapter \ref{chap:numerical}. 

The investigations in this chapter are performed for a waterbag initial condition defined by:
\begin{equation}
  \label{eq:hmf-wbic}
  f(\theta,p) = \left\{\begin{array}{l l}
      f_0 & \mbox{if } |p|\leq \Delta p,\cr
          & \mbox{\ \ \ } |\theta|\leq \Delta\theta, \cr
      0 & \mbox{otherwise,}
  \end{array}\right.
\end{equation}
where $\Delta\theta$ and $\Delta p$ characterize the extent of the waterbag in both directions of phase space. Normalization imposes $f_0 = \frac{1}{4 \Delta\theta\ \Delta p}$. We can
 also characterize the initial condition uniquely by $M_0=\frac{\sin\Delta\theta}{\Delta\theta}$ and $U=\frac{\Delta p^2}{6}+\frac{1}{2}(1-M_0^2)$.

\begin{figure}[ht]
  \centering
  \includegraphics[width=.8\linewidth]{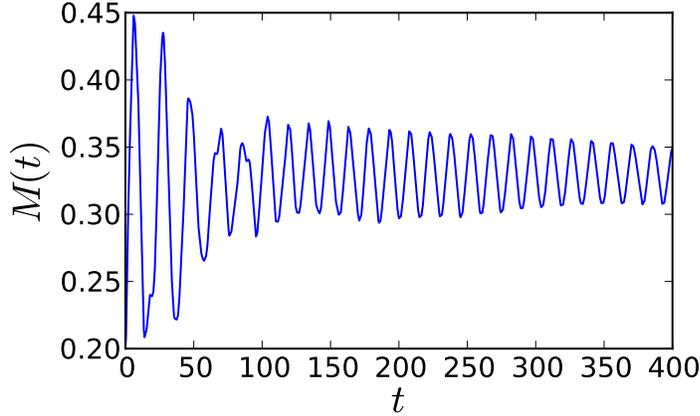}
  \caption{The magnetization $M$ in function of time for a waterbag initial condition with $U=0.6$ and $M_0=0.2$.}
  \label{fig:hmf-M_of_t}
\end{figure}

We use throughout this chapter the waterbag initial condition with $U=0.6$ and $M_0=0.2$. Figure~\ref{fig:hmf-M_of_t} displays the magnetization $M$ as a function of time. The parameters used in Fig.~\ref{fig:hmf-M_of_t} are: $N_\theta=N_p=512$ and $\Delta t=0.1$.
$M(t)$ begins with strong oscillations. After $t\approx 100$, the oscillations become regular and damp slowly.

\section{Mixing in fluids and in phase space}
\label{sec:hmf-mix-fluids}

The region of phase space defined by the initial waterbag can be regarded as a fluid. The Vlasov equation presents the form of a complicated advection equation deforming the said fluid. This deformation leads to a filamentary structure even in the simplest setup of free streaming (see section~\ref{sec:vlasov-filamentation}).

The Vlasov equation possesses the property of keeping levels constant: the one-particle probability distribution function, starting with $0$ and $f_0$ as the only two possible values, will keep $0$ and $f_0$ as the only accepted values during the whole time evolution. The conservation of the normalization in the course of time implies that the surface occupied by the fluid in phase space also remains a constant.
The perimeter of that region can grow in time as the shape of the fluid evolves and develops filaments.
We expect it to behave similarly to a 2D fluid submitted to an external perturbation, a situation giving rise to mixing in fluid dynamics \cite{ottino_book_1989}.

The length of the contour is monitored in the course of time and serves as a quantitative indication of the deformation of the waterbag.

For practical purposes, we define the perimeter $P_f(t)$ as the length of the contour of $f(\theta,p)$ whose height is $f_0/2$. This allows a convenient numerical computation in simulations through the use of the CONREC Fortran subroutine by Bourke \cite{bourke_conrec_1987}. We implemented this routine in the Vlasov simulation code presented in chapter~\ref{chap:numerical}.

\subsection{The free-streaming case}
\label{sec:hmf-free-streaming}

An analytical computation of $P_f(t)$ is possible in the case of free streaming. $P_f(t)$ grows linearly in time, except for a small deviation for short times: 
\begin{equation}
  \label{eq:hmf-free-streaming}
  P_f(t) = 4 \Dth + 4\Dp \sqrt{1+t^2} \quad .
\end{equation}

\begin{figure}[ht]
  \centering
  \includegraphics[width=.8\linewidth]{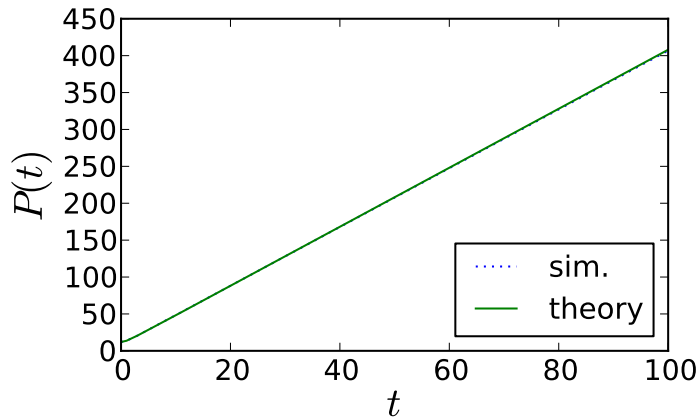}
  \caption{Evolution in the free streaming case of the perimeter $P_f(t)$ for an initial waterbag with $\Dth=2$ and $\Dp=1$. The simulation matches exactly the expected theoretical behaviour.}
  \label{fig:hmf-check}
\end{figure}

This can be readily checked through numerical simulation. Figure~\ref{fig:hmf-check} displays a perfect agreement between Eq.~(\ref{eq:hmf-free-streaming}) and the numerical computation.

We expect that a linear behavior of $P_f(t)$ will characterize a regular dynamics, as in the case of integrable systems.

\subsection{Dynamics in the presence of interactions}
\label{sec:hmf-with-int}

\begin{figure}[ht]
  \centering
  \includegraphics[width=.8\linewidth]{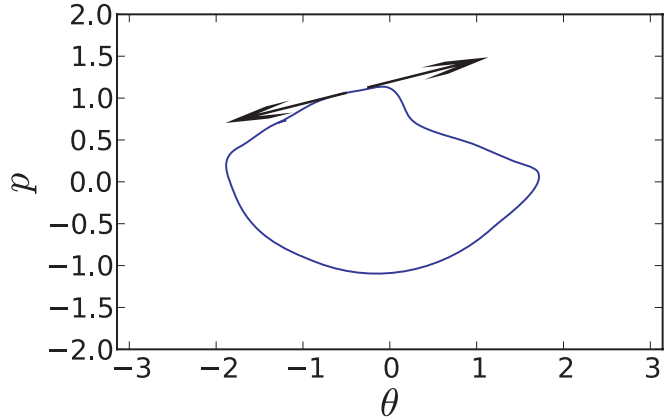}
  \caption{Illustration of the deformation of a waterbag. Between the two arrows, the contour experiences stretching, giving rise to a growth of $P_f(t)$.}
  \label{fig:hmf-stretch}
\end{figure}

\begin{figure}[ht]
  \centering
  \includegraphics[width=.8\linewidth]{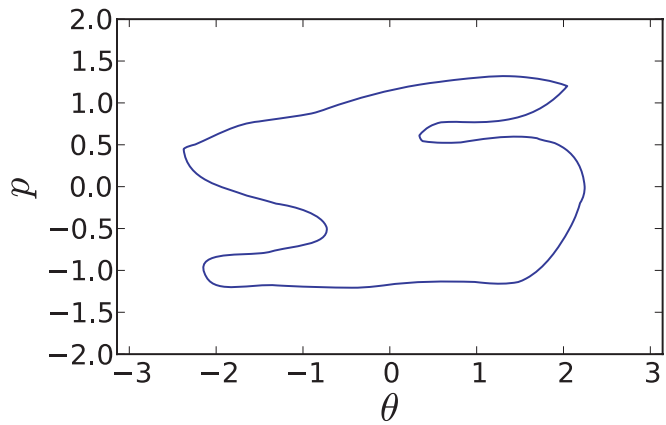}
  \caption{Illustration of the deformation of a waterbag. After the stretching, the contour folds on itself.}
  \label{fig:hmf-fold}
\end{figure}

The contour of the waterbag experiences complicated deformations in phase space. We give a sketch of this idea in Figs.~\ref{fig:hmf-stretch} and \ref{fig:hmf-fold}.

The repeated action of stretching and folding modifies considerably the original waterbag. It is recognized as a signature of a chaotic flow in fluid dynamics \cite{ottino_book_1989}.

In Fig.~\ref{fig:hmf-stretch}, the deformation pointed by the arrows leads to a stretching of the contour. The distance between two nearby points on the contour increases, causing the length of the contour to increase. If the separation between these points grows exponentially, the behavior of $P_f(t)$ will contain an exponential contribution~:
%. This contribution, if it is significant enough, allows us to write:
\begin{equation}
  P_f(t) \propto e^{\lambda t} \quad .
\end{equation}
If the average magnetization becomes a constant after some time $t_C$, we expect that the perimeter acquires a linear behavior after $t_C$, similar to the case of the pendulum (see below and in Fig.~\ref{fig:mix-pen-P}).

$P_f(t)$ being accessible in simulations, we dispose of a method to quantify the deformation of the waterbag in the course of time. The observation of an exponential behavior would then provide the {\it a posteriori} validation of our hypothesis.

\section{The pendulum}
\label{sec:hmf-pendulum}

The Hamiltonian for the HMF model is very similar to the one of a pendulum. In the pendulum, the field is set to a constant value while in the HMF model it depends on the positions of the particles.
This similarity is exploited in chapter~\ref{chap:pendulum} with great detail and we discuss here only what is needed to allow the interpretation of the fluid analogy.

For the initial waterbag given by $U=0.2$ and $M_0=0.6$, $M_y = 0$ at all times. The dynamics of the HMF model is thus similar to the one of a pendulum whose field intensity varies in time, given by the following Hamiltonian:
\begin{equation}
  \label{eq:hmf-pen-H}
  H = \sum_i \frac{p_i^2}{2} - M(t) \sum_i \cos\theta_i \quad .
\end{equation}

The phase space of a pendulum is divided by the separatrix in two qualitatively different parts~: particles are trapped if $|p| \lesssim 2 \sqrt{M}$, otherwise they travel periodically with a velocity of constant sign.

\begin{figure}[ht]
  \centering
  \includegraphics[width=0.8\linewidth]{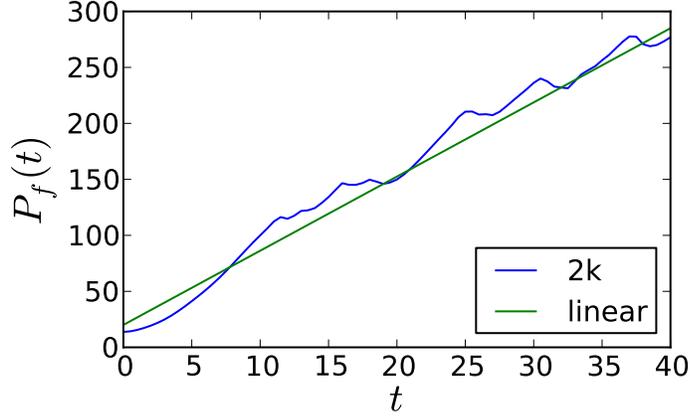}
  \caption{Evolution of $P_f(t)$ in the case of a pendulum with a constant field intensity. ``2k'' is a Vlasov simulation with parameters $N_\theta=N_p=2048$ and $\Delta t=0.1$, the initial condition is a waterbag with $\Dth\approx 2.6$ and $\Dp \approx 0.85$. $P_f(t)$ is approximated by a linear fit (``linear'' in the legend) around which there are small variations.}
  \label{fig:mix-pen-P}
\end{figure}

The case of a constant field intensity $M(t)=M(0)$ already induces a deformation of the initial waterbag. However, the separation between two nearby points of the contour remains bounded linearly. Figure~\ref{fig:mix-pen-P} displays the evolution of the perimeter in the case of the pendulum alongside with a linear fit.

\section{Structures and relaxation in phase space}
\label{sec:hmf-struct}

We now turn to the study of the dynamical evolution of the Hamiltonian Mean-Field (HMF) model.

\begin{figure}[ht]
  \centering
  \includegraphics[width=\linewidth]{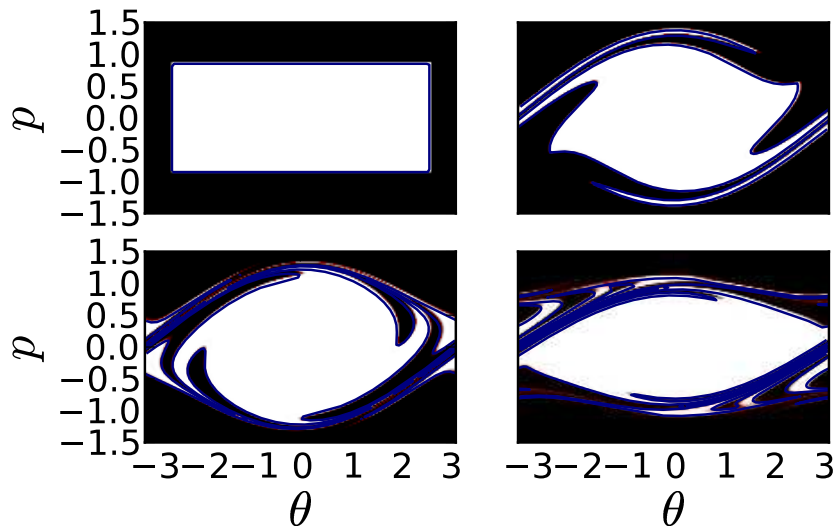}
  \caption{Evolution of the waterbag in phase space, in the case of the HMF model. Parameters of the simulation: $N_\theta=N_p=512$, $\Delta t = 0.1$, $U=0.2$ and $M_0=0.6$. The usual pseudo-color representation is completed by the contour of $f(\theta,p)$ for the value $f_0/2$. The central part of the waterbag remains constant but the contour of the waterbag displays filamentation.}
  \label{fig:hmf-panels-ps}
\end{figure}

The evolution of the initial waterbag is displayed in Fig.~\ref{fig:hmf-panels-ps}.
The central part of the fluid remains constant (the white region), experiencing a rotation in phase space. The deformation appears only on the contour, with the formation of filaments.
The length of the filaments increases in time while their thickness decreases.

In the region containing the filaments, a coarse-grained probability distribution function $\tilde f(\theta,p)$ can take any value between $0$ and $f_0$. The phase space is thus separated qualitatively in three regions:
\begin{enumerate}
\item The central region where $f(\theta,p)$ is constant at the value $f_0$.
\item An intermediate region where filaments are formed and where an effective mixing takes place.
\item The remainder of phase space, where $f(\theta,p)$ is equal to $0$.
\end{enumerate}

\begin{figure}[ht]
  \centering
  \includegraphics[width=.8\linewidth]{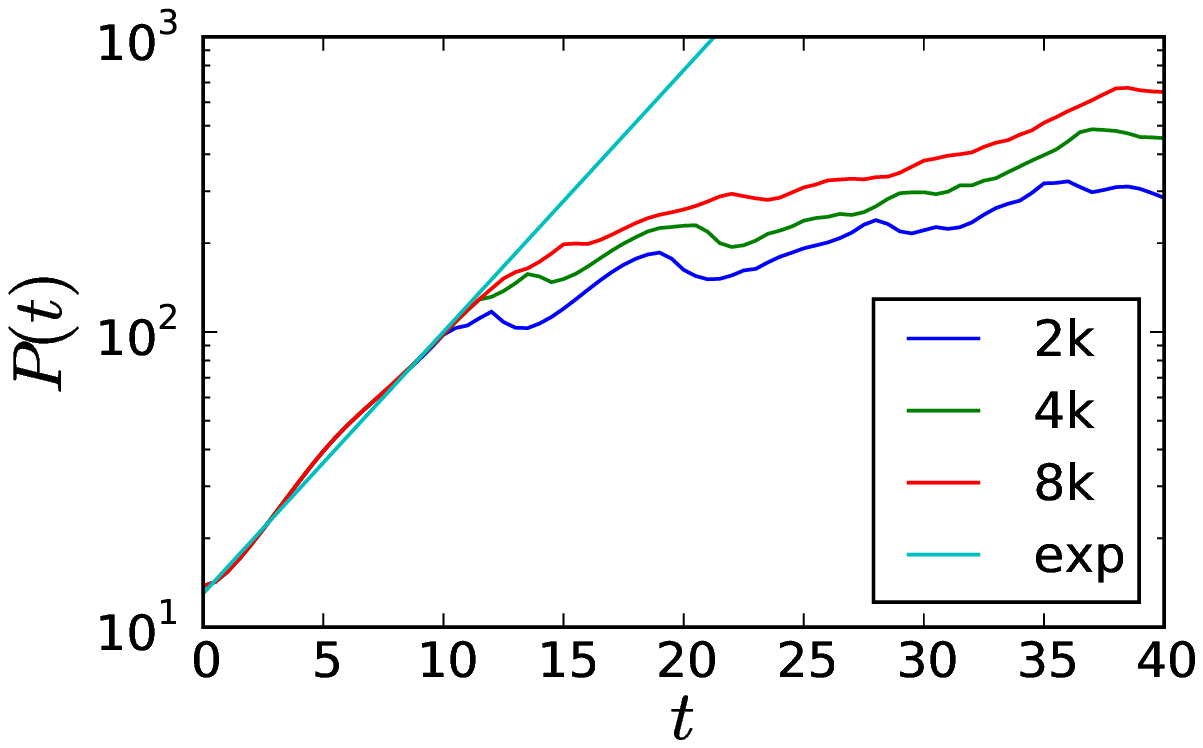}
  \caption{Evolution of $P_f(t)$ for the HMF model, same conditions as in Fig.~\ref{fig:hmf-panels-ps} but with different numerical accuracies. The number of grid points in runs ``2k'', ``4k'' and ``8k'' is $N_\theta=N_p=2048, 4096$ and $8192$ respectively. ``exp'' is an exponential fit of the region $0 \leq t \lesssim 12$}
  \label{fig:mix-f-P}
\end{figure}

The computation of $P_f(t)$ is performed in the course of the simulation. The result is displayed in Fig.~\ref{fig:mix-f-P} and exhibits an exponential behavior (the scale of the figure is logarithmic). A fit (``exp'' in the legend) to an exponential is superimposed on the figure. After a short time, the level of detail required to follow accurately the filamentary structure exceeds the accuracy of the numerical grid. The exponential behavior is then lost.

Successive simulations with higher accuracies have been run and are also displayed in Fig.~\ref{fig:mix-f-P}. The time during which $P_f(t)$ agrees with the exponential fit increases slightly with the grid accuracy. We expect successive increments of the number of grid points to lengthen that time of validity. This could be checked with either a parallelization of the simulation code or with the help of alternative algorithms.

\begin{figure}[ht]
  \centering
  \includegraphics[width=\linewidth]{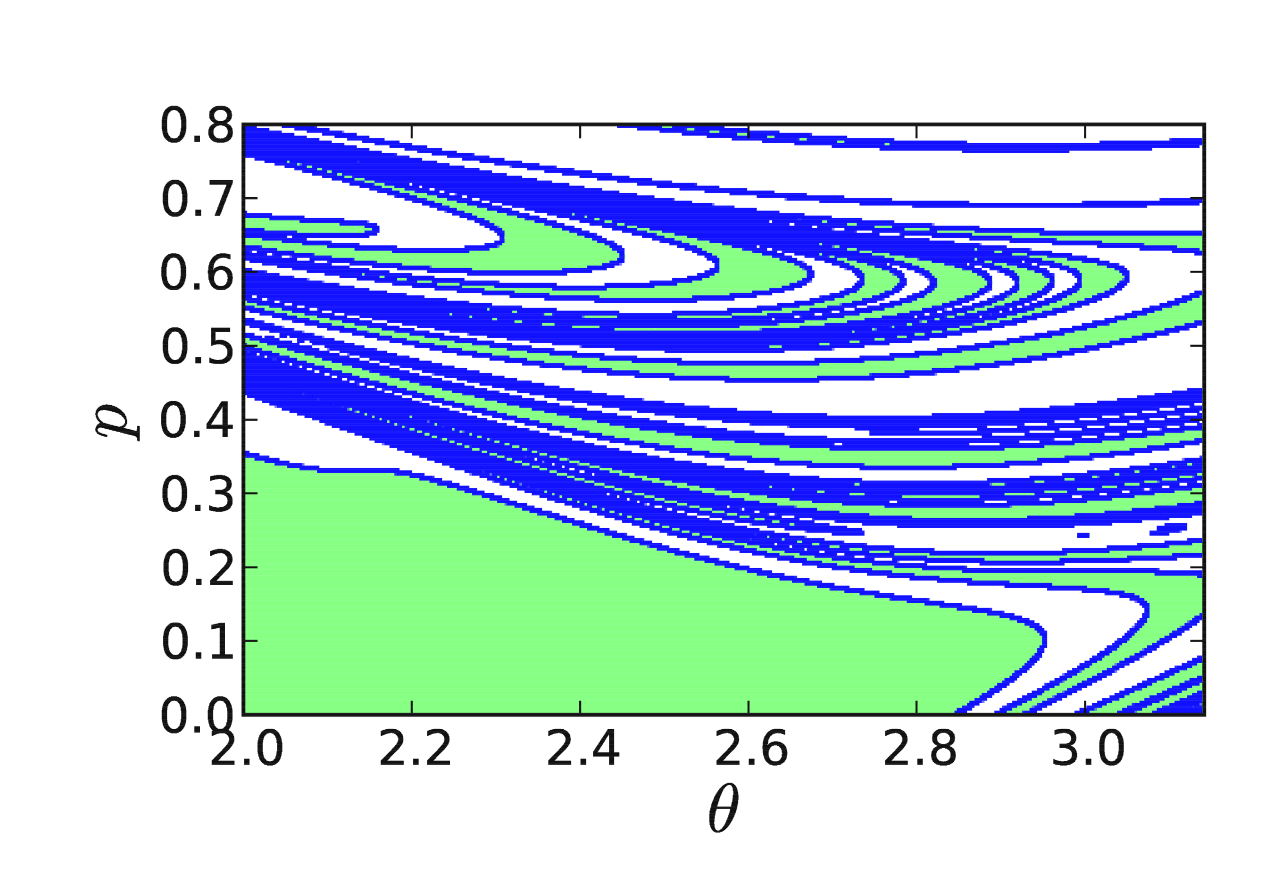}
  \caption{A zoom on a region of phase space for the HMF model. Parameters of the simulation: $N_\theta=N_p=8192$, $\Delta t = 0.1$, $U=0.2$ and $M_0=0.6$ (same conditions as in Fig.~\ref{fig:mix-f-P}). The value of $f(\theta,p)$ in the shaded region is $f_0$ and in the white region $0$. The blue lines indicate the contour computed by the CONREC subroutine. The upper part of the figure displays a series of intertwined contour lines resulting of the stretching and folding mechanism.}
  \label{fig:mix8_zoom}
\end{figure}

We display for the highest number of grid points a zoom on the contour lines in Fig.~\ref{fig:mix8_zoom}. A folding structure is clearly apparent in the upper region of the figure. This is a feature that is also found in fluid dynamics \cite{ottino_book_1989}.

A study of the $N$-body dynamics of the HMF model \cite{sguanci_et_al_apparent_fractal_2005} reveals that the dynamical evolution for high value of $N$ ($10^6$) does not lead to fractal structures. The reference~\cite{sguanci_et_al_apparent_fractal_2005} uses a singular initial condition that is not tractable by numerical simulation for the Vlasov equation, preventing a direct comparison of numerical results.

\section{Summary}
\label{sec:hmf-summary}

We provided in this chapter original contributions to the understanding of the approach to equilibrium in collisionless systems. In the Vlasov equation for the Hamiltonian Mean-Field model, we considered an analogy with fluid dynamics. The analysis of phase space shows a separation in different regions, one of which displaying mixing.

The presence of stretching and folding structures in phase space has been observed in numerical simulations. The deformation of the contour of an initial waterbag has been quantitatively monitored in the simulations and supports the hypothesis of an exponential elongation linked to the stretching and folding phenomenon.

A pursuit of this work is possible through the use of more advanced computational techniques. In addition, a systematic study of the dynamics is necessary in order to perform useful comparisons to the existing body of literature on the relaxation of collisionless systems.

% Local Variables:
% TeX-master: "main"
% End:

\cleardoublepage
\chapter{Dynamics of the Hamiltonian Mean-Field model on the basis of the equivalent non-interacting system}
\label{chap:pendulum}

\myabstract{With the aim of understanding the mechanisms of violent relaxation and of out-of-equilibrium phase transitions, we introduce a Vlasov equation for a set of uncoupled pendula and the corresponding Lynden-Bell theory. The asymptotic evolution of the system is found to match the prediction of the theory for a range of parameters. We go on by computing under an ergodic-like hypothesis the exact solution for the asymptotic evolution of the system. The stationary regime of this system is identical to the one of the Hamiltonian Mean-Field model, allowing us to improve our understanding of this latter model. We find an out-of-equilibrium phase transition under a self-consistent constraint.}{ uncoupled pendula, ergodic-like solution, stirring.}

The evolution of the Vlasov equation displays complex nonlinear dynamics in the presence of interactions. We propose to study a Vlasov equation that does not include interactions, providing an integrable example of Vlasov dynamics.

The steady state of the Hamiltonian Mean-Field (HMF) model corresponds to a constant value of the magnetization vector. The equations of motion are then equal to the ones of the set of pendula. This similarity motivates the investigation of the Vlasov dynamics for a set of uncoupled pendula. Lynden-Bell's theory aims at predicting the outcome of collisionless evolution and supposes the existence of a mixing mechanism in the microscopic dynamics. We develop it for our integrable system to test that particular hypothesis.

The investigation is pushed a step further by the computation of an exact solution to the asymptotic evolution of the set of pendula. Imposing a self-consistency condition explicitly specializes our study to situations comparable to the HMF model.
The idea to study the dynamics of a set of uncoupled pendula has been put forward recently in the literature \cite{leoncini_et_al_epl_2009,firpo_epl_2009} to construct a class of stationary solutions for the HMF model. Our approach differs in the sense that we detail the computation explicitly with the aim to understand the occurrence of a phase transition.

This chapter is structured as follows: the system of pendula and its associated Vlasov equation are introduced in section~\ref{sec:pendulum-vlasov}. 
Lynden-Bell's theory is developed in section~\ref{sec:pendulum-LB} and the exact solution to the asymptotic dynamics is given in section~\ref{sec:pendulum-exact}.

\section{Vlasov equation for the pendula}
\label{sec:pendulum-vlasov}

The system we consider consists of uncoupled pendula subjected to a fixed external field $\H$~:
\begin{equation}
  \label{eq:Nbody_pen}
  H = \sum_{j=1}^N \frac{p_j^2}{2} - \H \sum_{j=1}^N \cos \theta_j
\end{equation}
where the $p_j$ are the momenta, the $\theta_j$ the positions and $\H$ is the external field applied to the system. We introduce the bunching parameter
\begin{equation}
  \label{eq:pen_b}
  {\bf b} = (b_x,b_y) = \frac{1}{N} \left( \sum_j \cos\theta_j, \sum_j \sin\theta_j \right)
\end{equation}
and its modulus $b = \sqrt{b_x^2+b_y^2}$. In this system, a symmetric initial condition with $b_y=0$ leads to $b_y=0$ at all times. We consider only such situations in this chapter. The interaction part of the Hamiltonian, rescaled by $N$, can be rewritten as $ \frac{H_V}{N} = -\H b$. We also introduce the one-particle Hamiltonian $H_1(\theta,p) = \frac{p^2}{2} - \H \cos\theta$.

The bunching parameter $b$ is computed with the same expression as the magnetization in the HMF model but does not enter in the force acting on each particle. This fundamental difference implies that the dynamics of system~(\ref{eq:Nbody_pen}) is integrable.
The value of $b$ indicates the homogeneity of the system: $b=0$ implies that the system is homogeneous and $b\neq 0$ indicates an inhomogeneous system. $b=1$ means that all particles are located at $\theta=0$. We make use of $b$ to quantify the state of the system or track its evolution in time.

Stationary regimes of the HMF model (constant value of the magnetization $\bf M$) possess the same equations of motion as the ones of system~(\ref{eq:Nbody_pen})~:
\begin{equation}
  \label{eq:eom_stationary}
  \left\{
    \begin{array}{l l}
      \dot\theta_j &= p_j \cr
      \dot p_j     &=  - \H \sin\theta_j
    \end{array}
\right.
\end{equation}
allowing a comparison of QSSs of the HMF model with the ones of the uncoupled system.

In the limit $N\to\infty$, we describe the evolution of the system from the kinetic point of view, using the one-particle probability distribution function (PDF) in phase space. The PDF, $f(\theta,p;t)$, gives the probability $f(\theta,p;t) d\theta\ dp$ to find at time $t$ a particle in the phase space region centered on $(\theta,p)$ and of measure $d\theta\times dp$. We write the following Vlasov equation
\begin{equation}
  \label{eq:vlasov_pendula}
  \frac{\partial f}{\partial t} + p \frac{\partial f}{\partial \theta} - \frac{dV(\theta)}{d\theta} \frac{\partial f}{\partial p} = 0
\end{equation}
where $V(\theta) = - \H \cos\theta$ is the potential. This formulation, although similar to other Vlasov equations, is different in the sense that it does not include interactions, but only an external field.

A consequence of the fact that the potential does not depend of $f$ is the time independence of the energy level distribution function defined as:
\begin{equation}
  \label{eq:edp}
  p(e) = \int d\theta\ dp\ f(\theta,p) \delta(e - H_1(\theta,p))
\end{equation}

\begin{figure}[ht]
  \centering
  \includegraphics[width=0.7\linewidth]{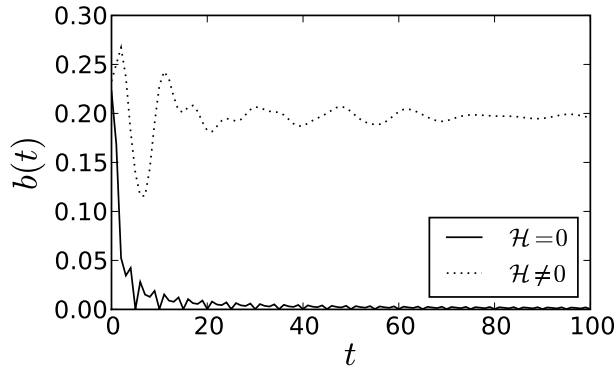}
  \caption{Evolution of the bunching parameter $b$ in function of time for the two situations $\H = 0$ and $\H \neq 0$.}
  \label{fig:b_of_t}
\end{figure}

\begin{figure}[ht]
  \centering
  \includegraphics[width=\linewidth]{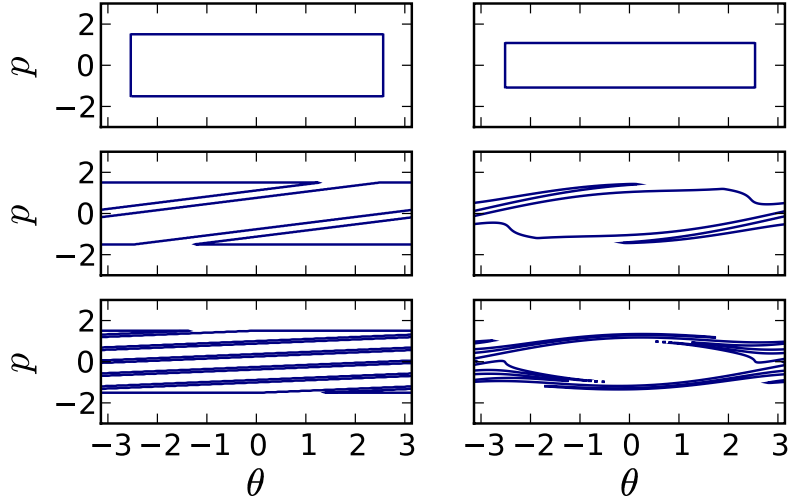}
  \caption{Evolution of a waterbag initial condition in phase space. The left column represents the $\H=0$ case, from top to bottom chronologically. The right column represents the $\H\neq 0$ case.}
  \label{fig:phase_space_pendulum}
\end{figure}

Two qualitatively separated regimes take place whether $\H$ is equal to $0$ or not. The evolution of $b(t)$ for these two situations is given in Fig.~\ref{fig:b_of_t} and we depict the evolution of the waterbag initial condition by the Vlasov dynamics in Fig.~\ref{fig:phase_space_pendulum}.

We notice from Figs.~\ref{fig:b_of_t} and \ref{fig:phase_space_pendulum} that in the free-streaming case ($\H=0$), the bunching parameter $b$ goes to zero, corresponding to a spread of the initial waterbag to the whole $[-\pi;\pi]$ interval, whereas in the $\H\neq 0$ case a finite value of $b$ is attained.
The spreading of the initial waterbag still occurs but follows equal energy lines in the phase space of the pendulum instead of straight lines.

The complete range of waterbag initial conditions are given equivalently by the parameters $\Dth$ and $\Dp$ or $U=\frac{\Delta p^2}{6} - \H b_0 $ and $b_0 = \frac{\sin\Delta\theta}{\Delta\theta}$. The waterbag is defined as
\begin{equation}
  \label{eq:pen_wb}
  f(\theta,p) = \left\{\begin{array}{l l}
      f_0 & \mbox{if } |p|\leq \Delta p,\cr
          & \mbox{\ \ \ } |\theta|\leq \Delta\theta, \cr
      0 & \mbox{otherwise.}
  \end{array}\right.
\end{equation}
with $f_0=\left(4 \Delta\theta \Delta p\right)^{-1}$.

The $\H\neq 0$ regime allows one to rescale the system in order to explore the possible regimes more easily.
We perform the following changes:
\begin{eqnarray}
  H\to& H'=H/\H , \label{eq:rescale_H}\\
  p_i\to& p_i' = p_i/\sqrt{\H} , \label{eq:rescale_p}\\
  \theta_i\to&\theta_i'=\theta_i \label{eq:rescale_th}
\end{eqnarray}
and obtain a system in which $\H$ is effectively set to $1$.

The rescaled system will be used when appropriate with explicit mention. For the comparison with HMF stationary states, we however must find a solution where $b=\H$, which is trivial under the inverted rescaling. As $\theta$ is not affected by the rescaling, the bunching is the same in both systems and to find $\H=b$, we only have to find a set of $\Delta\theta$ and $\Delta p$ for which the final $b$ is the desired one and rescale the system to adjust $\H$.
We remark however that homogeneous situations with $\H=0$ are not amenable to the rescaling.

\section{Lynden-Bell's theory for the pendula}
\label{sec:pendulum-LB}

Lynden-Bell devised a theory whose aim is the prediction of the outcome of the dynamics of the Vlasov equation \cite{lynden-bell_1967}. Its original purpose was to explain observations made on the light distribution from elliptical galaxies.

The theory is based on the maximization of an entropic functional under appropriate macroscopic constraints (see section~\ref{sec:vlasov-LB}). The underlying hypothesis is that the dynamical evolution of a collisionless system possesses a mechanism to explore the ensemble of states compatible with the constraints.

The set of uncoupled pendula allows, in principle, a formulation of the theory.
Its formulation is indeed very similar to the one for the Hamiltonian Mean-Field (HMF) model as it describes only stationary regimes.
The conservation of the energy distribution function implies that relaxation is not likely to occur in the system of pendula. We propose to check the possible application of the theory to this system in order to deepen our understanding of collisionless dynamics.

\subsection{Formulation of the theory}
\label{sec:pendulum-LB-eq}

We write here the expression of the entropy as applicable to a two-step initial one-particle probability distribution function (1-PDF) $f(\theta,p)$ of heights $0$ and $f_0$:
\begin{equation}
  \label{eq:LBentropy}
  s(\bar f) = - \int dp\ d\theta\ \left[ \frac{\bar f}{f_0}\ln\frac{\bar f}{f_0} + \left(1-\frac{\bar f}{f_0}\right)\ln\left(1-\frac{\bar f}{f_0}\right)  \right] \textrm{ . }
\end{equation}
where $\bar f(\theta,p)$ is the coarse grained 1-PDF. This coarse grained distribution can take values in the whole interval $[\ 0\ ; f_0]$. 
The solution $\bar f$ of the maximization problem for the entropy (\ref{eq:LBentropy}) is a prediction for the asymptotic outcome of the dynamics.
It is a prediction of the 1-PDF if one neglects the microscopic filamentary structures that develop in phase space.

Under the constraints of mass, momentum and energy conservations, the optimization of the entropy (\ref{eq:LBentropy}) for an initial waterbag, 
\begin{equation}
  \label{eq:wb}
  \left\{\begin{array}{l l l}
      f(\theta,p) &=& f_0 \textrm{ if } |\theta|\leq\Delta\theta \textrm{ and } |p|\leq\Delta{}p\cr
      &=&  0 \textrm{ , else}
    \end{array}\right. ,
\end{equation}
\begin{equation}
  f_0 = \frac{1}{4\ \Delta\theta\ \Delta{}p }
\end{equation}
leads to the following $\bar f(\theta,p)$:
\begin{equation}
  \label{eq:LBf}
  \bar f(\theta, p) = f_0 \frac{e^{-\beta (p^2/2-\H\cos\theta) -\mu }}{1+e^{-\beta (p^2/2-\H\cos\theta) -\mu }}
\end{equation}
where $\mu$ and $\beta$ are Lagrange multipliers associated respectively to mass and energy conservations. The resulting set of equations to be solved is:
\renewcommand{\arraystretch}{2}
\begin{equation}
  \label{eq:pendulum_LB}
  \begin{array}{r l l}
      \frac{f_0 x}{\sqrt{\beta}} \int d\theta\ e^{\beta \H\cos\theta} F_0(x e^{\beta \H\cos\theta}) &=& 1\cr
      \frac{f_0 x}{2 \beta^{2/3}} \int d\theta\ e^{\beta \H\cos\theta} F_2(x e^{\beta \H\cos\theta}) + \H(1-b) &=& U\cr
      \frac{f_0 x}{\sqrt{\beta}} \int d\theta\ \cos\theta\ e^{\beta \H\cos\theta} F_0(x e^{\beta \H\cos\theta}) &=& b
  \end{array}
\end{equation}
where $x=e^{-\mu}$, $b$ is the previously defined bunching parameter and the $F_n$'s are defined as:
\begin{equation}
  F_n(x) = \int dv\ v^n \frac{e^{-v^2/2}}{1+x\ e^{-v^2/2}} \textrm{ .}
\end{equation}
We solve the system of equations~(\ref{eq:pendulum_LB}) with the Newton-Raphson method \cite{NR_in_f90}.

The values of $b$ and $\H$ are not determined a priori. We solve the system of equations under the additional constraint that
\begin{equation}
  \label{eq:pen-LBselfc}
  b = \H
\end{equation}
that we call the self-consistency relation.
Taking into account Eq.~(\ref{eq:pen-LBselfc}), we set $\Delta\theta$, $\Delta p$ and the numerical procedures gives $b=\H$ , $\beta$ and $x$.

We remark that the system of equations (\ref{eq:pendulum_LB}) is similar to the equivalent problem for the HMF \cite{antoniazzi_et_al_pre_2007}. The value of $U$ needs to be adapted to accommodate the definitions of both systems.

%The possibility to rescale Hamiltonian Hamiltonian (\ref{eq:Nbody_pen}), setting $\H$ to $1$, has not been used because it prevents the solution $\H=b=0$.

\subsection{$\Dth = 1.66$}
\label{sec:pen_M0_6}

A first comparison is drawn for initially bunched systems. $\Dth$ is set to $1.66$ and $\Dp$ is varied. Eqs.~(\ref{eq:pendulum_LB}) are solved with the self-consistency relation $b=\H$.
A simulation of the Vlasov equation for the pendula with the given $\Dth$, $\Dp$ and the computed $\H$ provides a test of the validity of the theory.

\begin{figure}[ht]
  \centering
  \includegraphics[width=0.75\linewidth]{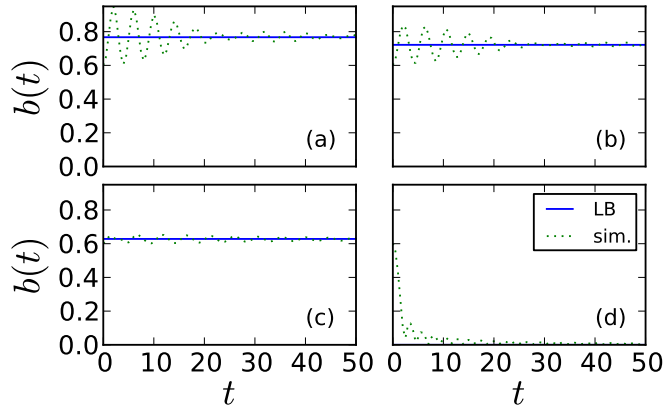}
  \caption{For $\Dth=1.66$, evolution of the bunching parameter in Vlasov simulations in the course of time for $\Dth=1.66$ and $\Dp=0.41,0.72,1.03$ and $1.34$ for panels (a),(b),(c) and (d) respectively. The corresponding value predicted by Lynden-Bell's theory is plotted for reference.}
  \label{fig:pen-b_of_t-a}
\end{figure}

The stirring causes variations in $b$ that diminish quickly as the initial waterbag evolves in phase space. This behaviour is displayed in Fig.~\ref{fig:pen-b_of_t-a}. The value attained by $b$ is in all situations in perfect agreement with the predicted one. This first result on the simulation of the pendulum bears an important signification with respect to Vlasov dynamics: the stirring mechanism by itself plays an important role in the evolution of the initial condition towards a ``most probable state'' as expected by Lynden-Bell's theory for a system without collisions.

\begin{figure}[ht]
  \centering
  \includegraphics[width=0.75\linewidth]{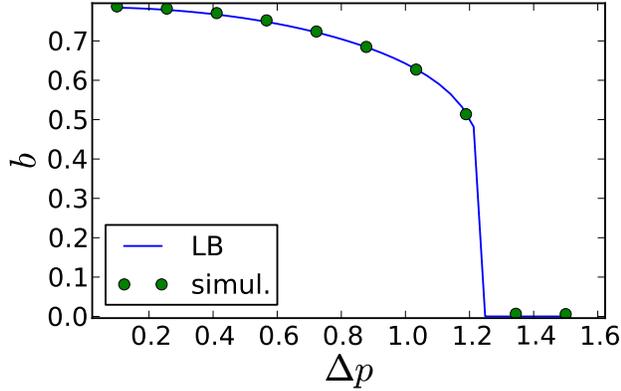}
  \caption{For $\Dth=1.66$, evolution of the bunching parameter predicted by Lynden-Bell's theory when $b=\H$ in function of $\Dp$. Data points correspond to Vlasov simulations with identical $\Dth$, $\Dp$ and $\H$.}
  \label{fig:pen-b_of_Dp_a}
\end{figure}

Fig.~\ref{fig:pen-b_of_Dp_a} summarizes this result for a wider range of $\Dp$'s. The points indicating the result of Vlasov simulations match everywhere the theory.
This is trivial when $\H$ is equal to $0$: the system evolves under free streaming only and goes to a homogeneous state. It is more unexpected when $\H\neq 0$ because Lynden-Bell's theory, as statistical mechanics, is based on the assumption that the complicated dynamical evolution of the system allows an effective exploration of the states allowed in its state space. This is obviously not the case when the dynamics is integrable.

An additional result visible in Fig.~\ref{fig:pen-b_of_Dp_a} is that for growing $\Dp$ a decrease of the resulting $b$ is observed. It is followed by an abrupt transition when the $b\neq 0$ solution ceases to exist.
Below the transition, both solutions to Lynden-Bell's theory exist, and the choice between any of the two can be made.

\begin{figure}[p]
  \centering
  \includegraphics[width=\linewidth]{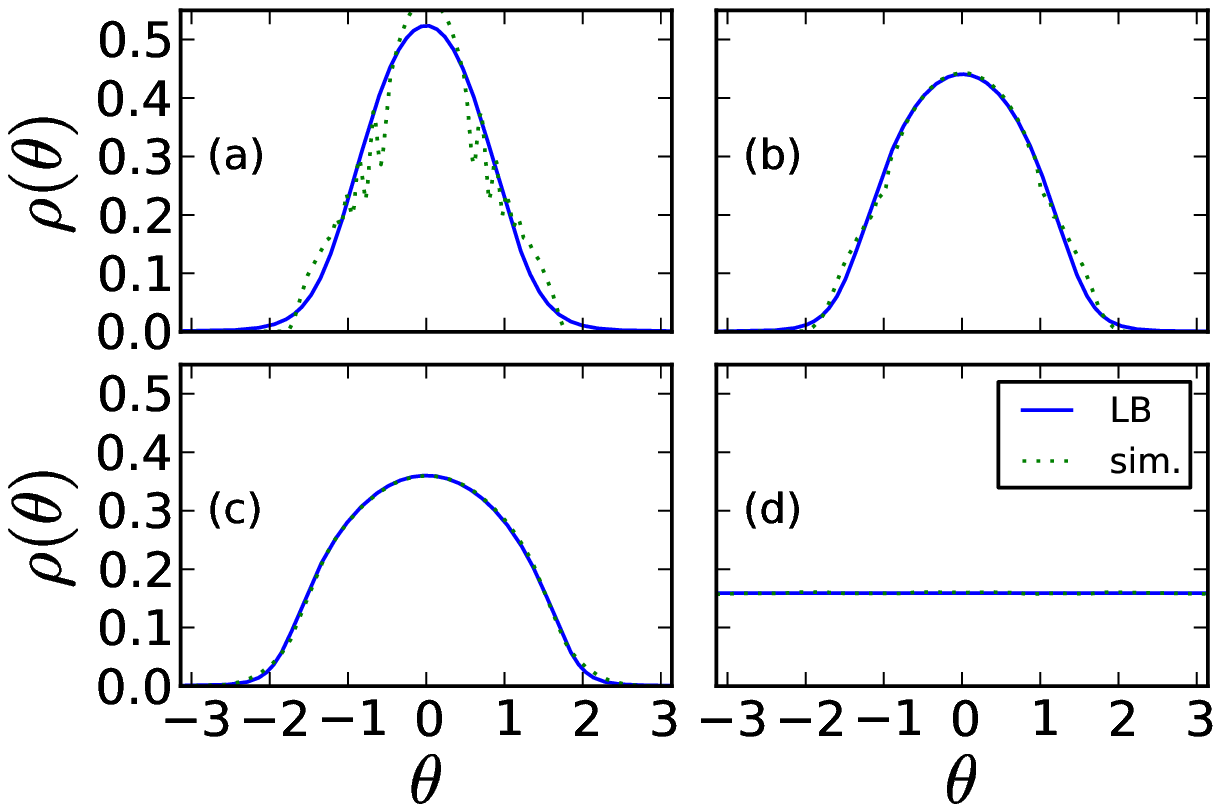}
  \caption{The $\theta$ marginal, $\rho$, for $\Dth=1.66$ and $\Dp=0.41,0.72,1.03$ and $1.34$ for panels a,b,c and d respectively. Apart from the remaining filamentary structure, the agreement is very good.}
  \label{fig:pen-rho-a}
  \includegraphics[width=\linewidth]{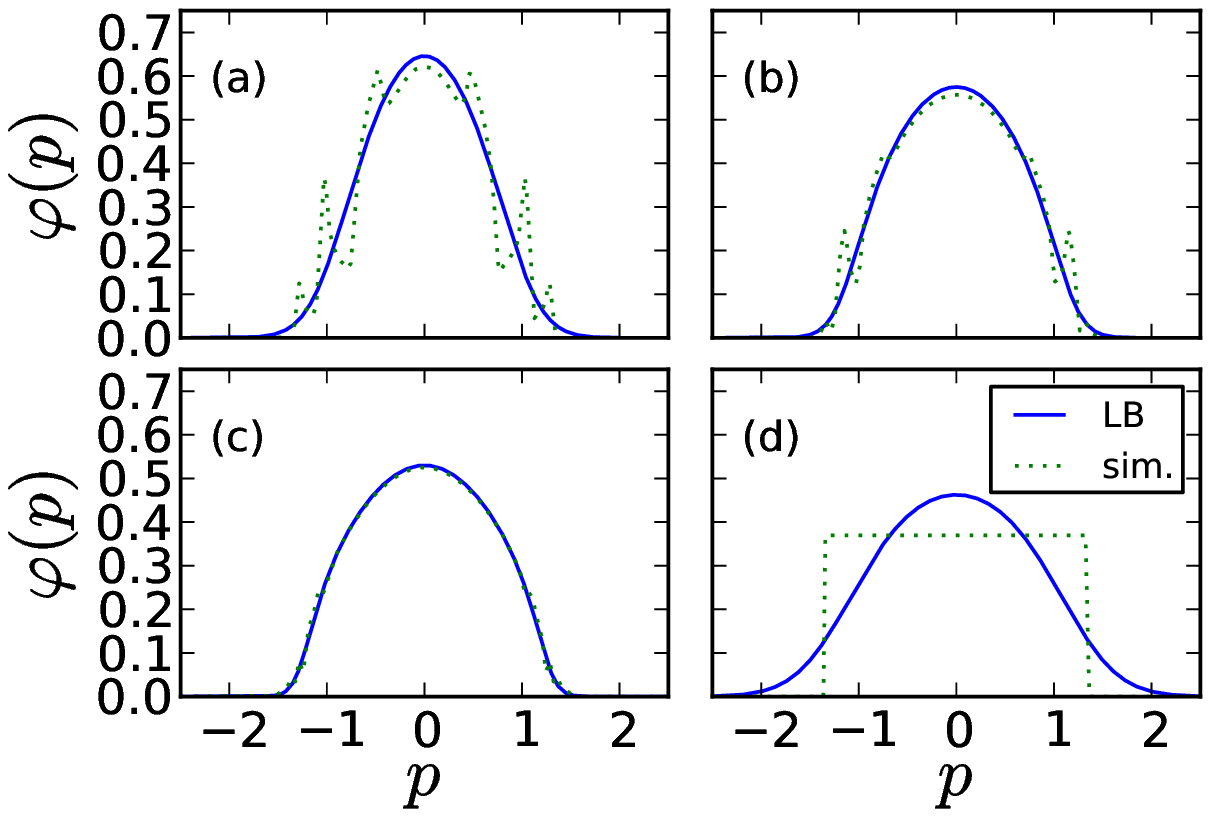}
  \caption{The $p$ marginal, $\varphi$, for $\Dth=1.66$ and $\Dp=0.41,0.72,1.03$ and $1.34$ for panels a,b,c and d respectively. The filamentary character is strong in panel a. Panel d displays a significant deviation from the theory for $\varphi$ whereas it was exact for $\rho$ (see Fig.~\ref{fig:pen-rho-a}).}
  \label{fig:pen-phi-a}
\end{figure}

To understand the effectiveness of our prediction, we detail the results of Fig.~\ref{fig:pen-b_of_Dp_a} with a comparison of the marginal distributions. Fig.~\ref{fig:pen-rho-a} displays $\rho(\theta)$ for some of the simulations. Lynden-Bell's theory clearly does not take into account the filaments, which leads to a disagreement strongly visible only in panel (a). Else, the predicted $\rho$ matches very closely the result of the simulations.

A complementary approach is given by the velocity marginal $\varphi(p)$ given in Fig.~\ref{fig:pen-phi-a}. There is still a marked effect of the filamentation in panel (a) ($\Dp=0.41$) but the most striking difference is in panel (d) ($\Dp=1.34$). The predicted field is $0$ and the velocity distribution is left unchanged from the original waterbag.
The discontinuous character of the waterbag is completely ignored by Lynden-Bell's theory. In the system of pendula, $p(e)$ does not change, which translates for $\H=0$ in the invariance of $\varphi(p)$ (because $e=\frac{p^2}{2}$).

The observed agreement is best in panel (c) ($\Dp=0.72$). In that situation, we note that the final value of $b$ is very close to the initial one.
While we find that the theory predicts with a good accuracy the outcome of the simulations, it makes sense that it works optimally when no strong changes are required between the start of the simulation and its end.
More detail on this particular aspect will be given in the next section, where a stronger disagreement between the theory and the simulation motivates an deeper analysis.

\subsection{$\Dth=\pi$}
\label{sec:pen_M0_0.6}

We now turn to homogeneous initial conditions. They differ from bunched ones in the sense that, in order to attain a finite bunching value, the system goes from a homogeneous state towards a qualitatively different inhomogeneous state.

\begin{figure}[ht]
  \centering
  \includegraphics[width=0.75\linewidth]{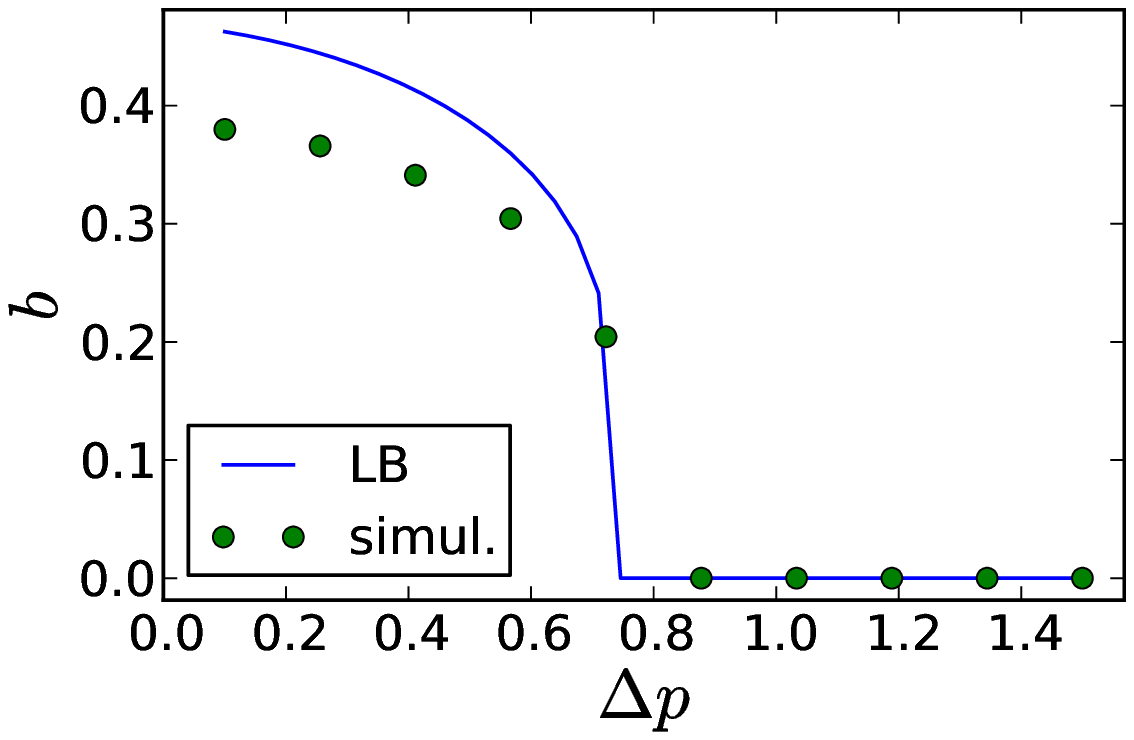}
  \caption{For $\Dth=\pi$, evolution of the bunching predicted by Lynden-Bell's theory when $b=\H$ in function of $\Dp$. Data points correspond to Vlasov simulations run with identical $\Dth$, $\Dp$ and $\H$.}
  \label{fig:pen-b_of_Dp_b}
\end{figure}

Figure~\ref{fig:pen-b_of_Dp_b} shows the dependence on $\Dp$ of the self-consistent bunching for $\Dth=\pi$, for Lynden-Bell's theory and Vlasov simulations. For sufficiently high $\Dp$ the bunching is zero and the system exactly evolves under free streaming. For lower values of $\Dp$, the bunching is finite but the two approaches do show a quantitative agreement.

\begin{figure}[p]
  \centering
  \includegraphics[width=\linewidth]{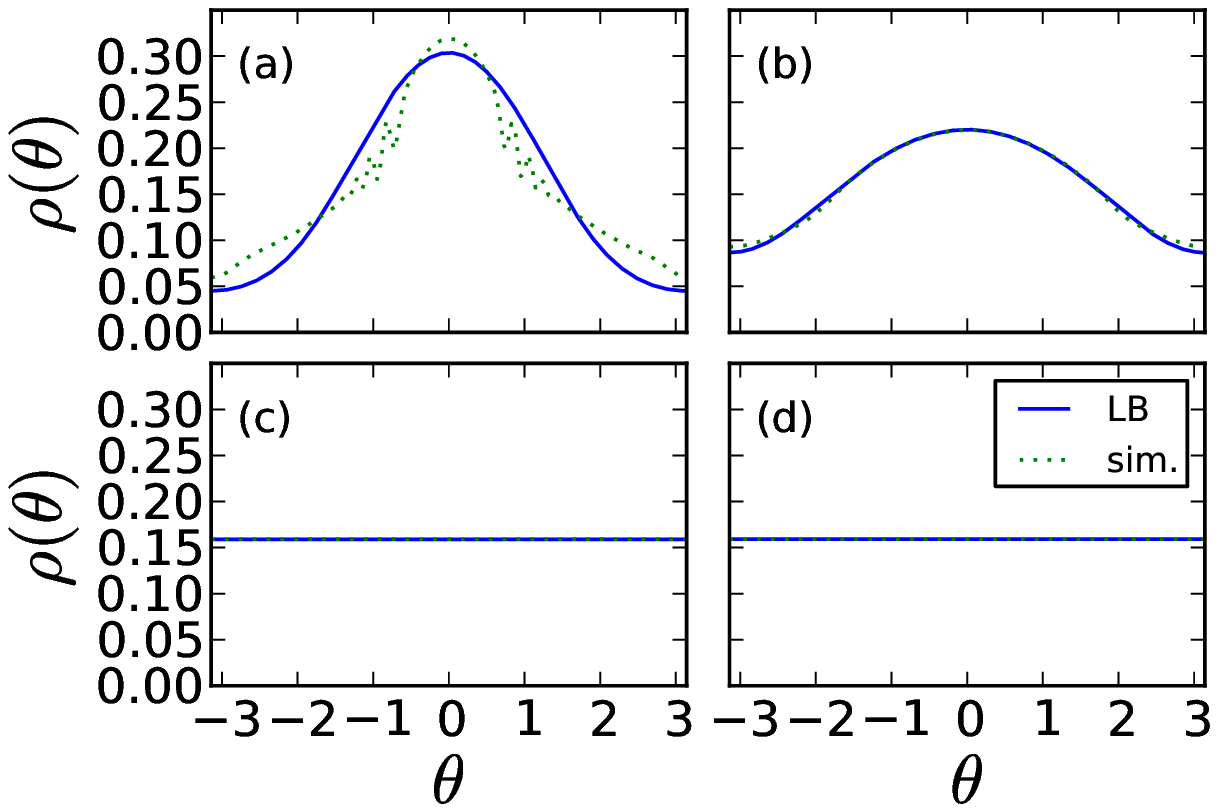}
  \caption{The $\theta$ marginal, $\rho$, for $\Dth=\pi$ and $\Dp=0.41,0.72,1.03$ and $1.34$ for panels (a),(b),(c) and (d) respectively. Apart from the remaining in panel (a), the agreement is very good.}
  \label{fig:pen-rho-b}
  \includegraphics[width=\linewidth]{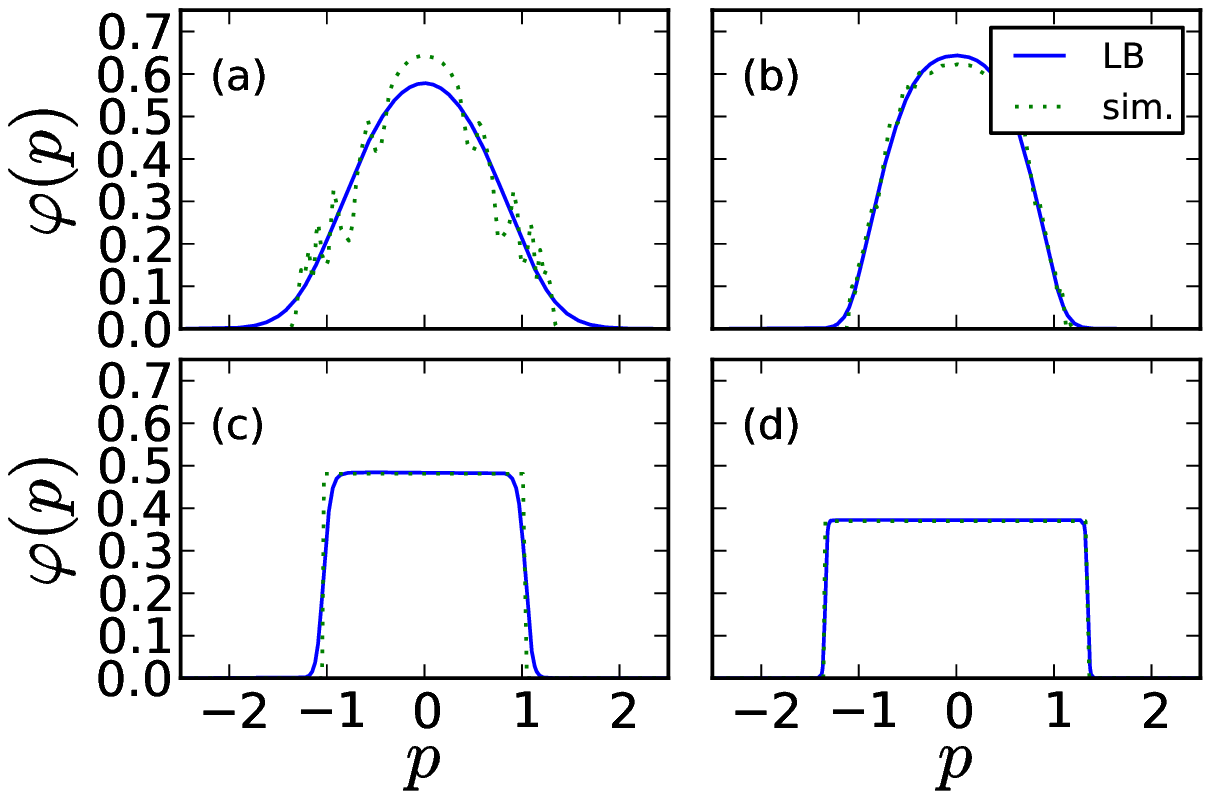}
  \caption{The $p$ marginal, $\varphi$, for $\Dth=\pi$ and $\Dp=0.41,0.72,1.03$ and $1.34$ for panels (a),(b),(c) and (d) respectively. The filamentary character is strong in panel (a). Panel (d) displays a small significant deviation from the theory for $\varphi$ whereas it was exact for $\rho$ (see Fig.~\ref{fig:pen-rho-b}).}
  \label{fig:pen-phi-b}
\end{figure}

At variance with what happens for $\Dth=1.66$, there is no satisfactory agreement anymore between the results of Lynden-Bell's theory and the simulation. 
This appears in the panel (a) of Figs.~\ref{fig:pen-rho-b} and \ref{fig:pen-phi-b}. $\rho(\theta)$ and $\varphi(p)$ show a disagreement between the theory and the simulation, corresponding to the left region in Fig.~\ref{fig:pen-b_of_Dp_b} (low values of $\Dp$).
A slight disagreement is also observed in panel (a) of Fig.~\ref{fig:pen-rho-a}, but the resulting value of $b$ is similar to the one predicted by the theory in that situation.

A quantity that we have not explored yet is the energy probability distribution function (energy PDF) $p(e)$. For the system of uncoupled pendula, $p(e)$ does not vary in time, which should prevent relaxation from happening.

\begin{figure}[h!]
  \centering
  \includegraphics[width=.8\linewidth]{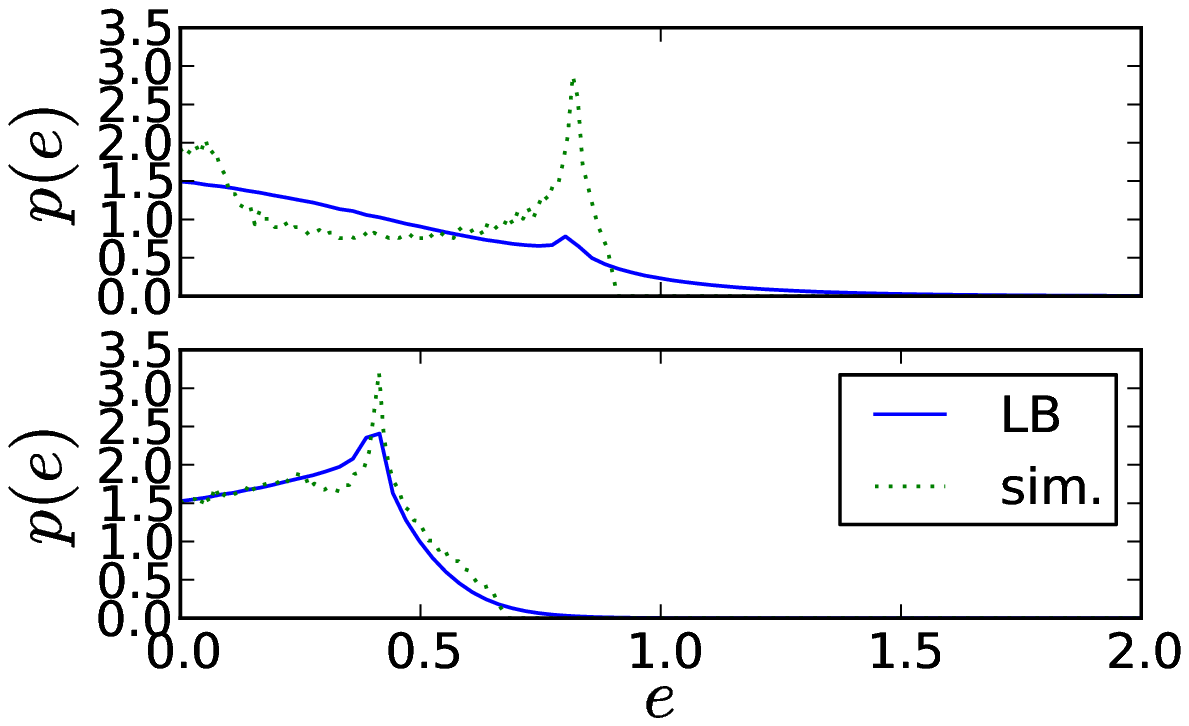}
  \caption{The energy probability distribution function for $\Dth=\pi$ (top) and $\Dp=0.41$ (bottom) and $0.72$ (corresponding to panels (a) and (b) of Fig.~\ref{fig:pen-rho-a} respectively), computed by Lynden-Bell's theory and directly on the simulation grid. Both results are in qualitative agreement for $\Dp=0.72$ (lower panel) but differ strongly for $\Dp=0.41$ (upper panel).}
  \label{fig:pen-LB-p_of_e-b}
\end{figure}

We display in Fig.~\ref{fig:pen-LB-p_of_e-b} the distribution $p(e)$~: directly computed on the simulation grid and compared with the distribution predicted by Lynden-Bell's theory.

The observation of $p(e)$ sheds light on the effectiveness of Lynden-Bell's theory as discussed in section~\ref{sec:pen_M0_0.6} and to a smaller extent in this section (panel (b) of Figs.~\ref{fig:pen-rho-b} and \ref{fig:pen-phi-b}). The theory and the simulation agree when $p(e)$ is close enough to the energy PDF predicted by Lynden-Bell's theory. Then, stirring in phase space induces a spread of the initial waterbag along equal-energy lines in phase space, giving oscillations of $b(t)$ followed by a relaxation to the value predicted by the theory (see Fig.~\ref{fig:pen-b_of_t-a}).

As the energy PDF is constant, we conclude that Lynden-Bell's theory is effective in the case of the pendula only when the initial condition is close enough to the energy distribution of the theory. When that condition is met, no relaxation is needed in order to attain the predicted bunching but only a repartition of the initial waterbag on equal energy lines in phase space.

\subsection{The role of the conserved quantities}

The constraints on the macroscopic quantities play an important role in Lynden-Bell's theory.
The system of uncoupled pendula offers the perspective to modify the theory~: the energy distribution function (DF) is conserved. The consequence is that any moment of the energy DF, $E^i[f]$, is expected to be a constant.
\begin{equation}
  \label{eq:pen-E^i}
  E^i[f] = \int d\theta\ dp\ \left(H_1(\theta,p)\right)^i f(\theta,p)
\end{equation}
Equation~(\ref{eq:pen-E^i}) provides the theory with an arbitrary number of conserved quantities that can be taken into account.
This task is not possible in the case of the HMF model, where there is no {\it a priori} conserved quantity that remains to be included.
The model that is introduced in this chapter thus presents an interest for the understanding of the role of the conserved quantities in Lynden-Bell's theory.

\section{Exact solution}
\label{sec:pendulum-exact}

On the basis of the uncoupled character of the system described by Eq.~(\ref{eq:vlasov_pendula}), we propose an exact solution to the asymptotic dynamics. This solution is based on an ergodic-like hypothesis that is justified by phase space stirring instead of chaotic dynamics.

The procedure allows to compute the bunching resulting from an arbitrary initial waterbag. A self-consistent requirement allows to discuss the similarity with stationary solutions of the Hamiltonian Mean-Field model.

We explain the principle of the procedure in section~\ref{sec:pen_ergodic} and its explicit computations in section~\ref{sec:pen_explicit}.

\subsection{The ergodic-like hypothesis}
\label{sec:pen_ergodic}

While ergodicity is usually considered to hold when a system experiences chaotic or mixing dynamics, we use this hypothesis in the following sense: every particle or fluid element experiences a motion that is constrained on its initial energy manifold. The non-isochronic character of the pendulum gives different periods to two nearby orbits, leading to stirring of the phase space. Especially, the system does not come back close to its initial condition.

The energy distribution $p(e)$ is constant and we make the hypothesis that the asymptotic state of the dynamics $f(\theta,p)$ is defined only on the basis of $p(e)$.

This is justified in our model because of phase space ``stirring'', that is, the evolution of the initial waterbag with no interaction leads the evolution towards a steady state. It is true on a coarse-grained level, similarly to Lynden-Bell's theory. Everywhere in phase space, a cell of volume $d\theta \times dp$ will contains after a time long enough, a filamentary structure. The value of $f$ is still two-step valued (at the microscopic level) because the original waterbag is two-step valued. However, the volume is occupied by filaments and the average value of $f$ in the cell will be proportional to $p(e)$.

\subsection{Explicit computation of the solution}
\label{sec:pen_explicit}

For the waterbag given by Eq.~(\ref{eq:pen_wb}), the energy distribution is given by
\begin{equation}
P_{\epsilon}(\epsilon,p_0)=\frac{1}{4\Dth\Dp}\int d\theta dp
~\delta(\frac{1}{2}p^2-\epsilon-\H\cos\theta) ~.
\end{equation}
Carrying out the integral over $p$, one gets
\begin{equation}
P_{\epsilon}(\epsilon,p_0)=\frac{2}{4\Dth\Dp} \int d \theta
~\frac{1}{\sqrt{2(\epsilon+H\cos\theta)}}.
\end{equation}

\begin{figure}[ht]
  \centering
  \includegraphics[width=0.9\linewidth]{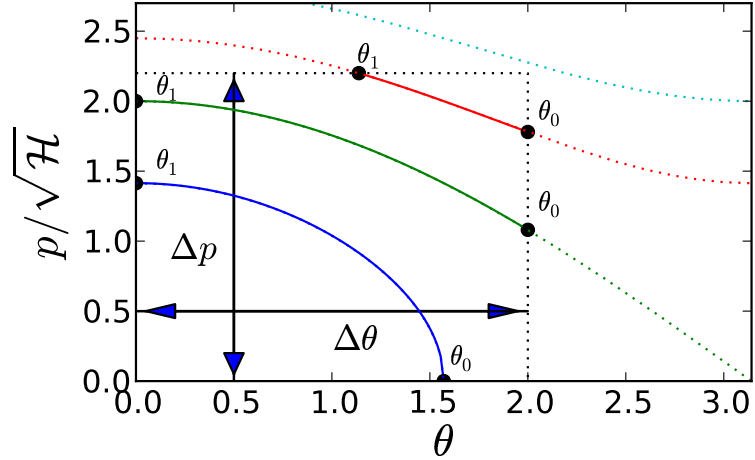}
  \caption{Graphical computation of Eqs.~(\ref{P(e>H)}) through (\ref{eq:pen-theta0}). The full lines represent the domain on which the integral is performed in order to compute $P_\epsilon$ for values of the energy $\epsilon=0,\H$ and $2\H$. The boundaries $\theta_0$ and $\theta_1$ are displayed for each energy.}
  \label{fig:pen-integ-wb}
\end{figure}

The integration over $\theta$ needs to be done on a domain enclosed in the original waterbag, or explicitly:
\begin{equation}
0 \le \epsilon+\H\cos\theta \le \frac{1}{2}\Dp^2 \quad \textrm{,} \quad |\theta| \leq \Delta\theta \quad \textrm{and} \quad |p| \leq \Dp
\end{equation}
These conditions lead to two solutions, depending on the ordering of $\epsilon$ with respect to $\H$. For $\epsilon \ge \H$, the distribution function is given by~:
\begin{equation}\label{P(e>H)}
P_{\epsilon}(\epsilon,\Dth,\Dp)=\frac{\sqrt{2}}{4\Dth\Dp}
\int_{-\Delta\theta}^{\Delta\theta} d \theta
~\frac{1}{\sqrt{(\epsilon+\H\cos\theta)}}-\frac{\sqrt{2}}{4\Dth\Dp}
\int_{-\theta_1}^{\theta_1} d \theta
~\frac{1}{\sqrt{(\epsilon+\H\cos\theta)}}~,
\end{equation}
while for $\epsilon \le \H$ it is
\begin{equation}\label{P(e<H)}
P_{\epsilon}(\epsilon,\Dth,\Dp)=\frac{\sqrt{2}}{4\Dth\Dp}
\int_{-\theta_0}^{\theta_0} d \theta
~\frac{1}{\sqrt{(\epsilon+\H\cos\theta)}}-\frac{\sqrt{2}}{4\Dth\Dp}
\int_{-\theta_1}^{\theta_1} d \theta
~\frac{1}{\sqrt{(\epsilon+\H\cos\theta)}}~.
\end{equation}
Here $\theta_1$ satisfies
\begin{equation}
  \label{eq:pen-theta1}
  \left\{\begin{array}{r l}
      \epsilon+\H\cos\theta_1=\Dp^2/2 &, \textrm{ for } \epsilon \ge \Dp^2/2 - \H \\
      \theta_1 = 0 &, \textrm{ else.}
  \end{array}\right.
\end{equation}
%for $\epsilon + \H \ge \Dp^2/2$, otherwise $\theta_1=0$.
Similarly,
$\theta_0$ satisfies
\begin{equation}
  \label{eq:pen-theta0}
  \left\{\begin{array}{r l}
      \epsilon+\H\cos\theta_0=0 &, \textrm{ for } -\H < \epsilon < \H\cos\Delta\theta \\
      \theta_0 = 0 &, \textrm{ for } \epsilon \leq -\H \\
      \theta_0 = \Dth &, \textrm{ else.}
  \end{array}\right.
\end{equation}
The graphical computations corresponding to Eqs.~(\ref{P(e>H)}) through (\ref{eq:pen-theta0}) are represented in Fig.~\ref{fig:pen-integ-wb}. $P_\epsilon$ corresponds to the length of the full lines, whose boundaries $\theta_1$ and $\theta_0$ are also displayed in the figure.

% In order to calculate the marginal distributions of $p$ and $\theta$
% we first note that the initial joint distribution $P^0(\theta, p)$
% is given by
% %
% %
% \begin{equation}
% P^0(\theta, p)=\bar P_\epsilon(\epsilon(\theta,p),\Dth,\Dp)~.
% \end{equation}
In the steady state, the $(\theta, p)$ distribution is such that for
any given energy, all the micro-states corresponding to that energy are
equally probable, with the boundaries given by the initial waterbag 
no longer taken into account . Thus the steady state distribution $P(\theta, p)$
may be expressed as
\begin{equation}
P(\theta, p)=\frac{1}{4\Dth\Dp}\frac{P_\epsilon(\epsilon(\theta,p),\Dth,\Dp)}{Q_{\epsilon}(\epsilon(\theta,p),\Dth,\Dp)}
\equiv\bar{P}_\epsilon(\epsilon(\theta,p),\Dth\Dp)~,
\end{equation}
where $Q_{\epsilon}(\epsilon(\theta,p),\Dth,\Dp)$ is given by
$P_\epsilon(\epsilon(\theta,p),\Dth,\Dp)$, as evaluated by Eq.~(\ref{P(e<H)}), except that one takes
\begin{equation}
  \theta_1=0
\end{equation}
and
\begin{equation}
  \theta_0 = \left\{\begin{array}{l l}
      \pi & \textrm{if } \epsilon \geq \H, \\
      \arccos(\Dth) & \textrm{else}, 
    \end{array}\right.
\end{equation}
in these integrals. The graphical computation of $Q_\epsilon$ is depicted in Fig.~\ref{fig:pen-integ-wb}. $Q_\epsilon$ correspond to the length of the dotted lines.

The value of $\bar P_\epsilon$ may be interpreted, with reference to Fig.~\ref{fig:pen-integ-wb}, as the ratio of the length of the full line by the length of the dotted line, computed for each value of $\epsilon$.
One can observe that for small values of the energy $\epsilon$, the line is full on the whole length, therefore $P_\epsilon=\left(4\Dth\Dp\right)^{-1}$. On the other hand, for $\epsilon$ ``above'' the waterbag, $\bar P_\epsilon=0$.

The $\theta$ marginal may be evaluated by integration over $p$.
\begin{equation}
P_\theta(\theta)= \int dp ~P(\theta, p)~=~\int d\epsilon ~ P(\theta,
p) \frac{dp}{d\epsilon}~=~\int d\epsilon ~
\bar{P}_\epsilon(\epsilon) \frac{dp}{d\epsilon}~.
\end{equation}
Hence, by expressing the derivative in terms of $\epsilon$ and
$\theta$, one finds
\begin{equation}
P_\theta(\theta)=\sqrt{2} \int_{-H\cos\theta}^{\infty} d \epsilon ~
\frac{1}{\sqrt{(\epsilon+H \cos
\theta)}}~\bar{P}_\epsilon(\epsilon)~.
\end{equation}

Similarly, the $p$ marginal is given by
\begin{equation}
P_p(p)=\int
d\epsilon~\bar{P}_\epsilon(\epsilon)\frac{d\theta}{d\epsilon}~.
\end{equation}
Expressing the derivative in terms of $\epsilon$ and $p$ one finally
obtains
\begin{equation}
P_p(p)=\int_{p^2-H}^{p^2+H} d\epsilon~
\frac{1}{H\sqrt{1-\frac{1}{H^2}(\epsilon-\frac{p^2}{2})}}~\bar{P}_\epsilon(\epsilon)
\end{equation}

The elliptic integrals are performed numerically using the specialized routine of Ref.~\cite{NR_in_f90}.

\subsection{Check against a simulation}
\label{sec:pendulum_check}

The only hypothesis made in the derivation of the exact solution is that the filaments will be smoothed out by a simulation. We perform a Vlasov simulation to check the adequacy of our theory.

\begin{figure}[ht]
  \centering
  \includegraphics[width=0.8\linewidth]{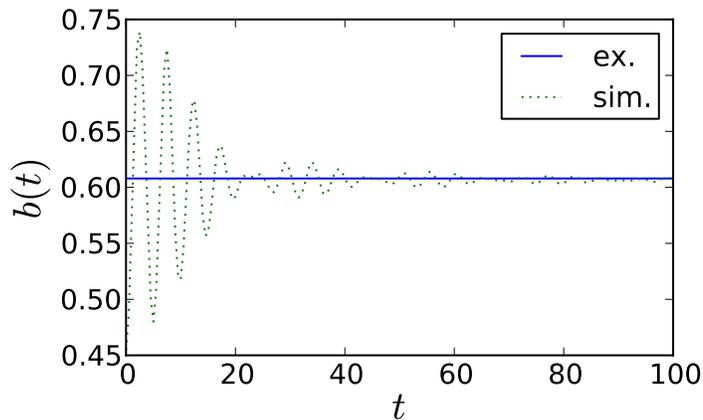}
  \caption{The bunching $b$ in the course of time for a Vlasov simulation with initial conditions $\Dth=2.0$, $\Dp=0.78$ and $\H=0.608$. After a transitory oscillations, $b$ relaxes to the predicted value.}
  \label{fig:pen_check_b}
\end{figure}

The bunching shows in Fig.~\ref{fig:pen_check_b} strong oscillations in a first stage, followed by a relaxation to the predicted value.

\begin{figure}[ht]
  \centering
  \begin{minipage}[h]{0.49\linewidth}
    \includegraphics[width=\linewidth]{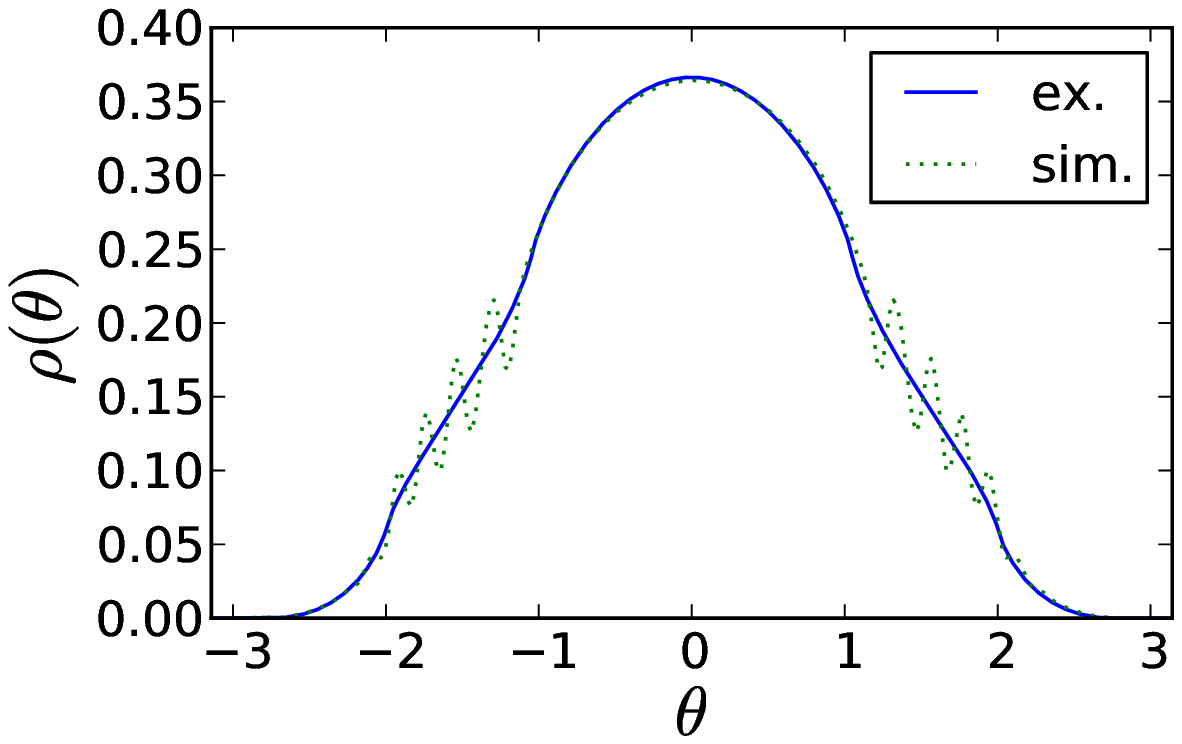}
    \caption{$\rho$ for the same simulation as Fig.~\ref{fig:pen_check_b}.}
    \label{fig:pen_check_rho}
  \end{minipage}
  \begin{minipage}[h]{0.49\linewidth}
    \includegraphics[width=\linewidth]{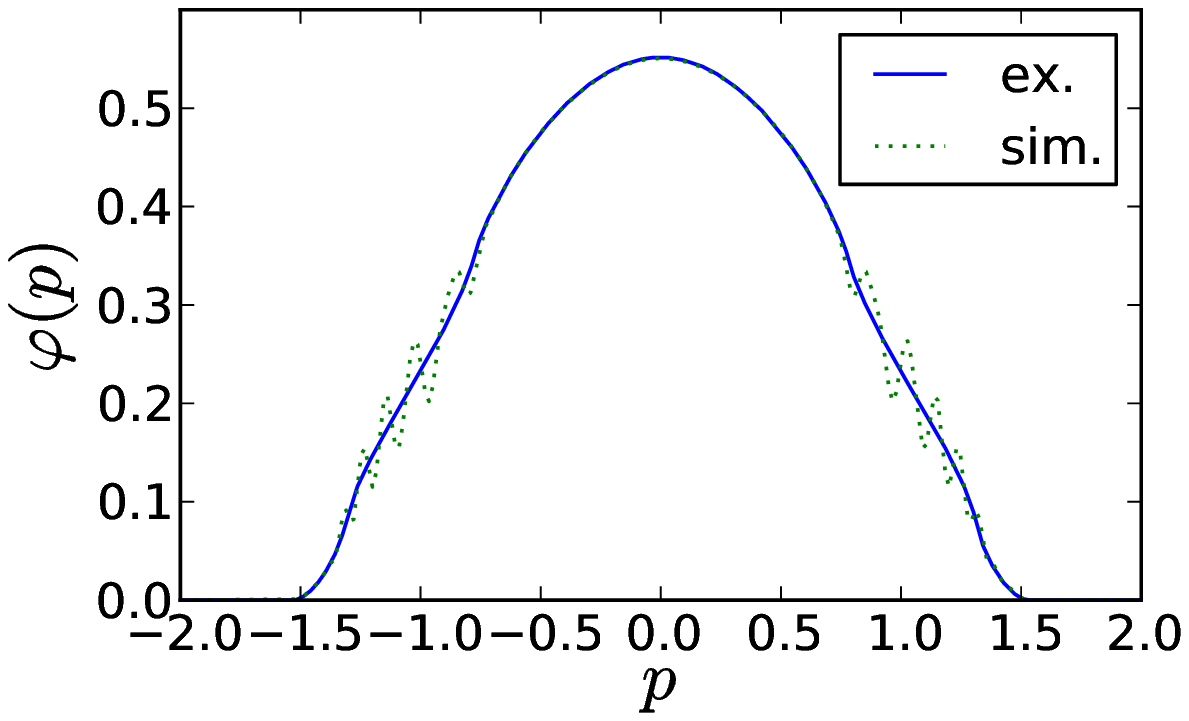}
    \caption{$\varphi$ for the same simulation as Fig.~\ref{fig:pen_check_b}.}
    \label{fig:pen_check_phi}
  \end{minipage}
\end{figure}

Figures~\ref{fig:pen_check_rho} and \ref{fig:pen_check_phi} depict $\rho$ and $\varphi$ respectively. As expected, apart from the signs of filamentation, the agreement is excellent.

\subsection{Back to the HMF model}
\label{sec:pendulum-linkHMF}

In the stationary regime for the HMF model, a comparison is possible with the set of pendula.
In order to proceed, one should evaluate the bunching
\begin{equation}
  \label{eq:pen_selfc}
b=\int_{-\pi}^{+\pi}d\theta~ \cos \theta P_{\theta}(\theta)~,
\end{equation}
and set $b=\H$, obtaining a self-consistent relation. Clearly $\H=b=0$ is a solution of this equation.
However one can in principle find another solution of this equation with non-zero magnetization $m$. To obtain these results, we use an iterative root finding routine (Brent's method, see Ref. \cite{NR_in_f90}), setting $\Dth$ and $\Dp$. The routine then seeks a root of $\int_{-\pi}^{+\pi}d\theta~ \cos \theta P_{\theta}(\theta) -\H$, where $\H$ is the variable.

The properties of both solutions are then cast in the HMF setting (where $\H=b$ is then called the magnetization $M$), allowing to use our exact results.
Especially, the behavior of Eq.~(\ref{eq:pen_selfc}) can possibly display a transition between regions of parameters with two solutions ($b=0$ and $b\neq 0$) and regions where only $b=0$ is possible.
\begin{figure}[ht]
  \centering
  \includegraphics[width=0.7\linewidth]{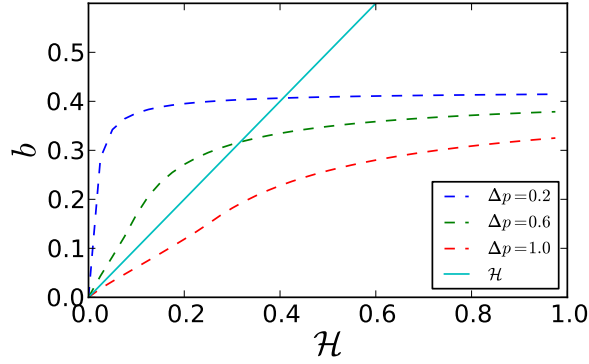}
  \caption{Graphical explanation of Eq.~(\ref{eq:pen_selfc}) in the case $\Dth=3.0$. The intersection between the diagonal and the computed $b$ gives the solution(s) to Eq.~(\ref{eq:pen_selfc}). For low $\Dp$ there can be two solutions ($\H =0$ and $\H\neq 0$) while above a threshold only one of the solutions exists.}
  \label{fig:pen_trans_300}
\end{figure}
Figure~\ref{fig:pen_trans_300} represents the computation of Eq.~(\ref{eq:pen_selfc}) for $\Dth=3.0$ and several values of $\Dp$. The intersection between $b$ and $\H$ gives the solution to the self-consistent condition.

The value of $\H$ given by Eq.~(\ref{eq:pen_selfc}) allows the computation of the kinetic and interaction contributions to the energy in the initial state:
\begin{equation}
  U = \frac{\Dp^2}{6} - \H b_0 \quad , \quad K_0 = \frac{\Dp^2}{6} ,
\end{equation}
and in the final state:
\begin{equation}
  U = K_\infty - \H^2 \quad .
\end{equation}
The conservation of $U$ allows us to compute the kinetic part in the final state, $K_\infty$, explicitly:
\begin{equation}
  K_\infty = K_0 - \H b_0 + \H^2 \quad .
\end{equation}

To obtain an identical asymptotic state for the HMF model, its kinetic energy must be equal to $K_\infty$ and its magnetization to $\H$. The energy for the HMF model is thus~:
\begin{equation}
  U_{HMF} = K_\infty + \undemi \left( 1 - \H^2 \right) \quad .
\end{equation}
In addition to that energy, we impose that the height of the waterbag, $f_0$, be the same for the two systems, as this quantity is an invariant of the dynamics.

Explicitly, finding the set of $\Dth$ and $\Dp$ amounts to solve the following equation:
\begin{equation}
  \frac{\Delta p^2}{6} + \frac{1}{2} (1 - \frac{\sin\Delta\theta}{\Delta\theta}) = U_{HMF} \quad ,
\end{equation}
that transforms, making use of $f_0 = \frac{1}{4 \Delta\theta \ \Delta p}$, into
\begin{equation}
  \label{eq:pen-Ur}
  \frac{1}{96 f_0^2 \Delta\theta^2} + \frac{1}{2} (1 - \frac{\sin\Delta\theta}{\Delta\theta}) = U_{HMF} \quad .
\end{equation}

We now present a comparison of the exact solution for the uncoupled pendula with a simulation of the HMF model.
We consider a pendulum with an initial condition given by $\Dth=1.66$ and $\Dp=0.5$. The resolution of the self-consistent problem for the exact solution gives $b=0.767$.

The initial condition for the HMF model, computed according to Eq.(\ref{eq:pen-Ur}), is defined by $\Dth=0.523$ and $\Dp=1.390$. The parameters for the simulation are $N_\theta=N_p=256$ and $\Delta t=0.1$.

\begin{figure}[ht]
  \centering
  \includegraphics[width=.8\linewidth]{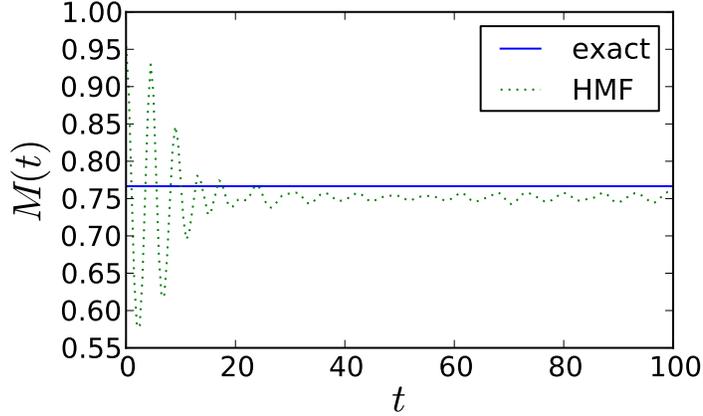}
  \caption{Evolution of the magnetization $M(t)$ in the HMF model in the course of time for the initial condition $\Dth=0.523$, $\Dp=1.390$. The value predicted for the corresponding exact solution is displayed for reference.}
  \label{fig:pen-HMF-M}
\end{figure}
Figure~\ref{fig:pen-HMF-M} displays $M(t)$ for the HMF simulation. $M(t)$ begins by oscillating strongly, then attains a value close to the one predicted by the theory. This similarity between the uncoupled pendula system and the HMF model comes a little bit as a surprise, given the difference between the two systems. We note that the value of $M$ attained in the simulation is far from the original value $M_0=0.955$, meaning that we can expect a better agreement when the final and the initial magnetizations are closer, in a similar way to the results for Lynden-Bell's theory.

\begin{figure}[ht]
  \centering
  \begin{minipage}[h]{0.49\linewidth}
    \includegraphics[width=\linewidth]{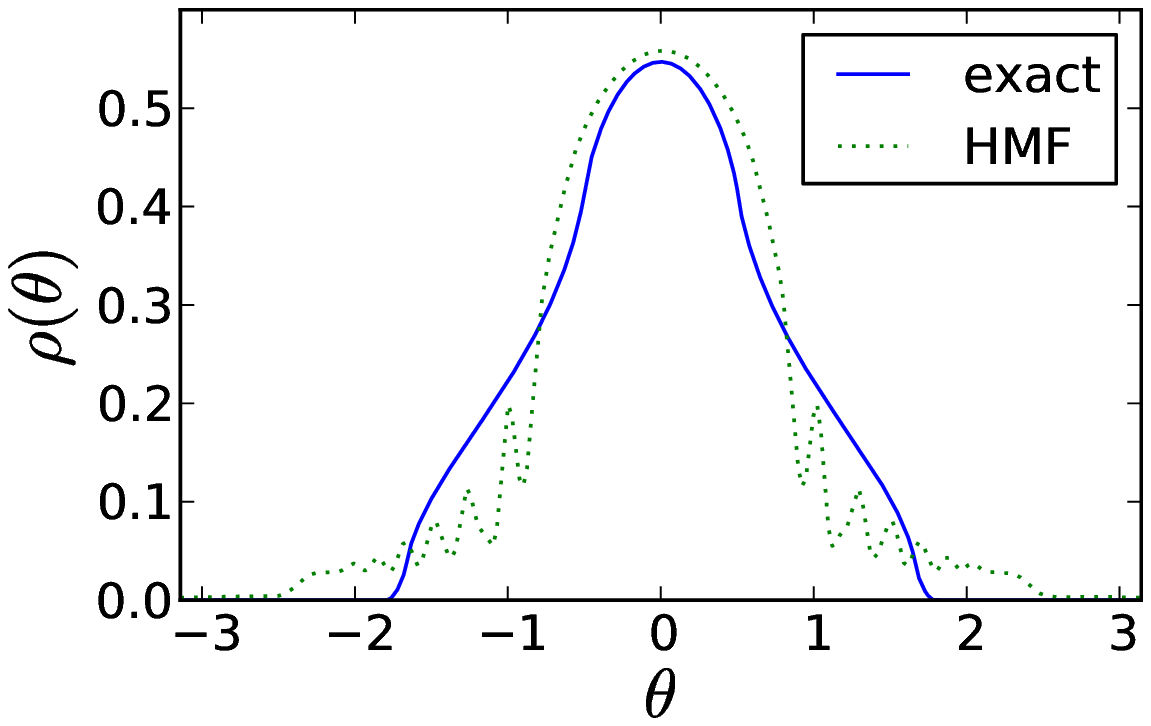}
  \end{minipage}
  \begin{minipage}[h]{0.49\linewidth}
    \includegraphics[width=\linewidth]{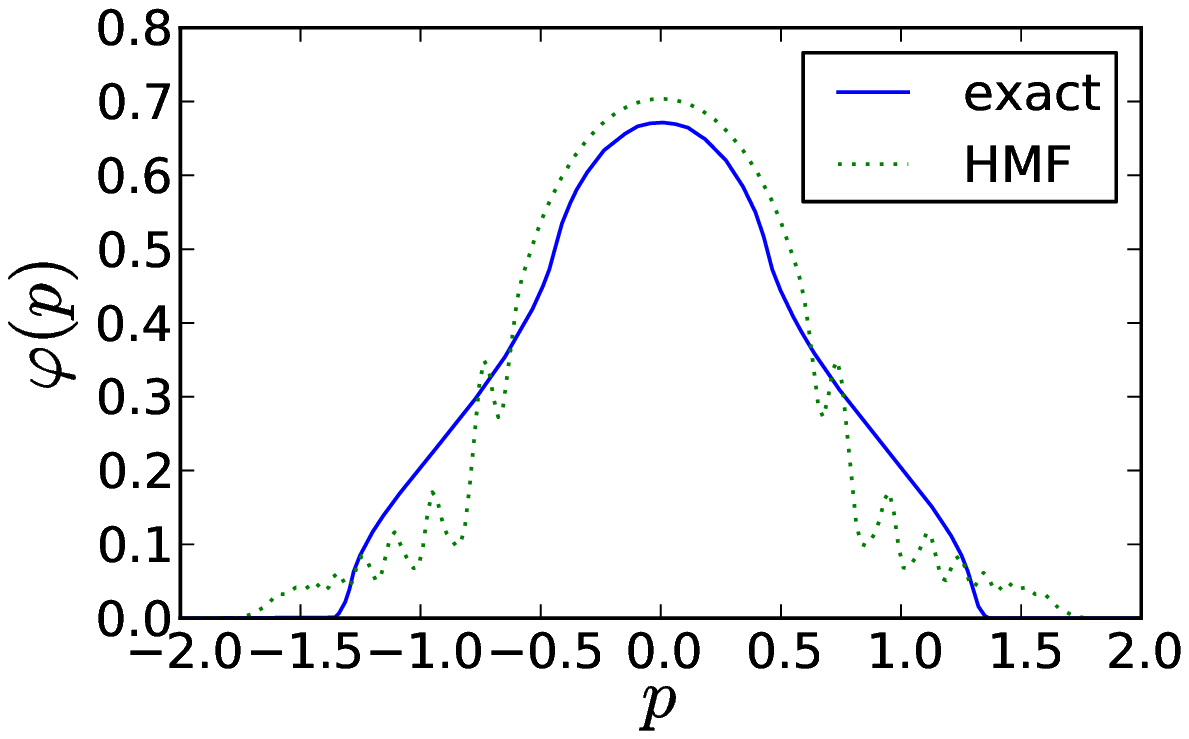}
  \end{minipage}
%\end{figure}
%\begin{figure}[ht]
  \centering
  \begin{minipage}[h]{0.49\linewidth}
    \caption{(left panel) Comparison of $\rho(\theta)$ for a simulation of the HMF model and the exact solution for the corresponding system of pendula. See text for details.}
    \label{fig:pen-rho-HMF}
  \end{minipage}
  \begin{minipage}[h]{0.49\linewidth}
    \caption{(right panel) Same as Fig.~\ref{fig:pen-phi-HMF}, but for $\varphi(p)$.\bigskip\bigskip\smallskip\smallskip\smallskip}
    \label{fig:pen-phi-HMF}
  \end{minipage}
\end{figure}

For completeness, we also display the marginals for the two systems in Figs.~\ref{fig:pen-rho-HMF} and \ref{fig:pen-phi-HMF}. The general aspect of the curves is similar but a quantitative agreement is not present.

\subsection{Occurrence of a phase transition}
\label{sec:pen-ex-trans}

Equation~(\ref{eq:pen_selfc}) always admit a solution where $\H=b=0$. It may possess a solution where $\H$ and $b$ are finite. The passage from a region of parameters $\Dth$ and $\Dp
$ where two solutions exist to one where only the homogeneous solution is present is similar to a phase transition.
This transition is found for instance in Ref.~\cite{leoncini_et_al_epl_2009}.

\begin{figure}[ht]
  \centering
  \includegraphics[width=.8\linewidth]{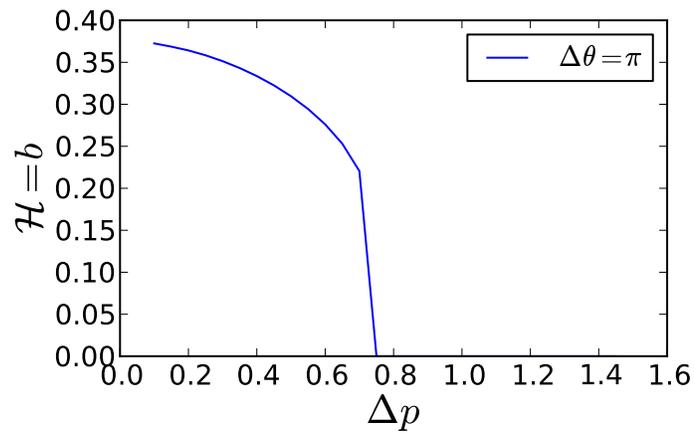}
  \caption{Representation of the solution of Eq.~(\ref{eq:pen_selfc}) for $\Dth=\pi$ in function of $\Dp$. We observe a transition from $\H=b$ finite to $\H=b=0$.}
  \label{fig:pen-trans}
\end{figure}

We illustrate this behavior in Fig.~\ref{fig:pen-trans}. Above a threshold in $\Dp$, $\H=b$ drops to $0$.

\clearpage
\subsection{Discussion}
\label{sec:pen-summary}

In this chapter, we have proposed a Vlasov equation for a system similar to the Hamiltonian Mean-Field (HMF) model but without interactions.
We elaborated Lynden-Bell's theory for that system and found a remarkable agreement for the predicted value of the bunching, in a range of parameters.

With the help of the energy probability distribution functions (energy PDF), we found that a factor explaining the effectiveness of the theory is the similarity between the initial energy PDF and the one predicted by the theory.
This finding is an important addition to the understanding of violent relaxation in which incomplete relaxation is observed, as it confirms the adequacy of Lynden-Bell's theory for a wider range of applications than expected.

Then we analyzed an exact solution of the asymptotic evolution of the Vlasov equation, under an ergodic-like hypothesis. This solution shows some similarity with the asymptotic state of the HMF model, especially with respect to the value of the bunching. Solving a self-consistency relation, we also displayed a transition between two regions of parameters in which the relation possesses either one or two solutions.
Understanding this transition with more detail is of great importance to the understanding of transitions in related mean-fields models such as the HMF model and the Free-Electron Laser.

% Local Variables:
% TeX-master: "main"
% End:

\cleardoublepage

\chapter{Conclusions and perspectives}
\label{chap:conclusions}

We performed in this thesis the investigation of toy models in which the interaction potential is long ranged, namely the Hamiltonian Mean-Field model (HMF) and the Colson-Bonifacio model for the free-electron laser (FEL). Their dynamical description in the thermodynamic limit is cast with the help of the Vlasov equation.

We reported the use of a numerical procedure to compute explicitly the evolution of these models. This represents the first detailed numerical computation for the Vlasov equation for simple models of long-range interactions.
The present work puts emphasis on the interest of these computations for both domains of research~: the numerical resolution of the Vlasov equation {\it and} the theoretical study of the Vlasov equation for models with long-range interactions.

The occurrence of quasi-stationary states (QSS) and out-of-equilibrium phase transitions presents a conceptual challenge to the understanding of kinetic theory. At the present time, the statistical theory of Lynden-Bell provides a correct predictive and interpretative framework for the Vlasov dynamics. It has been reported to describe successfully the asymptotic dynamics of the HMF model and of the FEL in a number of situations.

\bigskip

The study of the FEL via numerical simulations of the Vlasov equation provides a detailed description of the out-of-equilibrium dynamics. The intensity of the laser field produced by the FEL device takes significantly different values depending on the initial condition. Specializing on waterbag initial conditions, we described the out-of-equilibrium phase transition occurring between these regimes.
We interpreted further this transition in term of Lynden-Bell's theory, but acknowledge that in the case of the FEL it does not yet provides a fully predictive theory.
A direct observation of the structures in phase space allowed us to understand a shortcoming of the theory, as dynamical features cannot be understood within its framework. An original application of Vlasov simulations coupled to Poincar{\'e} section of discrete particles is introduced and is expected to yield accurate results on the phase space properties of the dynamics. The occurrence of period-2 and of higher periods orbits is anticipated.

\bigskip

We then turned to the investigation of the origin of relaxation in Vlasov dynamics for the HMF model. The use of an analogy with fluid dynamics allows to quantify the deformations experienced by an initial waterbag~: the evolution in time of the perimeter of the fluid gives information on the microscopic dynamics in phase space. The occurrence of stretching and folding structures, in relation with the filamentation, provides a new point of view on the collisionless evolution and its approach to equilibrium.
We performed the numerical computation of the perimeter in a situation and found it to be exponential. The high level of details required to compute the perimeter required the use of high accuracy simulations.

\bigskip

The analogy between the HMF model and a set of uncoupled pendula is cast in the last chapter. We introduce the Vlasov equation corresponding to that system.
Lynden-Bell's theory is developed and investigated in this context and is shown to predict an out-of-equilibrium phase transition under a self-consistency hypothesis. The results of Lynden-Bell's theory are encouraging, which is surprising for a system in which there is no interaction, hence no source of relaxation. A careful analysis reveals that the theory predicts good results when the initial condition presents an energy distribution function close to the one predicted by Lynden-Bell's theory.

An exact computation of the asymptotic dynamics for this Vlasov equation is then performed. It matches exactly the simulation results up to a small filamentary structure.
This system, under the self-consistency hypothesis, also displays a phase transition.
A comparison is drawn with the HMF model, and we expect that the full understanding of the exact system will yield more insight on the out-of-equilibrium phase transition occurring in the HMF model or in the FEL.

\bigskip

In the light of the results presented in this thesis, a number of questions remain open. The exact characterization of QSSs is yet to be achieved. Especially, the dynamical features appearing in theses QSSs are not interpreted in terms of kinetic theory. The fine structures found in phase space have an impact on the macroscopic dynamics that needs to be investigated.

The detailed presentation of numerical computations calls for comparisons with other methods with the hope that theoretical investigation on toy models may contribute to the progress of simulational methods.

\bigskip

Finally, we would like to mention a number of encouraging results that point to possible experimental realizations of systems with long-range interactions.

The out-of-equilibrium phase transition occurring in the Free-Electron Laser has the potential to be realized experimentally.
A model for collective atomic recoil lasing (CARL) possesses a mathematical description close to the one of the FEL \cite{bonifacio_et_al_CARL_pra_1994}. Experiments on the CARL may offer more flexibility and provide a new opportunity to investigate the effects of long-range interactions \cite{slama_et_al_assisi_2007,campa_et_al_phys_rep_2009}.

The dipolar interaction in spin structures is also a candidate towards the experimental realization of a system with long-range interaction. The authors of Ref.~\cite{campa_et_al_prb_2007} propose a setup in which the statistical mechanics of the system can display ensemble inequivalence, negative specific heat or temperature jumps.

% Local Variables:
% TeX-master: "main"
% End:

\appendix

\cleardoublepage
\chapter{Stability of the homogeneous waterbag}
\label{sec:stabilitywb}

This Appendix is devoted to reviewing the stability condition 
of system (\ref{eq:vlasovwave}) for an initial homogeneous distribution of the waterbag type. 
The derivation follows from a straightforward linear analysis which can be found for
instance in \cite{elskens_escande_book}. We shall hereafter make reference to the calculation detailed
in \cite{bachelard_fanelli_cnsns_2010}. Let us start by assuming a general 
equilibrium setting where the spatial distribution is homogeneous ($A_x=A_y=0$)
and $f=f_0(p)$, i.e. a generic function of the variable $p$. Then one can
linearize around the equilibrium and eventually derive an explicit 
solution which holds for a relatively short time. To this end we write:

\begin{eqnarray}\label{pertubrsolu}
  &f(\theta,p,t) &=f_0(p)+f_1(\theta,p,t),\cr
  &A_x(t)&=X_1(t)\quad\mbox{and}\quad A_y(t)=Y_1(t)\quad.
\end{eqnarray}
where the quantities labeled with the index $1$ stand for the linear 
perturbation. Introducing in system~(\ref{eq:vlasovwave}) and retaining the lowest order yields:

\begin{eqnarray}
 (\partial_{\bar{z}}+p\partial_\theta)f_1 -2\eta(X_1 \cos \theta&-Y_1 \sin \theta)&=0\label{deuxsolia}\\
 \int_{-\pi}^{\pi}\!\! d \theta\!\!\int_{-\infty}^{+\infty}\!\! d p\  f_1\cos \theta&
 -\frac{d X_1}{d \bar{z}} &=0\label{deuxsolib}\\
  \int_{-\pi}^{\pi}\!\! d \theta\!\!\int_{-\infty}^{+\infty}\!\! d p\  f_1\sin \theta&
 +\frac{d Y_1}{d \bar{z}} &=0\label{deuxsolic}
 \end{eqnarray}
where we introduced $\eta(p)=\partial_p f_0(p)$. The above linear system admits a
solution in terms of normal modes:

\begin{eqnarray}
f_1(\theta,p,\bar{z})&=&F_1(p)\, e^{i(\theta-\omega \bar{z})}+
F_1^*(p)\, e^{-i(\theta-\omega^* \bar{z})}\label{solf1}\\
X_1(\bar{z})&=&X_1\, e^{-i\omega \bar{z}}+X_1^*\, e^{i\omega^* \bar{z}}\\
Y_1(\bar{z})&=&iY_1 \, e^{-i\omega \bar{z}}-iY_1^*\, e^{i\omega^*
\bar{z}}\quad.\label{solY1}
\end{eqnarray}
where the symbol $*$ refers to the complex conjugate and in general 
$\omega\in \mathbf{C}$. Making use of the above ansatz in the linearized system
of equations returns the following consistency equation:

\begin{equation}\label{dispersionfinal}
\omega =  \int_{-\infty}^{+\infty}\!\! dp
\,\frac{\partial_p f_0}{p-\omega}\quad
\end{equation}
often referred to as the dispersion relation. To determine whether a given
distribution $f_0(p)$ is stable or unstable, one can solve the above dispersion
relation and estimate the sign of the imaginary part of  $\omega$. Depending
on the sign the field grows exponentially (instability) or 
oscillates indefinitely (stability). If the selected initial condition is
parametrized via an adjustable parameter, one can then calculate the
corresponding theshold value which discriminates between stable and 
unstable regimes. This analytical procedure can be persecuted in simple cases,
as the one addressed in this appendix (the waterbag). For more complicated
situations one can resort to the celebrated Nyquist method, first introduced in   
plasma physics \cite{nyquist_1932} (see also \cite{chavanis_delfini_epjb_2009}). For the case at hand,
$f_0(p)$ takes the form:

\begin{equation}\label{ic}
f_0(p)= \frac{1}{2 \pi} \frac{1}{2 \Delta p} \left[
\Theta(p+\Delta p) - \Theta(p-\Delta p) 
\right]
\end{equation}
where $\Theta$ stands for the Heaviside function. Inserting (\ref{ic})
into (\ref{dispersionfinal}), carrying out the integral explicitly and looking
for the value of $\Delta p$ which sets the transition between complex and real
$\omega$, leads to $\Delta p \simeq 1.37$ or equivalently 
$\epsilon = (\Delta p)^2 / 6 \simeq 0.315$.

% Local Variables:
% TeX-master: "main"
% End:

\cleardoublepage

\chapter{Routines}
\label{chap:routines}

We review in this appendix technical aspects of the computations presented in this thesis.

\section{The simulation program}
\label{sec:routines-prog}
\lstset{language=Fortran}

The simulation program used to perform the Vlasov simulations has been written in the course of the thesis.
It makes use of modern programming concept applied to the Fortran~90 language in the spirit of Ref.~\cite{decyk_et_al_inheritance_f90_1998}.

The program is based on a Fortran~90 module that contains the description of a grid in phase space.
The main element of the module is the derived type ``grid'' that contains the basic physical values needed to perform the simulation. The derived type grid is shown in listing~\ref{lst-grid}.
A number of subroutines makes use of ``grid'' and perform operations such as setting an initial condition, compute the marginal distributions, perform the advection steps in both directions of phase space on the basis of spline interpolation, \ldots

Specific modules for the Hamiltonian Mean-Field (HMF) model and the Free-Electron Laser combine ``grid'' with additional model-specific variables (the magnetization for the HMF model for instance).
The Fortran~90 module contains in addition subroutines that write the data in a systematic way in hdf5 files. hdf5 is a structured file format available at http://www.hdfgroup.org/HDF5/ .

\begin{lstlisting}[breaklines=true, breakindent=0.5\linewidth,float=ht,caption={Derived data type used in the Fortran~90 simulation program.},label=lst-grid]
type grid
  integer :: Nx, Nv           ! The number of grid points in the space and the velocity directions
  double precision :: xmin, xmax, vmin, vmax
  !                           ! The boundaries of phase space
  double precision :: dx, dv  ! The step of the grid in the space and the velocity directions
  double precision :: DT      ! The time step
  
  double precision, allocatable :: f(:,:)
  !                           ! The array containing f.
  double precision, allocatable :: f2(:,:), g(:,:)
  !                           ! The arrays containing the second derivatives of f needed to perform the spline interpolation (f2) and a copy of f (g).
  double precision, allocatable :: rho(:), phi(:)
  !                           ! The arrays containing the marginals in the x and in v.
end type grid
\end{lstlisting}

\section{Simulated annealing}
\label{sec:routines-SA}
\lstset{language=Fortran}

The simulated annealing algorithm (SA) is a general optimization algorithm \cite{kirkpatrick_et_al_simulated_annealing}. It consists of a random sampling in the space of parameters in which each moves is accepted or rejected on the basis of a cost function.
It is used in this thesis to find ``candidate roots'' to the set of equations given by Lynden-Bell's theory.

The SA algorithm accepts any move that decreases the cost function. For the moves that increase the cost function, the acception is based on the following criterion~:
\begin{equation}
  \exp \frac{- \Delta \textrm{cost\_function}}{T} \geq \xi
\end{equation}
where $\textrm{cost\_function}$ is the cost function, $T$ is the effective temperature and $0<\xi\leq 1$ is a random number. The structure of such a program is given in listing~\ref{lst-SA}.

\begin{lstlisting}[breaklines=true, breakindent=0.5\linewidth,float=ht,caption={Principle of the simulated annealing algorithm.},label=lst-SA]
do i=1,n_steps
  T = T_program(i)            ! T_program is the programmed temperature.
  oldx = x                    ! We keep the old parameter in case of rejection.
  call move(x, T)             ! We perform a random move of the parameter x.
  cost = cost_function(x)     ! cost_function is a cost function defined by the user
  if (cost.leq.oldcost) then
    ! We accept the move if the cost decreases.
    oldcost = cost
  else
    if (exp( - (cost - oldcost) / T).gt.random_number())
      then
        oldcost = cost
      else
        ! else we reject the move
        x = oldx
    end if
  end if
end do
\end{lstlisting}

\begin{figure}[ht]
  \centering
  \includegraphics[width=.8\linewidth]{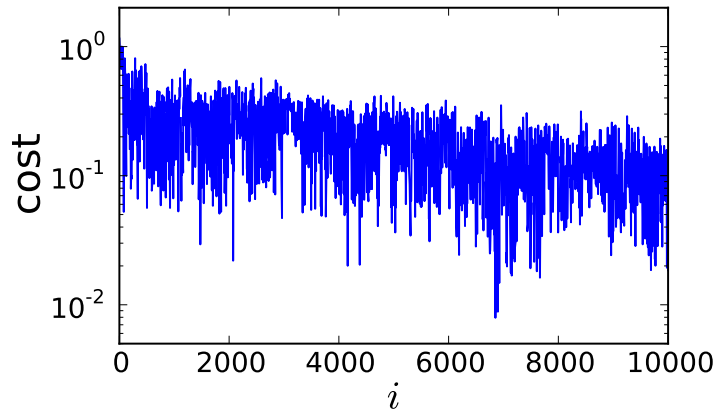}
  \caption{Evolution of the minimum cost in a series of 16 runs of simulated annealing. The computation is performed for the set of equations~(\ref{eq:pendulum_LB}) with parameters $\Dth=2.5$ and $\Dp=0.4$.}
  \label{fig:rout-i-cost}
\end{figure}

Performing a series of SA runs allows one to find candidates for the solution of the set of equations defined by Lynden-Bell's theory.
An example is given in Fig.~\ref{fig:rout-i-cost}~: The minimum of 16 runs for the same parameters is displayed. This minimum decreases in the course of the iterations. The programmed temperature is defined as $T(i) = \exp{-\frac{2.5\ i}{n\_steps}}$.

% \section{Technical acknowledgments}

% From a technical point of view, the author of this thesis wishes to acknowledge~:
% \begin{itemize}
% \item The open source libraries numpy, scipy, matplotlib and h5py. They have proven of high quality and indispensible to my work.
% \item The eternal gnuplot.
% \item ``This research has made use of NASA's Astrophysics Data System''.
% \item The arXiv.org preprint servers that allows a fast communication of scientific results.
% \item The webpage ``http://www.cs.rpi.edu/\~{}szymansk/oof90.html'' that gave me a fresh point of view on the Fortran~90 programming language.
% \end{itemize}

% Local Variables:
% TeX-master: "main"
% End:

\cleardoublepage\addcontentsline{toc}{chapter}{Bibliography}\bibliographystyle{apalike}\bibliography{books,confs,these}\end{document}